\documentclass{jfm}
\usepackage{graphicx}
\usepackage{amsmath}
\usepackage{relsize}
\usepackage{amssymb}  
\usepackage{dcolumn}
\usepackage{bm}
\usepackage{bigints}
\usepackage{enumitem}
\usepackage{float}
\usepackage{verbatim}
\usepackage{makecell}
\usepackage{enumitem}
\usepackage[export]{adjustbox}
\usepackage{scalerel}[2016-12-29]
\usepackage[dvipsnames]{xcolor}
\usepackage{multirow,bigdelim}
\usepackage{lscape}
\usepackage{tikz}
\usepackage{siunitx}
\usepackage{cancel}
\usepackage{lineno}

\usepackage[colorlinks=true, linkcolor=blue, urlcolor=blue, citecolor=blue]{hyperref}

\newcount\ndots
\def\drwln#1#2{\raise 2.5pt\vbox{\hrule width #1pt height #2pt}}
\def\spc#1{\hskip #1pt}
\def\solid{\drwln{18}{1}\ }

\def\dashed{\hbox {\drwln{4}{1}\spc{2}
                   \drwln{4}{1}\spc{2}\drwln{4}{1}}\nobreak\ }
\def\dasheddotted {\hbox {\drwln{4}{1.0}\spc{2}
                   \drwln{1}{1.0}\spc{2}\drwln{4}{1.0}}\nobreak\ }

\def\bdot{\hbox{\drwln{1}{.5}\spc{2}}}
\def\dotted{\hbox{\leaders\bdot\hskip 24pt}\nobreak\ }

\def\circle   {$\circ$\nobreak\ }

\def\filsqr   {${\vcenter{\hrule height 2pt
                          \hbox{\vrule width 2.2pt height 0.2pt \kern 0.1pt
                                \vrule width 2.2pt}
                                \hrule height 2.2pt}}$\nobreak\ }

\definecolor{forestgreen}{rgb}{0.13, 0.5, 0.13}

\definecolor{mygray}{gray}{0.6}
\definecolor{mypink3}{cmyk}{0, 0.7808, 0.4429, 0.1412}

\definecolor{battleshipgrey}{rgb}{0.52, 0.52, 0.51}                                
\definecolor{hotmagenta}{rgb}{1.0, 0.11, 0.81}
\definecolor{aqua}{rgb}{0.0, 1.0, 1.0}

\makeatletter
\newcommand{\ostar}{\mathbin{\mathpalette\make@circled{\color{aqua}\times}}}
\newcommand{\ostarr}{\mathbin{\mathpalette\make@circled{\color{magenta}\times}}}
\newcommand{\make@circled}[2]{%
  \ooalign{$\m@th#1\smallbigcirc{#1}$\cr\hidewidth$\m@th#1#2$\hidewidth\cr}%
}
\newcommand{\smallbigcirc}[1]{%
  \vcenter{\hbox{\scalebox{0.77778}{$\m@th#1\bigcirc$}}}%
}

\makeatother

\begin{document}
\setlength{\unitlength}{1.2cm}

\title{Spanwise wall forcing can reduce turbulent heat transfer more than drag}
\shortauthor{Rouhi, Hultmark and Smits}

\shorttitle{Spanwise forcing reduces heat transfer more than drag}

\author{Amirreza Rouhi$^{1}$ \corresp{\email{amirreza.rouhi@ntu.ac.uk}}, Marcus Hultmark$^{2}$, \and Alexander J. Smits$^2$}
\affiliation{$^1$Department of Engineering, School of Science and Technology\\
                Nottingham Trent University, Nottingham NG11 8NS, United Kingdom\\[5pt]                
             $^2$Department of Mechanical and Aerospace Engineering,
             Princeton University,\\ Princeton, NJ 08544, USA}
\maketitle


\begin{abstract}
Direct numerical simulations are performed of turbulent forced convection in a half channel flow with wall oscillating either as a spanwise plane oscillation or to generate a streamwise travelling wave. The friction Reynolds number is fixed at $Re_{\tau_0} = 590$, but the Prandtl number $Pr$ is varied from 0.71 to 20.  For $Pr>1$, the heat transfer is reduced by more than the drag, 40\% compared to 30\% at $Pr=7.5$. This outcome is related to the different responses of the velocity and thermal fields to the Stokes layer. It is shown that the Stokes layer near the wall attenuates the large-scale energy of  the turbulent heat-flux and the turbulent shear-stress, but  amplifies their small-scale energy. At higher Prandtl numbers, the thinning of the conductive sublayer means that the energetic scales of the turbulent heat-flux move closer to the wall, where they are exposed to a stronger Stokes layer production, increasing the contribution of the small-scale energy amplification.  A predictive model is derived for the Reynolds and Prandtl number dependence of the heat-transfer reduction
based on the scaling of the thermal statistics. The model agrees well with the computations for Prandtl numbers up to 20.

\end{abstract}




\section{Introduction}\label{sec:intro}
\begin{figure}
  \centering
 \includegraphics[width=0.66\textwidth,trim={{0.0\textwidth} {0.0\textwidth} {0.6\textwidth} {0.0\textwidth}},clip]{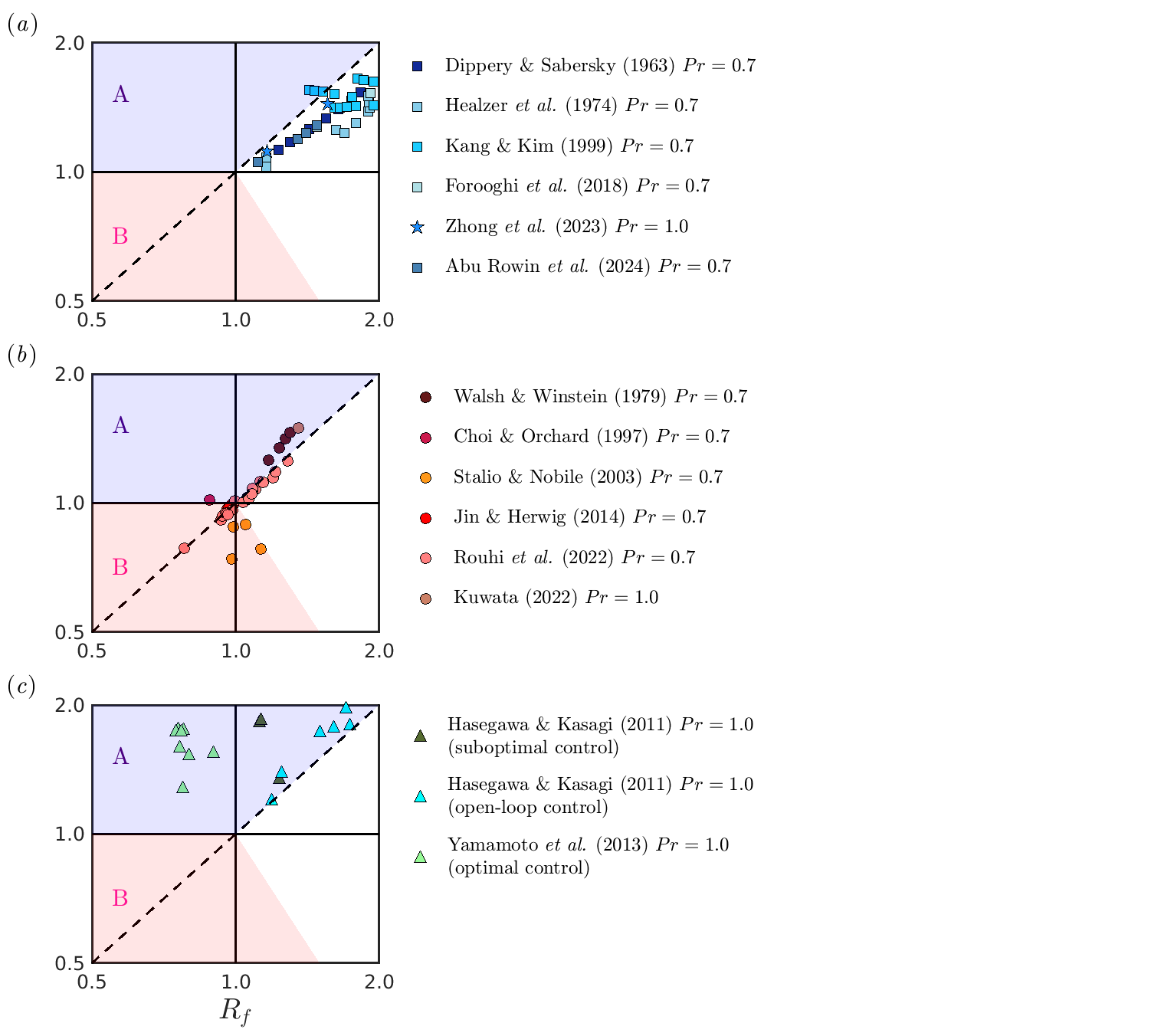}
 \includegraphics[width=0.32\textwidth,trim={{0.0\textwidth} {-0.1\textwidth} {0.0\textwidth} {0.0\textwidth}},clip]{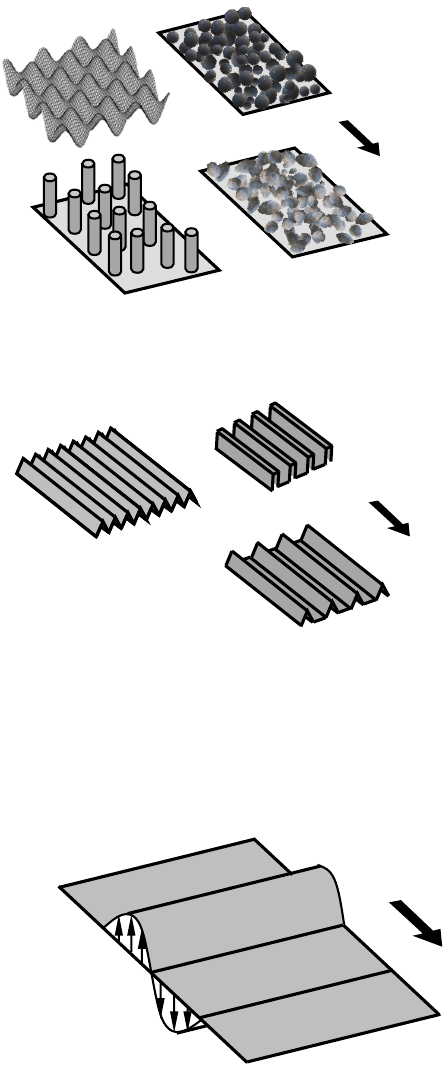}
 \put(-10.98,7.9){$R_h$}
 \put(-10.98,4.8){$R_h$}
 \put(-10.98,1.7){$R_h$}
 \put(-3.7,9.2){\scriptsize{egg-carton}}
 \put(-2.5,9.4){\scriptsize{spherical}}
 \put(-2.6,6.7){\scriptsize{pin fins}}
 \put(-1.5,6.9){\scriptsize{grit/irregular}}
 \put(-2.5,4.4){\scriptsize{riblets}}
 \put(-3.2,2.6){\scriptsize{blowing/suction}}
  \caption{Performance plots showing the fractional changes in the skin-friction coefficient $R_f=C_f/C_{f_0}$  and  Stanton number $R_h=C_h/C_{h_0}$ (log-log scale). The diagonal dashed line represents $R_h = R_f$. (\textit{a}) Grit or irregular roughness~\citep{dipprey1963,forooghi2018dns}, spherical roughness~\citep{healzer1974}, pin fins~\citep{kang1999effect} and egg-carton roughness~\citep{zhong2023heat,rowin2024modelling}. (\textit{b}) Riblets~\citep{walsh1979,choi1997,stalio2003,jin2014,rouhi2022riblet,kuwata2022dissimilar}.  (\textit{c}) Wall blowing/suction~\citep{hasegawa2011dissimilar,yamamoto2013}. All data points are at $Pr = 0.7 - 1.0$. 
  }
  \label{fig:RA_intro}
\end{figure}

In many industrial applications, controlling heat transfer is as important as controlling drag.  
Some examples of the kinds of methods  to control both drag and heat transfer are given in figure~\ref{fig:RA_intro}.  The performance of any given method is commonly based on the changes achieved in the skin-friction coefficient $C_f$ and the Stanton number $C_h$, where
\begin{equation}
 C_f \equiv \frac{2\overline{\tau_w}}{\rho U_\mathrm{ref}^2}, \quad C_h \equiv \frac{\overline{q_w}}{\rho c_p U_\mathrm{ref} (T_w - T_\mathrm{ref})}. \label{eq:cf_ch_def}
\end{equation}
Here $\overline{\tau_w}$, $\overline{q_w}$, $\rho$, $c_p$ and $T_w$ are the wall shear-stress, wall heat-flux, fluid density, its specific heat capacity and wall temperature, respectively. The overbar in $\overline{\tau_w}$ and $\overline{q_w}$ represents averaging over the homogeneous directions and time. In a boundary layer, $U_\mathrm{ref}$ and  $T_\mathrm{ref}$ are  the free-stream velocity and temperature \citep{walsh1979}, and in a channel flow, $U_\mathrm{ref}$ and $T_\mathrm{ref}$ are  the bulk velocity and either the bulk temperature \citep{stalio2003} or the mixed-mean temperature \citep{dipprey1963, macdonald2019, rouhi2022riblet}.

Figure~\ref{fig:RA_intro} also shows the type of performance plot that is often used to measure the efficacy of  control methods \citep{fan2009performance,bunker2013,huang2017numerical,rouhi2022riblet}.   Here $C_f$ and $C_h$ are measured in the flow with control (the target case), and $C_{f_0}$ and $C_{h_0}$ are measured in the absence of control (the reference case), at matched Reynolds and Prandtl numbers. Some boundary layer studies match Reynolds numbers based on the free-stream velocity and the distance from the inlet~\citep{walsh1979,choi1997}. Some channel flow studies match Reynolds numbers based on the channel height and the bulk velocity~\citep{stalio2003,jin2014,rouhi2022riblet} or the friction velocity~\citep{macdonald2019,zhong2023heat}.
The diagonal dashed-line represents equal fractional change in drag $R_f$ ($=C_f/C_{f_0}$) and fractional change in heat transfer $R_h$ ($=C_h/C_{h_0}$). In type A applications (shaded in purple), the objective is to maximise the heat transfer between the fluid and the solid surface, while minimising the drag. Examples include air passages in the gas turbine blades~\citep{baek2022flow,otto2022heat} or solar air collectors~\citep{vengadesan2020review,RANI2022107597}. In both applications, the air needs to absorb heat from the surface (turbine blade or absorber tube of the collector), yet the friction loss needs to be minimised through the passages. In type B applications (shaded in pink), the objective is to simultaneously minimise the heat transfer and drag between the fluid and the solid surface. Examples include transportation of crude oil~\citep{yu2010numerical,han2015fast,yuan2023investigation} or transportation of industrial waste heat~\citep{hasegawa1998analysis,ma2009review,xie2017absorption}. In these applications, a heated fluid is transported through pipelines to a long-distance demand site, and requires heat loss and friction loss to be minimised.

From figure~\ref{fig:RA_intro}(\textit{a}), we see that rough surfaces and pin fins  augment the heat transfer ($R_h>1$), but for most data points the drag increase exceeds the heat-transfer increase  ($R_f>R_h > 1$), and so they fall outside the objective space for either type A or B applications. Several phenomena contribute to the imbalance between $R_f$ and $R_h$. The flow recirculations behind the roughness elements trigger pressure drag in $C_f$, in addition to the viscous drag. However, $C_h$ consists of the wall heat-flux only~\citep{macdonald2019}, which is the thermal analogue of the viscous stress. Additionally, the recirculation zones break the analogy between the viscous stress and the wall heat-flux~\citep{macdonald2019}, and the analogy between the turbulent shear-stress and the turbulent heat-flux~\citep{kuwata2021direct}. Also, by increasing the roughness size, the conductive sublayer becomes thinner than the viscous sublayer~\citep{macdonald2019,zhong2023heat}.


In figure~\ref{fig:RA_intro}(\textit{b}), we show the performance of riblets, as two-dimensional streamwise-aligned surface protrusions which do not induce pressure drag. Most data points are scattered near $R_h = R_f$, but a few data points fall into the regions of interest for type A or B applications. \cite{rouhi2022riblet} and \cite{kuwata2022dissimilar} relate such behaviour to the formation of Kelvin-Helmholtz rollers and secondary flows by certain riblet designs.

Figure~\ref{fig:RA_intro}(\textit{c}) shows the data points for wall blowing/suction in a turbulent channel flow. All data points fall into the objective space for type A applications. The data points that noticeably yield $R_h > R_f$ are from the application of a suboptimal control framework~\citep{lee1998suboptimal} or an optimal control framework~\citep{bewley2001dns}. Especially, the optimal control application results in simultaneous heat-transfer augmentation and drag reduction ($R_h > 1 > R_f$). \cite{hasegawa2011dissimilar} and \cite{yamamoto2013} discovered that both the optimal and suboptimal control inputs exhibit characteristics similar to a travelling wave. When the coherent part of the suboptimal control inputs are applied as an open-loop control framework, the data points fall closer to $R_h = R_f$.

\begin{figure}
  \centering
 \includegraphics[width=0.90\textwidth,trim={{0.0\textwidth} {0.0\textwidth} {0.0\textwidth} {0.0\textwidth}},clip]{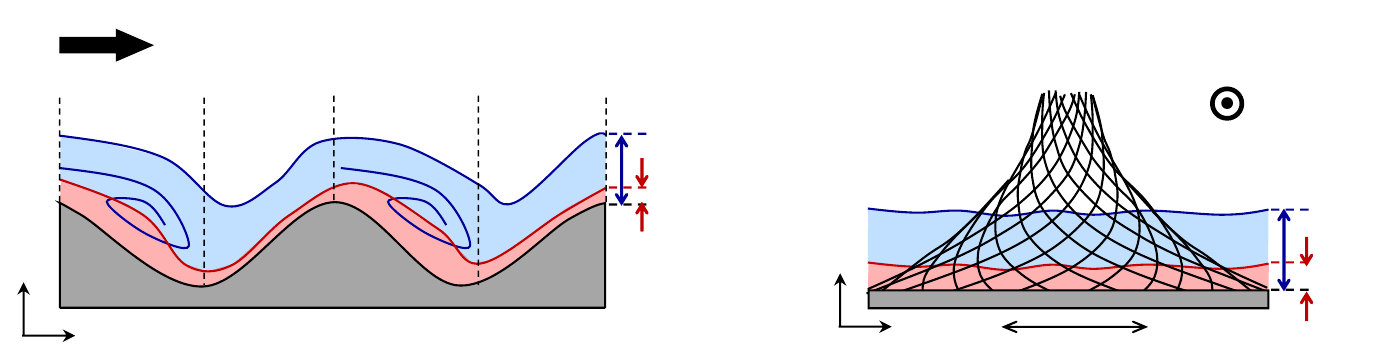}
 \put(-10.1,2.7){(\textit{a})}
 \put(-4.0,2.2){(\textit{b})}
 \put(-9.7,2.5){\scriptsize{Flow}}
 \put(-9.7,0.0){\scriptsize{$x$}}
 \put(-10.15,0.5){\scriptsize{$y$}}
 \put(-9.6,1.8){\scriptsize{sheltered}}
 \put(-8.55,1.8){\scriptsize{exposed}}
 \put(-7.6,1.8){\scriptsize{sheltered}}
 \put(-6.55,1.8){\scriptsize{exposed}}
 \put(-5.55,1.75){\scriptsize{$\delta_\mathrm{vis. \: subl.}$}}
 \put(-5.35,1.16){\scriptsize{$\delta_\mathrm{cnd. \: subl.}$}}
 \put(-2.9,2.1){\scriptsize{Stokes layer}}
 \put(-1.4,2.05){\scriptsize{Flow}}
  \put(-3.75,0.05){\scriptsize{$z$}}
 \put(-4.20,0.55){\scriptsize{$y$}}
 \put(-2.4,0.00){\scriptsize{$W_s$}}
  \put(-0.75,1.14){\scriptsize{$\delta_\mathrm{vis. \: subl.}$}}
 \put(-0.55,0.54){\scriptsize{$\delta_\mathrm{cnd. \: subl.}$}}
  \vspace{.2cm}  
 \includegraphics[width=0.55\textwidth,trim={{0.05\textwidth} {0.05\textwidth} {0.9\textwidth} {0.6\textwidth}},clip]{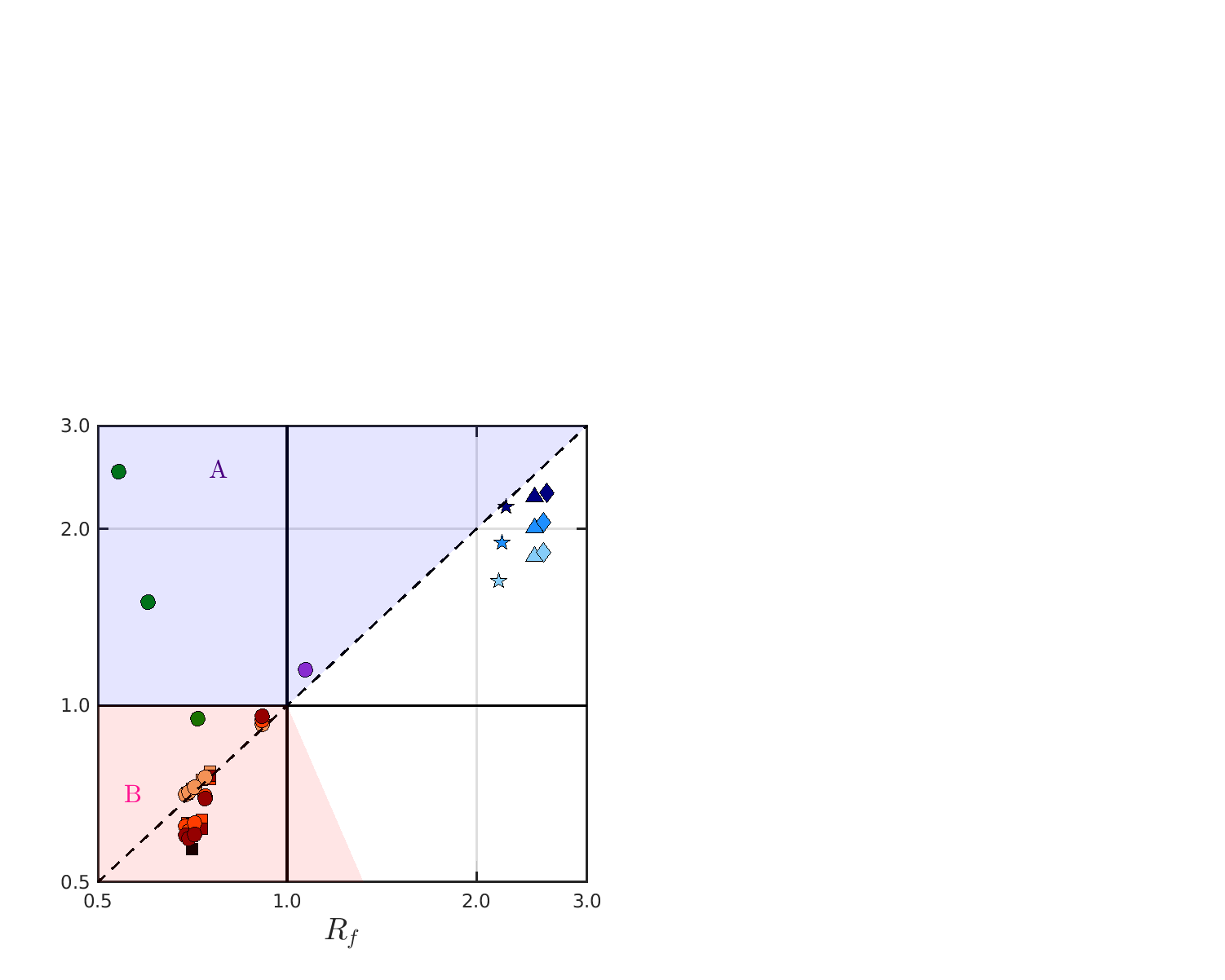}\hspace{.4cm}
 \includegraphics[width=0.36\textwidth,trim={{0.0\textwidth} {-0.10\textwidth} {0.0\textwidth} {0.0\textwidth}},clip]{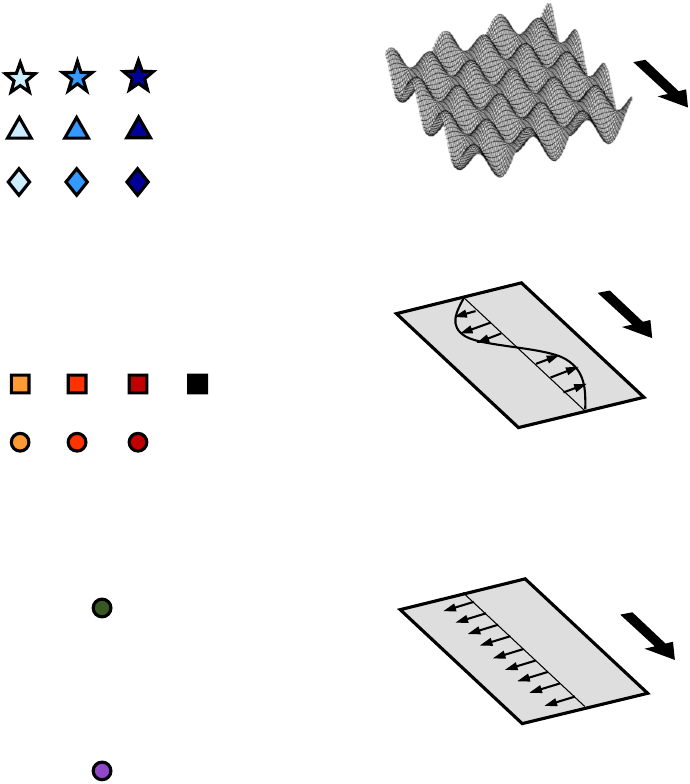}
 \put(-10.7,5.6){(\textit{c})}
 \put(-10.5,3){$R_h$}
 \put(-4.4,5.3){\scriptsize{\cite{zhong2023heat}}}
 \put(-3.8,5.0){\scriptsize{$Pr$}}
 \put(-4.15,4.75){\scriptsize{$0.5$ $1.0$ $2.0$}}
 \put(-3.1,4.52){\scriptsize{$395$}}
 \put(-3.1,4.22){\scriptsize{$590$}}
 \put(-3.1,3.92){\scriptsize{$720$}}
 \put(-2.65,4.25){\scriptsize{$Re_\tau$}}
 \put(-4.2,3.45){\scriptsize{Present study}}
 \put(-3.6,3.15){\scriptsize{$Pr$}}
 \put(-4.2,2.95){\scriptsize{$0.7$ $4.0$ $7.5$ $20.0$}}
 \put(-4.2,1.9){\scriptsize{\cite{fang2009large}}}
 \put(-3.9,1.65){\scriptsize{$Pr=0.7$}}
 \put(-4.3,0.95){\scriptsize{\cite{guerin2023breaking}}}
 \put(-3.9,0.70){\scriptsize{$Pr=1.0$}}
 \put(-1.8,5.1){\scriptsize{egg-carton}}
 \put(-1.8,3.55){\scriptsize{travelling wave}}
 \put(-1.9,2.35){\scriptsize{$W_s = A \sin(\kappa_x x + \omega t)$}}
 \put(-1.8,1.8){\scriptsize{plane oscillation}}
 \put(-1.5,0.55){\scriptsize{$W_s = A \sin(\omega t)$}}
  \caption{Effect of $Pr$ on the performance of egg-carton roughness by \cite{zhong2023heat}, plane oscillation and travelling wave from the present study (table~\ref{tab:runs}), and plane oscillation by \cite{fang2009large} and \cite{guerin2023breaking}. Star, triangle and diamond symbols represent egg-carton roughness, square symbols represent streamwise travelling wave ($\kappa_x \ne 0$ in \ref{eq:wall_motion}), and circle symbols represent plane oscillation ($\kappa_x = 0$ in \ref{eq:wall_motion}). (\textit{a,b}) Illustrate the different physics underlying the wall heat-transfer over roughness and wall oscillation for $Pr > 1$; $\delta_\mathrm{vis. \: subl.}$ and $\delta_\mathrm{cnd. \: subl.}$ respectively indicate the viscous sublayer and conductive sublayer thicknesses. (\textit{c}) compiles all the data points in the performance plot.}
  \label{fig:RA_travelling_wave}
\end{figure}



In the present study, we focus on wall oscillation either as a spanwise oscillating plane, or a streamwise travelling wave, as described by
\begin{equation}
W_s (x,t) = A \sin(\kappa_x x + \omega t) \label{eq:wall_motion}
\end{equation}
where $W_s$ is the instantaneous spanwise surface velocity that oscillates with amplitude $A$ and frequency $\omega$.  With $\kappa_x \ne 0$, the mechanism generates a streamwise travelling wave, and with $\kappa_x = 0$ the motion is a simple spanwise oscillating plane (as illustrated in figure~\ref{fig:RA_travelling_wave}). We denote the former case as travelling wave, and the latter one as plane oscillation. The drag performance of (\ref{eq:wall_motion}) has been extensively investigated in the literature~\citep{jung1992suppression, quadrio2000numerical, quadrio2009, viotti2009, quadrio2011, quadrio2011b, gatti2013, hurst2014, gatti2016, marusic2021, rouhi2022turbulent, chandran2022turbulent}. 
Dimensional analysis for the drag yields 
\begin{equation}
R_f=C_f/C_{f_0} = f_f (A^+,\kappa^+_x,\omega^+,Re_{\tau_0})
\label{Cf_dim}
\end{equation}
 \citep{marusic2021,rouhi2022turbulent}. The superscript `$+$' indicates normalisation using the viscous velocity ($u_{\tau_0} = \sqrt{\overline{\tau_{w_0}}/\rho}$) and length ($\nu/u_{\tau_0}$) scales, where $\nu$ is the fluid kinematic viscosity. The superscript `$\ast$' indicates viscous scaling based on the actual friction velocity, that is, $u_{\tau}$ of the drag-altered flow for the actuated cases. For the heat transfer, we need to include the Prandtl number $Pr= \nu/\alpha$, where $\alpha$ is the fluid thermal diffusivity, and we find
\begin{equation}
R_h=C_h/C_{h_0} = f_h (A^+,\kappa^+_x,\omega^+,Re_{\tau_0}, Pr).
\label{Ch_dim}
\end{equation}
The  drag reduction is given by $DR = 1- R_f$, and the heat-transfer reduction is given by $HR = 1 - R_h$.
For $HR$ we have a five-dimensional parameter space (\ref{Ch_dim}) that is investigated over a limited extent~\citep{fang2009large,fang2010heat,guerin2023breaking}, as shown in figure~\ref{fig:RA_travelling_wave}. These studies consider spanwise plane oscillation $(\kappa^+_x = 0)$ in a turbulent channel flow at $Re_{\tau_0} = h u_{\tau_0}/\nu \simeq 180$ and $Pr \simeq 0.7 - 1.0$. \cite{fang2009large,fang2010heat} consider three cases with $T^+_{osc} = 2\pi/\omega^+ = 100$ and $A^+ = 6, 13, 19$, and \cite{guerin2023breaking} consider one case with $T^+_{osc} = 500$ and $A^+ = 30$. These cases fall into the objective space for type A applications (except one case with $R_f < 1, R_h \simeq 1$).

Figure~\ref{fig:RA_travelling_wave} highlights that we are at an early stage in terms of our knowledge of $HR$ for the surface actuation (\ref{eq:wall_motion}), and just like the investigations on $DR$, we need to explore different parameters of (\ref{Ch_dim}) as a build up process. Comparing (\ref{Ch_dim}) with (\ref{Cf_dim}), $Pr$ is a key parameter that differentiates $R_h$ from $R_f$, and it leads to their significant disparity (figure~\ref{fig:RA_travelling_wave}). Furthermore, the fluids for type B applications have $Pr \gg 1$ (e.g.\ crude oil). Therefore, in the present study we focus on $Pr$ and $\omega^+$ as our parameters of interest (figure~\ref{fig:RA_travelling_wave} and table~\ref{tab:runs}). We conduct DNSs of turbulent forced convection in a half channel flow with fix $A^+ = 12$ and $Re_{\tau_0} = 590$, and we consider two values for $\kappa^+_x$ to simulate plane oscillation ($\kappa^+_x = 0$) and travelling wave ($\kappa^+_x = 0.0014$, $\lambda^+ = 2\pi/\kappa^+_x \simeq 4500$). We consider $Pr = 0.71$ (air), $4.0$, $7.5$ (water) and $20$ (molten salt). At each $0.71 \le Pr \le 7.5$, we systematically vary $\omega^+$ from $0.022$ to $0.110$, corresponding to the regime where we expect $DR$ to reach $30\%$~\citep{gatti2016,rouhi2022turbulent}. Our preference for the present actuation parameters is motivated by an existing surface-actuation test bed~\citep{marusic2021,chandran2022turbulent}, that can operate under these conditions. This provides the opportunity to extend the present numerical study through experiment. As the first study that investigates $Pr$ effects for this problem, we conduct a thorough study of grid resolution requirements for our considered parameter space (Appendix~\ref{sec:grid_domain}).

 As an overview, we note that all our results give a decrease in drag and heat transfer, with maximum  $DR=30\%$ and $HR=40\%$ (see figure~\ref{fig:RA_travelling_wave}). For Prandtl numbers greater than one, however, $HR$ increases more than $DR$. The results fall into the objective space for type B applications, a space that has largely been neglected in the past  despite its industrial relevance. The surfaces that we reviewed in figure~\ref{fig:RA_intro} consider $Pr = 0.7 - 1.0$, however, we anticipate that their performances will change with increasing $Pr$. For instance, in figure~\ref{fig:RA_travelling_wave}, we overlay the data points of egg-carton roughness at $Pr = 0.5, 1.0, 2.0$ by \cite{zhong2023heat}. Unlike plane oscillation or travelling wave, the roughness data points approach $R_h = R_f$ with increasing $Pr$ because heat transfer over these two surface types is governed by different flow mechanisms that are unique to each surface, leading to different trends with $Pr$. Over the rough surface (figure~\ref{fig:RA_travelling_wave}\textit{a}), variation of $C_h$ with $Pr$ depends on the area fraction of the sheltered and exposed regions, associated with the recirculation and attached zones that follow different local $Pr$ scaling~\citep{zhong2023heat,rowin2024modelling}. Over the surface oscillation (figure~\ref{fig:RA_travelling_wave}\textit{b}), the variation of $C_h$ with $Pr$ is an aspect that we investigate in the present study. We discover that by increasing $Pr$ and thinning of the conductive sublayer, the Stokes layer, formed in response to the wall oscillation, transfers more energy to the near-wall temperature scales than the velocity scales. We also find that $HR$ yields different $Pr$ scaling depending on the actuation parameters. In \S\ref{sec:hr_model}, we derive a model that predicts that plane oscillation and travelling wave can reduce the heat loss beyond $50\%$ for $Pr \gtrsim \mathcal{O}(10^2)$, a regime that is important for crude oil transportation through pipelines.  Our model is supported by our DNS data point at $Pr = 20$ that yields $HR = 43\%$ (black square in figure~\ref{fig:RA_travelling_wave}).

 \begin{figure}
  \centering
 \includegraphics[width=\textwidth,trim={{0.07\textwidth} {0.15\textwidth} {0.0\textwidth} {0.1\textwidth}},clip]{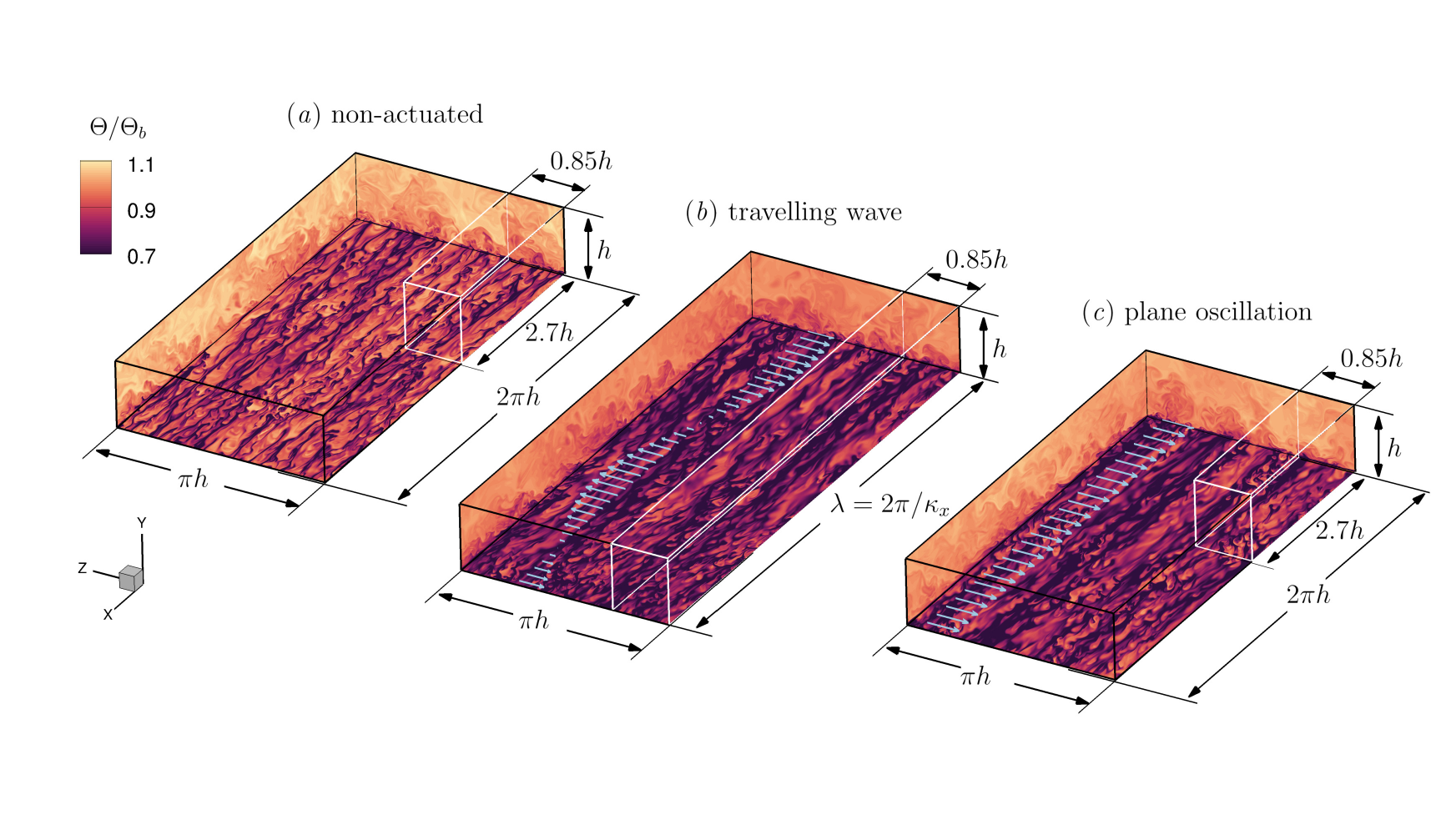}
  \caption{Computational configurations in a half channel flow for the present study. (\textit{a}) Non-actuated case as the reference. (\textit{b}) Actuated case with the travelling wave with $A^+ = 12$, $\kappa^+_x = 0.0014$; the domain length encompasses one wavelength $L_x = 2\pi/\kappa_x \simeq 7.6h$. (\textit{c}) Actuated case with spanwise plane oscillation with $A^+ = 12$. In (\textit{b,c}), we overlay the instantaneous wall motion as a vector field. Full domain shown by the large box with black edges; reduced domain shown by the small box with white edges. In (\textit{a,b,c}), we visualise the instantaneous fields of $\Theta$ at $y^+ = 15, Re_{\tau_0} = 590$ and $Pr = 7.5$, and for $\omega^+ = 0.088$ in (\textit{b,c}).}
  \label{fig:flowviz_setup}
\end{figure}


\section{Numerical flow setup}\label{sec:setup}

\subsection{Governing equations and simulation setup}
The governing equations for an incompressible fluid with constant $\rho, \nu$ and thermal diffusivity $\alpha$ are solved in a half channel flow (figure~\ref{fig:flowviz_setup}), where
\begin{eqnarray}
 \boldsymbol{\nabla}\boldsymbol{\cdot}\mathbf{U} &=& 0, \label{eq:cont} \\
 \frac{\partial \mathbf{U}}{\partial t} + \boldsymbol{\nabla}\boldsymbol{\cdot}\mathbf{(UU)} &=& -\frac{1}{\rho}\boldsymbol{\nabla}p + \nu \nabla^2 \mathbf{U} -\frac{1}{\rho}\frac{\mathrm{d}\overline{P}}{\mathrm{d}x}\hat{\mathbf{e}}_x, \label{eq:mom} \\
 \frac{\partial \Theta}{\partial t} + \boldsymbol{\nabla}\boldsymbol{\cdot}(\mathbf{U}\Theta) &=& \alpha \nabla^2 \Theta -U\frac{\mathrm{d}T_w}{\mathrm{d}x}, \label{eq:scalar}
\end{eqnarray}
and \eqref{eq:cont}, \eqref{eq:mom} and \eqref{eq:scalar} are the continuity, velocity and temperature transport equations, respectively. We ignore buoyancy, as appropriate for forced convection. In our notation, $\mathbf{U} = (U,V,W)$ is the velocity vector, and $x$, $y$ and $z$ are the streamwise, wall-normal and spanwise directions, respectively. In (\ref{eq:mom}),   the total pressure gradient was decomposed into the driving (mean) part $\mathrm{d}\overline{P}/\mathrm{d}x$ and the periodic part $\boldsymbol{\nabla} p$. 
Similarly, the total temperature $T = (dT_w/dx)x + \Theta$ was decomposed into the mean part $dT_w/dx$ and the periodic part $\Theta$. 
This thermal driving approach imposes a prescribed mean heat flux at the wall, making it a suitable boundary condition for a periodic domain, and it has been widely used for the simulation of forced convection in a channel flow~\citep{kasagi1992,watanabe2002,stalio2003,jin2014,alcantara2021directb,alcantara2021direct,rouhi2022riblet}. By averaging (\ref{eq:mom}) and (\ref{eq:scalar}) in time and over the entire fluid domain, we obtain
\begin{equation}
 -\frac{1}{\rho}\frac{\mathrm{d}\overline{P}}{\mathrm{d}x}h = \frac{\overline{\tau_w}}{\rho} \equiv u^2_\tau, \qquad -U_b \frac{\mathrm{d}T_w}{\mathrm{d}x}h = \frac{\overline{q_w}}{\rho c_p} \equiv \theta_\tau u_\tau \tag{2.4\textit{a,b}} \label{eq:global}
\end{equation}
where $\overline{\tau_w}$ and $\overline{ q_w }$ are respectively the $xz$-plane and time averaged wall shear-stress and wall heat-flux, $U_b$ is the bulk velocity, $h$ is the half channel height, $u_\tau$ and $\theta_\tau$ are the friction velocity and friction temperature, and $c_p$ is the specific heat capacity. We adjust $d\overline{P}/dx$ based on a target flow rate (namely, a bulk Reynolds number $Re_b \equiv U_b h / \nu$) that is matched between the non-actuated reference case (figure~\ref{fig:flowviz_setup}\textit{a}) and the actuated cases (figure~\ref{fig:flowviz_setup}\textit{b,c}).

Equations (\ref{eq:cont}-\ref{eq:scalar}) are solved using a fully conservative fourth-order finite difference code, employed by previous DNS studies of thermal convection~\citep{ng2015vertical,rouhi2021coriolis,zhong2023heat,rowin2024modelling}. The half channel flow has periodic boundary conditions in the streamwise and spanwise directions (figure~\ref{fig:flowviz_setup}). The bottom-wall velocity boundary conditions are $U=V=W = 0$ for the non-actuated case (figure~\ref{fig:flowviz_setup}\textit{a}), $U=V=0$, $W = A \sin(\kappa_x x + \omega t)$ for the actuated case with the travelling wave (figure~\ref{fig:flowviz_setup}\textit{b}), and $U=V=0$, $W = A \sin(\omega t)$ for the actuated case with the spanwise plane oscillation (figure~\ref{fig:flowviz_setup}\textit{c}). The top boundary conditions for the velocity are the free-slip and impermeable conditions $(\partial U/\partial y = \partial W/\partial y = V = 0)$. The boundary conditions for $\Theta$ at the bottom wall and top boundary are $\Theta = 0$ and $\partial \Theta / \partial y = 0$, respectively. In other words, the total temperature at the bottom wall increases linearly in the $x$-direction $T = (dT_w/dx)x$, and at the top boundary the boundary condition is adiabatic ($\partial T / \partial y = 0$). Past studies have employed half channel flow to study turbulent flow over complex surfaces~\citep{yuan2014estimation,macdonald2017,rouhi2019}. This configuration has a lower computational cost compared to the conventional poiseuille channel flow with no-slip condition at the bottom and top walls. Yet the two configurations have almost identical mean velocity profiles up to the logarithmic region, and similar turbulent stresses~\citep{yao2022direct}. This is also supported in Appendix~\ref{sec:grid_domain}, where we obtain very similar mean velocity and turbulent stress profiles between our non-actuated half channel flow and the DNS of poiseuille channel flow by \cite{moser1999direct} (figure~\ref{fig:grid_domain_non_actuated}\textit{a,b}), and the values of $C_{f_0}$ differ by less than $2\%$ (table~\ref{tab:non_act}).

Throughout this manuscript, we call $\Theta$ the temperature. We denote the $xz$-plane and time averaged quantities with overbars (e.g.\ $\overline{\Theta}$ is the plane and time averaged $\Theta$), the turbulent quantities with lowercase letters (e.g.\ $\theta$ is the turbulent temperature). Following this notation, $\overline{u^2}$ and $ \overline{uv}$ are the turbulent stress components, and $\overline{\theta^2}$ and $ \overline{\theta v}$ are their analogue for the turbulent temperature fluxes.

\subsection{Simulation cases}

Table~\ref{tab:runs} summarises the production calculations. All the calculations were performed at a fixed bulk Reynolds number $Re_b \simeq 11004$, equivalent to a friction Reynolds number $Re_{\tau_0} = 590$ for the non-actuated flow. The grid sizes are $\Delta^+_x \times \Delta^+_z \simeq 8 \times 4, \Delta^+_y = 0.28 - 8.5$, chosen based on extensive validation studies (Appendix~\ref{sec:grid_domain}). In comparison, \cite{alcantara2021directb} used $\Delta^+_x \times \Delta^+_z \simeq 8 \times 4, \Delta^+_y = 0.27 - 5.3$ for turbulent channel flow at $Re_{\tau_0} = 500$ and $1 \le Pr \le 7$, and \cite{pirozzoli2023prandtl} used $\Delta^+_x \times \Delta^+_z \simeq 10 \times 4.5$ with $30$ points within $y^+ \le 40$ for turbulent pipe flow at $Re_{\tau_0} \simeq 1100$ and $Pr \le 16$. Here, we have $60$ points within $y^+ \le 40$. As shown in Appendix~\ref{sec:grid_domain}, we obtain less than $2\%$ difference in $DR$ and $HR$ when using finer resolutions than our production calculation grids (tables~\ref{tab:non_act} to \ref{tab:act_pr}), and find very good agreement in the first and second order velocity and temperature statistics as well as their spectrograms (figures~\ref{fig:grid_domain_non_actuated} to \ref{fig:grid_stats_actuated}). 

 \begin{table}
\centering
 \begin{tabular}{cc|ccc|ccc|c}
    \multicolumn{2}{c|}{non-actuated} & \multicolumn{3}{c|}{travelling wave} & \multicolumn{3}{c|}{oscillating plane} &  \\
     \multicolumn{2}{c|}{}            & \multicolumn{3}{c|}{$A^+ =12, \kappa^+_x = 0.0014$} & \multicolumn{3}{c|}{$A^+ =12$}         &  \\
     \multicolumn{2}{c|}{}            & \multicolumn{3}{c|}{} & \multicolumn{3}{c|}{}         &  \\
     \multicolumn{2}{c|}{$L_x , L_z = 2.7 h , 0.85 h$}         & \multicolumn{3}{c|}{$L_x, L_z = 7.6 h, 0.85 h$} & \multicolumn{3}{c|}{$L_x , L_z = 2.7 h , 0.85 h$}         &  \\ 
     \multicolumn{2}{c|}{$\Delta^+_x, \Delta^+_z = 8.3, 3.9$}            & \multicolumn{3}{c|}{$\Delta^+_x, \Delta^+_z = 8.8, 3.9$} & \multicolumn{3}{c|}{$\Delta^+_x, \Delta^+_z = 8.3, 3.9$}         &  \\
     \multicolumn{2}{c|}{$\Delta^+_y = 0.28 - 8.5$}            & \multicolumn{3}{c|}{$\Delta^+_y = 0.28 - 8.5$} & \multicolumn{3}{c|}{$\Delta^+_y = 0.28 - 8.5$}         &  \\
     \multicolumn{2}{c|}{}            & \multicolumn{3}{c|}{} & \multicolumn{3}{c|}{}         &  \\
     $C_{f_0} \times 10^{-3}$ & $C_{h_0} \times 10^{-4}$ & $\omega^+$ & $DR\%$ & $HR\%$ & $\omega^+$ & $DR\%$ & $HR\%$ & \\ \hline
            $5.71$ & $33.82$ &  
            $0.022$ & $25.2$ & $23.6$ &            
            $0.022$ & $8.6$ & $7.3$ & \\
     & &  
            $0.044$ & $30.6$ & $29.3$ &
            $0.044$ & $25.9$ & $24.7$ & \\
    &  & 
            $0.066$ & $30.5$ & $29.1$ &
            $0.066$ & $31.1$ & $29.6$ & $Pr=0.71$ \\  
     & & 
            $0.088$ & $29.3$ & $27.9$ & 
            $0.088$ & $30.3$ & $28.9$ & \\
     &  & 
            $0.110$ & $26.8$ & $25.3$ & 
            $0.110$ & $28.7$ & $27.4$ & \\ \\
    $5.71$ & $14.14$ &   
    $0.022$ & $24.5$ & $25.3$ &
    $0.022$ & $9.5$ & $5.9$ & \\
    &  & 
     $0.044$ & $31.0$ & $37.0$ &  
     $0.044$ & $25.9$ & $30.1$ & \\
  &  & 
    $0.066$ & $30.7$ & $38.4$ &  
    $0.066$ & $30.4$ & $37.8$ & $Pr=4.0$ \\  
    &  & 
     $0.088$ & $29.1$ & $37.3$ & 
     $0.088$ & $30.6$ & $39.0$ & \\
    &  & 
     $0.110$ & $27.6$ & $36.2$ &  
     $0.110$ & $28.2$ & $36.9$ & \\ \\
    $5.72$ & $9.90$ &  
     $0.022$ & $25.6$ & $24.0$ &
    $0.022$ & $9.7$ & $4.3$ & \\
     &  & 
   $0.044$ & $31.8$ & $37.7$ & 
   $0.044$ & $26.5$ & $30.8$ & \\
    &  & 
   $0.066$ & $30.7$ & $39.8$ &  
   $0.066$ & $31.3$ & $40.0$ & $Pr=7.5$ \\  
     &  & 
   $0.088$ & $30.5$ & $40.4$ & 
   $0.088$ & $30.2$ & $40.7$ & \\
    &  & 
   $0.110$ & $27.5$ & $38.0$ & 
   $0.110$ & $28.7$ & $39.8$ & \\ \\
   $5.71$ & $5.47$ & 
  $0.088$ & $29.5$ & $43.5$ &  
     &   &   & $Pr=20.0$ \\    
    \end{tabular}
\caption{Production calculations at $Re_{\tau_0} = 590$   (reduced domain $L_x \ge 2.7h, L_z = 0.85h$). Non-actuated cases (left column, figure~\ref{fig:flowviz_setup}\textit{a}). Travelling wave cases (middle column, figure~\ref{fig:flowviz_setup}\textit{b}). Spanwise plane oscillation cases (right column, figure~\ref{fig:flowviz_setup}\textit{c}). Each row represents one simulation case. The domain and grid sizes for each configuration are reported at the top.}
\label{tab:runs}
\end{table}



\begin{figure}
  \centering
 \includegraphics[width=1.0\textwidth,trim={{0.0\textwidth} {0.15\textwidth} {0.0\textwidth} {0.0\textwidth}},clip]{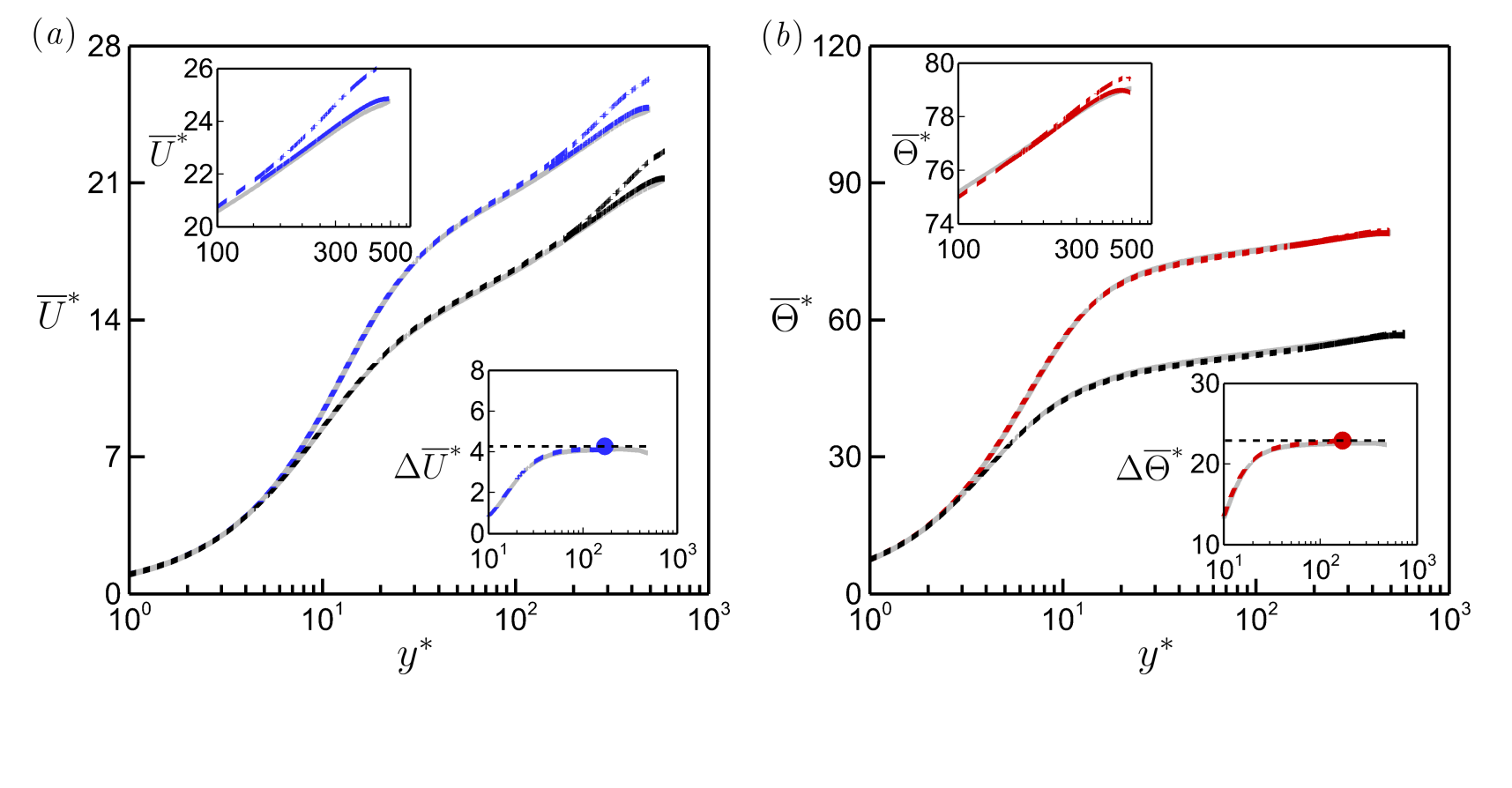}
  \caption{Comparison at $Re_{\tau_0} = 590$ of (\textit{a}) mean velocity $\overline{U}^*$, and (\textit{b}) mean temperature $\overline{\Theta}^*$ between the full-domain (grey lines) and the reduced-domain half channel flow (black, blue and red lines). Black lines: non-actuated case at $Pr = 7.5$. Blue and red lines: travelling wave case at  $Pr=7.5$ with $A^+ = 12$, $\kappa^+_x = 0.0014$, $\omega^+ = 0.088$. For the reduced-domain cases, the resolved portion is shown with a dashed line ($y^* \lesssim 170$), and the unresolved portion is shown with a dashed-dotted line ($y^* \gtrsim 170$). The unresolved portion is replaced by the reconstructed profiles (solid lines), as explained in \S\ref{sec:calculate_cf_ch}. The insets plot the velocity and temperature differences $\Delta \overline{U}^*$ and $\Delta \overline{\Theta}^*$;  bullets mark $y^*_\mathrm{res} = 170$, where we obtain the log-law shifts $\Delta \overline{U}^*_\mathrm{log}$ and $\Delta \overline{\Theta}^*_\mathrm{log}$.}
  \label{fig:profile_reconstruction}
\end{figure}

To reduce  the computational cost, the production calculations were performed in a reduced domain rather than a full-domain conventional channel flow (see figure~\ref{fig:flowviz_setup}). Because of  the domain truncation, the flow is fully resolved up to a fraction of the domain height $y_\mathrm{res} < h$ (figure~\ref{fig:profile_reconstruction}). In the past, the reduced-domain channel flow has been used for accurate calculation of the drag and the near-wall turbulence over various static and deforming surfaces, including egg-carton roughness with $y^+_\mathrm{res} \lesssim 250$~\citep{macdonald2017,macdonald2018direct}, riblets with $y^+_\mathrm{res} \simeq 100$~\citep{endrikat2020}, and travelling waves with $y^+_\mathrm{res} \lesssim 1000$~\citep{gatti2016,rouhi2022turbulent}.  The reduced-domain channel flow has also been employed for accurate calculation of the wall heat-flux and the near-wall thermal field over rough surfaces with $y^+_\mathrm{res} \lesssim 600$~\citep{macdonald2019,zhong2023heat,rowin2024modelling}, and riblets with $y^+_\mathrm{res} \simeq 100$~\citep{rouhi2022riblet}.  
In our case, as shown in Appendix~\ref{sec:grid_domain}, with $\Delta^+_x \times \Delta^+_z \simeq 8 \times 4$ the difference between the reduced and the full domains is less than $1\%$ in terms of $DR$ and $HR$ (full coarse and reduced coarse cases in table~\ref{tab:act}). Furthermore, the results for the two domain sizes agree well in terms of the statistics of velocity and temperature, and their spectrograms (figures~\ref{fig:grid_domain_actuated} and \ref{fig:grid_stats_actuated}). For the reduced domain sizes, we follow the prescriptions by \cite{chung2015} and \cite{macdonald2017}, that were extended to the travelling wave actuation case by \cite{rouhi2022turbulent}. The resolved height $y^+_\mathrm{res}$ needs to fall in the logarithmic region, and the  domain width $L^+_z$ and length $L^+_x$ are adjusted so that $L^+_z \simeq 2.5 y^+_\mathrm{res}$ and $ L^+_x \gtrsim \max{(3L^+_z,1000,\lambda^+)}$, where $\lambda = 2\pi/\kappa_x$ is the travelling wavelength. Here, we chose $y^+_\mathrm{res} = 200$, which resolves up to a third of the half channel height. Hence the domain sizes are $L_x \times L_z \simeq 2.7 h \times 0.85 h$ for the non-actuated and the plane oscillation cases (figure~\ref{fig:flowviz_setup}\textit{a,c}).  For the travelling wave cases, however,  $L_x$ cannot be truncated because it is constrained by the travelling wavelength $\lambda \simeq 7.6 h$, and so we use $L_x \times L_z \simeq 7.6 h \times 0.85 h$ (figure~\ref{fig:flowviz_setup}\textit{b}). 
   For the actuated cases, $y_\mathrm{res}$ scaled by the actuated (drag-reduced) friction velocity is $170 \lesssim y^*_\mathrm{res} \lesssim 190$ (that is, less than 200). Therefore, $y^*_\mathrm{res} = 170$ is taken to be the maximum resolved height for all our cases.

\subsection{Calculating the skin-friction coefficient and Stanton number}\label{sec:calculate_cf_ch}

We can write the skin-friction coefficient and Stanton number as $C_f = 2/{U^*_b}^2$ and $C_h = 1/(U^*_b \Theta^*_b)$. Although some studies use the mixed-mean temperature to define $C_h$~\citep{macdonald2019,rouhi2022riblet,zhong2023heat}, we use the bulk temperature $\Theta_b$ as  adopted by \cite{stalio2003} and \cite{kuwata2022dissimilar}, primarily because it is more straightforward to derive a predictive model for $HR$ with this definition (Appendix~\ref{sec:gq_model_hr}). For our cases, there is only a maximum $2\%$ difference between $C_h$ based on $\Theta_b$ and the one based on the mixed-mean temperature. 


For the full-domain half channel flow, the mean velocity $\overline{U}^*$ and temperature $\overline{\Theta}^*$ are fully resolved across the entire domain (grey profiles in figure~\ref{fig:profile_reconstruction}), and so $U^*_b$ and $\Theta^*_b$ can be obtained by direct integration of $\overline{U}^*$ and $\overline{\Theta}^*$. 
For the reduced-domain half channel flow, we first need to construct $\overline{U}^*$ and $\overline{\Theta}^*$ beyond $y^*_\mathrm{res}$ \citep{rouhi2022riblet,rouhi2022turbulent}.   In figure~\ref{fig:profile_reconstruction}, we see that the resolved portions of the $\overline{U}^*$ and $\overline{\Theta}^*$ profiles up to $y^*_\mathrm{res} \simeq 170$ are in excellent agreement with the full-domain profiles. 
  Beyond $y^*_\mathrm{res}$, however, the flow is unresolved due to the domain truncation, appearing as a fictitious wake in the $\overline{U}^*$ and $\overline{\Theta}^*$ profiles (dashed-dotted lines in figure~\ref{fig:profile_reconstruction}). The fictitious wake is weaker for the $\overline{\Theta}^*$ profile at $Pr = 7.5$ than the $\overline{U}^*$ profile, which could also be regarded as a $\overline{\Theta}^*$ profile at $Pr = 1.0$. In other words, the domain truncation affects $\overline{\Theta}^*$ to a lesser degree by increasing $Pr$. Nevertheless, even for the $\overline{U}^*$ profile, the fictitious wake is constrained to $y^* \gtrsim 170$. We replace these unresolved portions with the composite profiles of $\overline{U}^*$ and $\overline{\Theta}^*$ (\ref{eq:up_log}, \ref{eq:cp_log} in Appendix~\ref{sec:gq_model_hr}). 
For this, we need the log-law shifts $\Delta \overline{U}^*_\mathrm{log}$ and $\Delta \overline{\Theta}^*_\mathrm{log}$ in (\ref{eq:up_log}) and (\ref{eq:cp_log}), which we obtain by plotting $\Delta \overline{U}^* = \overline{U}^* - \overline{U}^*_0$ and $\Delta \overline{\Theta}^* = \overline{\Theta}^* - \overline{\Theta}^*_0$, the differences between the actuated and the non-actuated cases (see  figure~\ref{fig:profile_reconstruction}). 
Finally, we find $U^*_b$ and $\Theta^*_b$ by integrating the resolved portion of the profiles up to $y^*_\mathrm{res}$ and the reconstructed portion beyond $y^*_\mathrm{res}$.  By applying this profile reconstruction, we obtain less than $1\%$ difference in $C_f$ and $C_h$ between the full-domain case and the reduced domain case (table~\ref{tab:act} in Appendix~\ref{sec:grid_domain}).

\section{Results}\label{sec:results}

\begin{figure}
  \centering
 \includegraphics[width=\textwidth,trim={{0.0\textwidth} {0.25\textwidth} {0.0\textwidth} {0.0\textwidth}},clip]{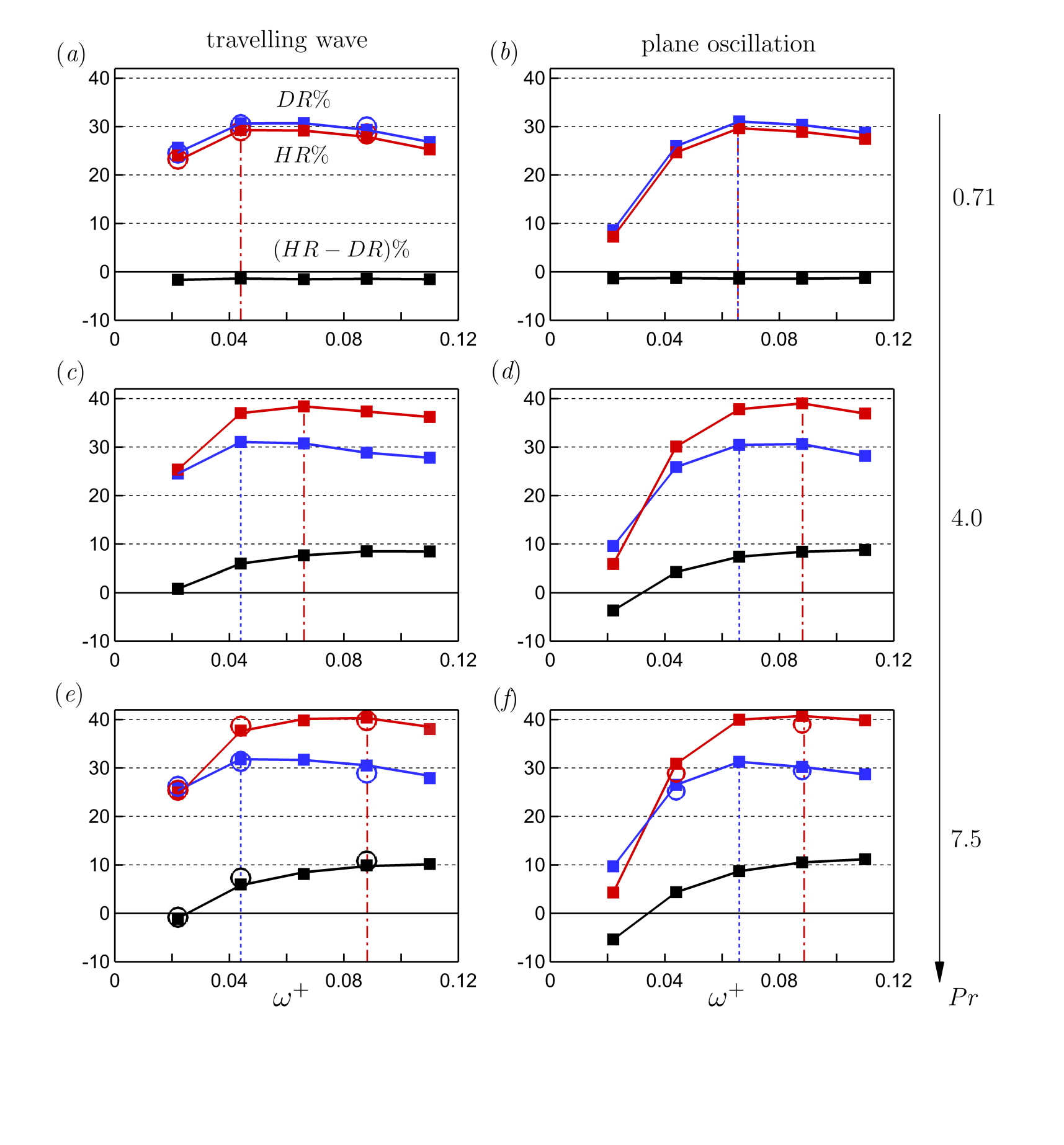} 
  \caption{Values of $DR\%$ (blue symbols), $HR\%$  (red symbols) and their difference  (black symbols) for the cases given in  table~\ref{tab:runs}. Filled squares: reduced domain simulations; empty circles:  full domain simulations.  Blue dashed line marks the maximum $DR$; red dashed-dotted line marks the maximum  $HR$. 
  }
  \label{fig:averaged_HR_DR_Pr}
\end{figure}

\subsection{Overall variations of $DR$ and $HR$}\label{sec:overall_dr_hr}

The values of $HR$,  $DR$ and the difference $HR - DR$ are shown in figure~\ref{fig:averaged_HR_DR_Pr}. For all cases considered, $HR$ and $DR$ initially increase with the frequency of forcing, but for $ \omega^+  \gtrsim 0.04$ they reach a maximum value before slowly decreasing at higher frequencies. The maximum drag reduction is about $30\%$ for all Prandtl numbers, but the maximum heat-transfer reduction increases from 30\% to  about $40\%$ with increasing Prandtl number, marking a significant disparity between $DR$ and $HR$. The increasing disparity towards $HR > DR$ occurs for $\omega^+ \ge 0.04$, however, at $\omega^+ = 0.022$, $HR \lesssim DR$. The comparison between the reduced-domain results (filled squares) and the full-domain results (open circles) supports the reliability of the production runs. At $Pr= 7.5$, the reduced domain grid  is more than twice as coarse as the full domain grid, yet there is less than $2\%$ difference in $DR$ and $HR$ (Appendix~\ref{sec:grid_domain} gives more details).  The trends in $DR$ seen here have been widely recorded in the previous literature, as the review by \cite{ricco2021review} makes clear.  However, to the authors' knowledge, the behaviour of  $HR$ has not been reported before. In \S\ref{sec:hr_dr_break}, we relate the observed trends between $HR$ and $DR$ to the attenuation in the turbulent shear-stress $\overline{uv}$ and turbulent scalar-flux $\overline{\theta v}$, followed by studying their interactions with the Stokes layer (\S\ref{sec:stokes}-\ref{sec:plane_oscillation}).

\begin{figure}
  \centering
 \includegraphics[width=.5\textwidth,trim={{0.0\textwidth} {0.15\textwidth} {0.0\textwidth} {0.0\textwidth}},clip]{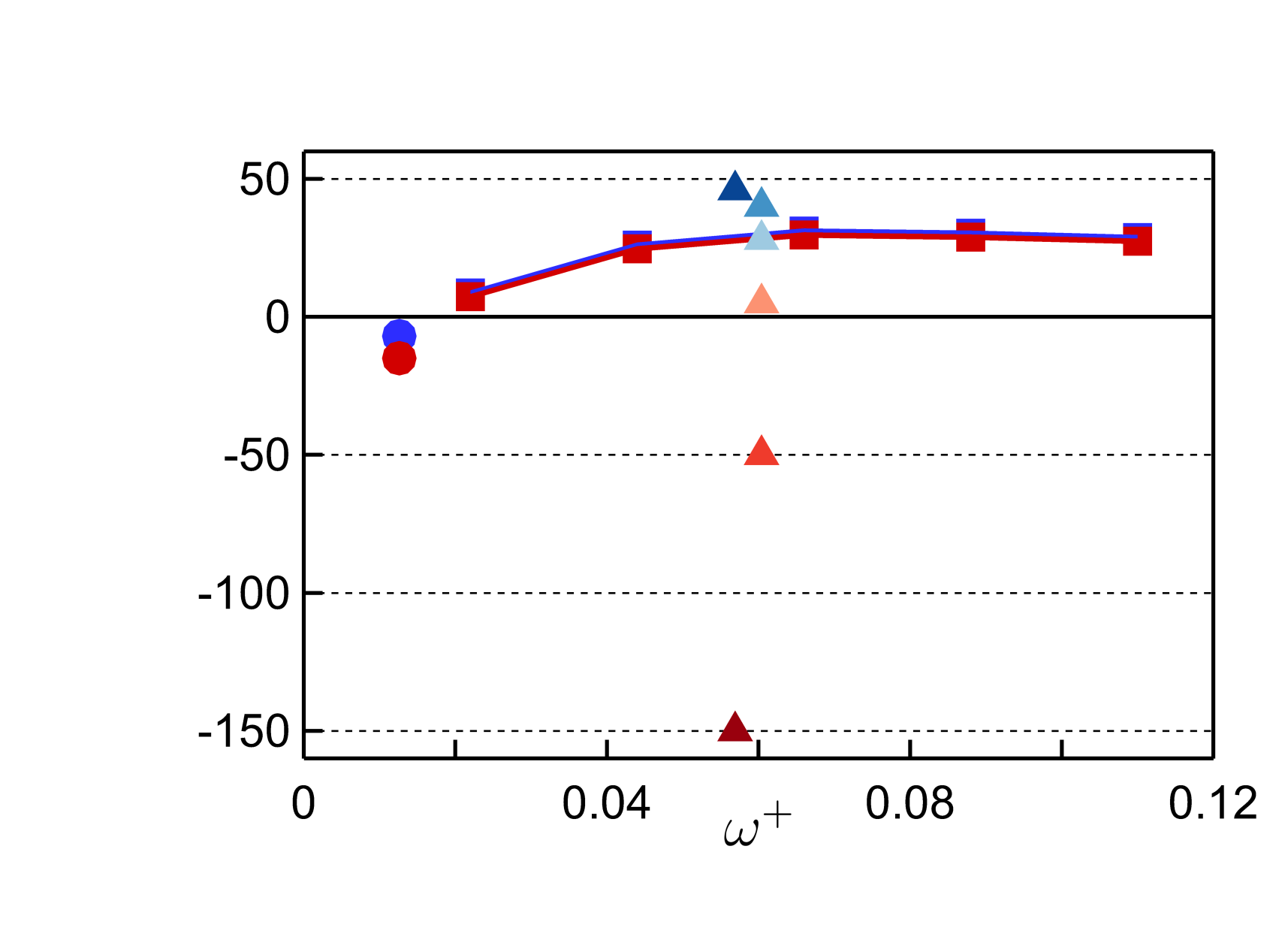} 
 \includegraphics[width=.48\textwidth,trim={{0.0\textwidth} {-0.15\textwidth} {0.0\textwidth} {0.0\textwidth}},clip]{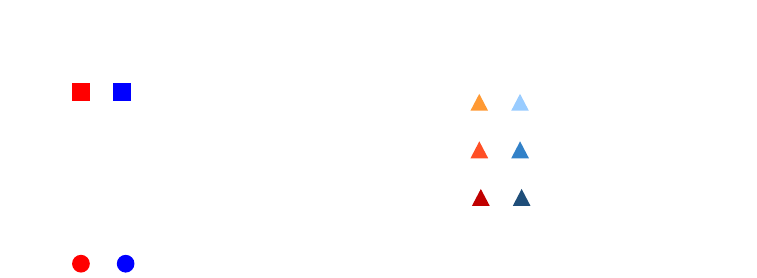}
 \put(-5.3,2.62){\scriptsize{Present study DNS}}
 \put(-5.4,2.38){\scriptsize{$Pr = 0.71, Re_{\tau_0} = 590$}}
 \put(-4.35,2.09){\scriptsize{$A^+ = 12$}}
\put(-5.3,1.42){\scriptsize{\cite{guerin2023breaking} DNS}}
 \put(-5.4,1.18){\scriptsize{$Pr = 1.0, Re_{\tau_0} = 180$}}
 \put(-4.35,0.84){\scriptsize{$A^+ = 30$}}
 \put(-2.45,2.82){\scriptsize{\cite{fang2009large} LES}}
 \put(-2.4,2.54){\scriptsize{$Pr = 0.72, Re_{\tau_0} = 180$}}
 \put(-2.0,2.28){\scriptsize{$Ma = 0.5$}}
 \put(-1.55,1.99){\scriptsize{$A^+ = 6.4$}} 
 \put(-1.55,1.67){\scriptsize{$A^+ = 12.7$}} 
 \put(-1.55,1.35){\scriptsize{$A^+ = 19.0$}}
  \caption{Plane oscillation data of $DR\%$ (blue symbols) and $HR\%$ (orange/red symbols) at $Pr = 0.7 - 1.0$ from our present study, \cite{guerin2023breaking}, and \cite{fang2009large}.}
  \label{fig:compare_HR_DR}
\end{figure}

In figure~\ref{fig:compare_HR_DR}, we plot the plane oscillation data of $DR$ and $HR$ at close $Pr = 0.71 - 1.0$ from our present study (figure~\ref{fig:averaged_HR_DR_Pr}\textit{b}), \cite{guerin2023breaking}, and \cite{fang2009large}. Unlike our data, the two cited studies obtain heat-transfer increase ($HR \lesssim 0$). Compared to \cite{guerin2023breaking}, our $A^+$ values differ by $2.5$ times, and $\omega^+$ values differ by $1.7$ times. Nevertheless, based on our findings from \S\ref{sec:stokes}-\ref{sec:plane_oscillation}, we speculate that $HR < DR < 0$ by \cite{guerin2023breaking} is related to the highly protrusive Stokes layer (beyond $50$ viscous units) for $\omega^+ < 0.04$ (figure~\ref{fig:DR_HR_duw_dcw_plane}). Compared to \cite{fang2009large}, we have close values of $A^+$ and $\omega^+$. At $A^+ \simeq 12.0$ and $\omega^+\simeq 0.06$, the $DR$ values differ by $9\%$ between \cite{fang2009large} ($DR = 40\%$, $Re_{\tau_0} = 180$) and our study ($DR = 31\%$ at $Re_{\tau_0} = 590$); this is due to the Reynolds number difference, as discussed in the literature~\citep{gatti2016,marusic2021,rouhi2022turbulent}. However, the values of $HR$ have opposite signs and are different by $80\%$, $HR = 30\%$ from our study versus $HR = -50\%$ from \cite{fang2009large}. In addition to $Re_{\tau_0}$, we speculate that such difference is related to the different computational setups. We conduct DNS of incompressible turbulent channel flow with $\Delta^+_x \times \Delta^+_z = 8.3 \times 3.9$, while \cite{fang2009large} conduct large-eddy simulation (LES) of compressible turbulent channel flow (Mach number $Ma = 0.5$) with $\Delta^+_x \times \Delta^+_z = 35 \times 12$.

\subsection{Optimal actuation frequency}\label{sec:optimal_omega}

Figure~\ref{fig:averaged_HR_DR_Pr} indicates that the optimal actuation frequency for $DR$ is $\omega^+ \approx 0.044$ for the travelling wave, and $\approx 0.066$  for the plane oscillation, regardless of the Prandtl number.   For $HR$, the optimal frequency coincides with that for $DR$ at $Pr = 0.71$, but it increases to $0.088$ for both types of actuation as the Prandtl number increases to 7.5. 

\begin{figure}
  \centering
 \includegraphics[width=1.0\textwidth,trim={{0.0\textwidth} {0.02\textwidth} {0.1\textwidth} {0.0\textwidth}},clip]{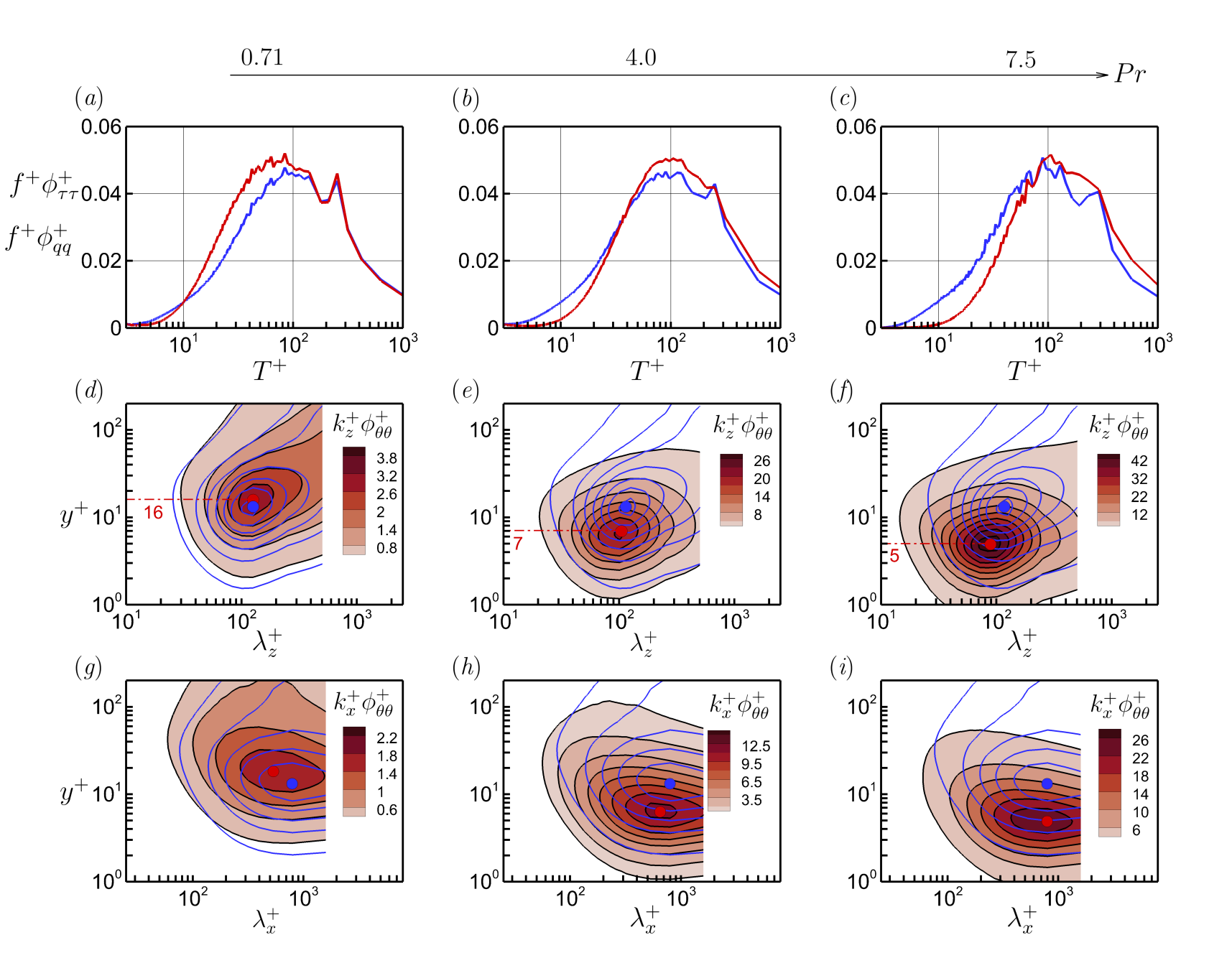}
  \caption{Non-actuated half channel flow with increasing Prandtl number. (\textit{a,d,g}) $Pr = 0.71$; (\textit{b,e,h}) $Pr = 4.0$; (\textit{c,f,i}) $Pr = 7.5$. (\textit{a--c}) Pre-multiplied frequency spectra of wall shear-stress $f^+ \phi^+_{\tau \tau}$ (blue line) and wall heat-flux $f^+ \phi^+_{qq}$ (red line), where $T^+ = 1/f^+$. (\textit{d--f}) Spectrograms of the streamwise velocity fluctuations (blue contour lines) and temperature fluctuations (filled contour) pre-multiplied by the spanwise wavenumber $k^+_z$. (\textit{g--i}) Same as (\textit{d--f}), pre-multiplied by the streamwise wavenumber $k^+_x$. Contour levels for $k^+_z \phi^+_{u u}$ start from $0.2$ to $3.8$ with an increment of $0.6$; contour levels for $k^+_x \phi^+_{u u}$ start from $0.2$ to $2.2$ with an increment of $0.4$. The blue and red bullets in (\textit{d--f}) locate the maximum in $k^+_z \phi^+_{u u}$ and $k^+_z \phi^+_{\theta \theta}$, respectively, and in (\textit{g--i}) locate the maximum in $k^+_x \phi^+_{u u}$ and $k^+_x \phi^+_{\theta \theta}$, respectively.}
  \label{fig:non_actuated_spectra}
\end{figure}

Previous studies relate the optimal frequency for drag reduction to the characteristic time-scale of the energetic velocity scales associated with the near-wall cycle of turbulence, $\mathcal{T}^+_u $ \citep{quadrio2009,chandran2022turbulent}. When the actuation period $T^+_{osc} \equiv 2 \pi/\omega^+$ matches this time scale, the wall oscillation becomes more effective in disrupting the near-wall scales, leading to the maximum $DR$~\citep{ricco2021review}. \cite{chandran2022turbulent} discuss this interaction using the pre-multiplied spectrum of wall shear-stress $f^+ \phi^+_{\tau \tau}$, and the corresponding spectra for our non-actuated case are shown in figure~\ref{fig:non_actuated_spectra}(\textit{a--c}) (blue lines).   In agreement with \cite{chandran2022turbulent}, the peaks in the shear-stress spectra occur at $\mathcal{T}^+_u \simeq 100$, corresponding to $\omega^+ \simeq 0.063$, 
which broadly matches the frequencies of actuation for maximum drag reduction found here. 

Figure~\ref{fig:non_actuated_spectra} also shows the corresponding spectra for the fluctuating wall heat-flux, $f^+ \phi^+_{qq}$ (red lines), as well as the pre-multiplied spectrograms of the fluctuating streamwise velocity and temperature ($k^+_z \phi^+_{uu}, \ k^+_x \phi^+_{uu}, \ k^+_z \phi^+_{\theta\theta}, \ k^+_x \phi^+_{\theta\theta}$).   Figures~\ref{fig:non_actuated_spectra}(\textit{a,b,c}) reveal that while $f^+ \phi^+_{qq}$ is sensitive to $Pr$ at small time-scales ($T^+ \lesssim 70$), the time scale of the energetic temperature scales $\mathcal{T}^+_\theta \simeq 100$ is insensitive to $Pr$, and coincides with $\mathcal{T}^+_u$.
The streamwise and spanwise lengths of the energetic temperature scales vary somewhat with Prandtl number (see figures~\ref{fig:non_actuated_spectra}\textit{d--i}), but they also remain close to the length-scales of the energetic velocity scales.   However, the peak in $k^+_z \phi^+_{\theta \theta}$ drops from $y^+ = 16$ at $Pr = 0.71$ to $y^+ = 5$ at $Pr = 7.5$, as the conductive sublayer thins with increasing Prandtl number \citep{alcantara2021directb,kader1981temperature,schwertfirm2007dns}. The $y^+$ location of the peak in $k^+_z \phi^+_{\theta \theta}$ scales with $\sim Pr^{-1/2}$, which is steeper than the scaling of $\sim Pr^{-1/3}$ for the conductive sublayer thickness for $Pr \gtrsim 1$~\citep{shaw1977turbulent,kader1981temperature,schwertfirm2007dns,alcantara2021directb,pirozzoli2023prandtl}. We also analysed the $\overline{\theta^2}^+$ profiles of \cite{pirozzoli2023prandtl} for the DNS of turbulent pipe flow at $Re_{\tau_0} \simeq 1100$ and $Pr = 0.5, 1, 2, 4, 16$, and the $y^+$ locations of their inner peaks scale with $\sim Pr^{-1/2}$.

For the plane oscillation, we expect that the maximum $DR$ and $HR$ is achieved when the actuation period is close to $\mathcal{T}^+_u$ and $\mathcal{T}^+_\theta$, respectively. 
For the travelling wave, however, the optimal actuation period cannot be directly compared with $\mathcal{T}^+_u$ or $\mathcal{T}^+_\theta$, owing to the relative streamwise motion between the travelling wave and the advecting near-wall scales. As discussed by \cite{quadrio2009}, for maximum $DR$, $\mathcal{T}^+_u$ must be compared with a relative oscillation period $2\pi/(\kappa^+_x \mathcal{U}^+_u + \omega^+)$ as seen by an observer travelling with the convection speed of the near-wall energetic velocity scales $\mathcal{U}^+_u$. Similarly, for maximum $HR$, $\mathcal{T}^+_\theta$ must be compared with $2\pi/(\kappa^+_x \mathcal{U}^+_\theta + \omega^+)$, where $\mathcal{U}^+_\theta$ is the convection speed of the near-wall energetic temperature scales. After some recasting (using $\mathcal{T}^+_u \simeq \mathcal{T}^+_\theta \simeq 100$, $\kappa^+_x = 0.0014$), we estimate the optimum frequencies of actuation for $DR$ and $HR$ to be, respectively, 
\begin{align}
 \omega^+_{\mathrm{opt}, DR} &= \frac{2\pi}{\mathcal{T}^+_u} - \kappa^+_x \mathcal{U}^+_u \simeq 0.046 \tag{3.1\textit{a}} \label{eq:t_u}, \\
 \omega^+_{\mathrm{opt}, HR} &= \frac{2\pi}{\mathcal{T}^+_\theta} - \kappa^+_x \mathcal{U}^+_\theta \simeq 0.063 - 0.0014 \mathcal{U}^+_\theta, \tag{3.1\textit{b}} \label{eq:t_c}
\end{align}
where we have assumed that $\mathcal{U}^+_u$  follows a universal curve \citep{kim1993propagation,del2009estimation,liu2020input}, where $\mathcal{U}^+_u \simeq 12$ at $y^+ \simeq 15$ (marked with blue bullets in figures~\ref{fig:non_actuated_spectra}\textit{d--i}).  

The estimated optimum frequency for $DR$ is independent of Prandtl number, at about $0.046$, consistent with our DNS results for the travelling wave case (figures~\ref{fig:averaged_HR_DR_Pr}\textit{a,c,e}). 
However, the optimum frequency for $HR$ depends on $Pr$, since we expect that $\mathcal{U}^+_\theta$ decreases as the energetic temperature scales move closer to the wall with increasing Prandtl number  (red bullets in figure~\ref{fig:non_actuated_spectra}\textit{d--i}).  
\cite{hetsroni2004convection} computed the profiles of $\mathcal{U}^+_\theta$ at $Pr = 1.0$, 5.4 and 54 in a configuration similar to that used in the present study, without actuation.  Using their results, we estimate that $\mathcal{U}^+_\theta$  decreases from $12$ to $6$ as $Pr$ increases from $0.71$ to $7.5$, which gives $\omega^+_{\mathrm{opt},HR} = 0.046$ at $Pr = 0.71$ and $0.055$ at $Pr = 7.5$. The estimated value at $Pr = 7.5$ is smaller than the value of 0.088 given by the DNS (figure~\ref{fig:averaged_HR_DR_Pr}\textit{e}), 
but the  trends with Prandtl number  are consistent.  In \S\ref{sec:stokes} and \ref{sec:stokes2}, we conduct a more quantitative justification for the trends in $\omega^+_{\mathrm{opt},HR}$ by considering  the interaction between the Stokes layer and the near-wall thermal field.

\begin{figure}
  \centering
 \includegraphics[width=1.0\textwidth,trim={{0.0\textwidth} {0.04\textwidth} {0.0\textwidth} {0.0\textwidth}},clip]{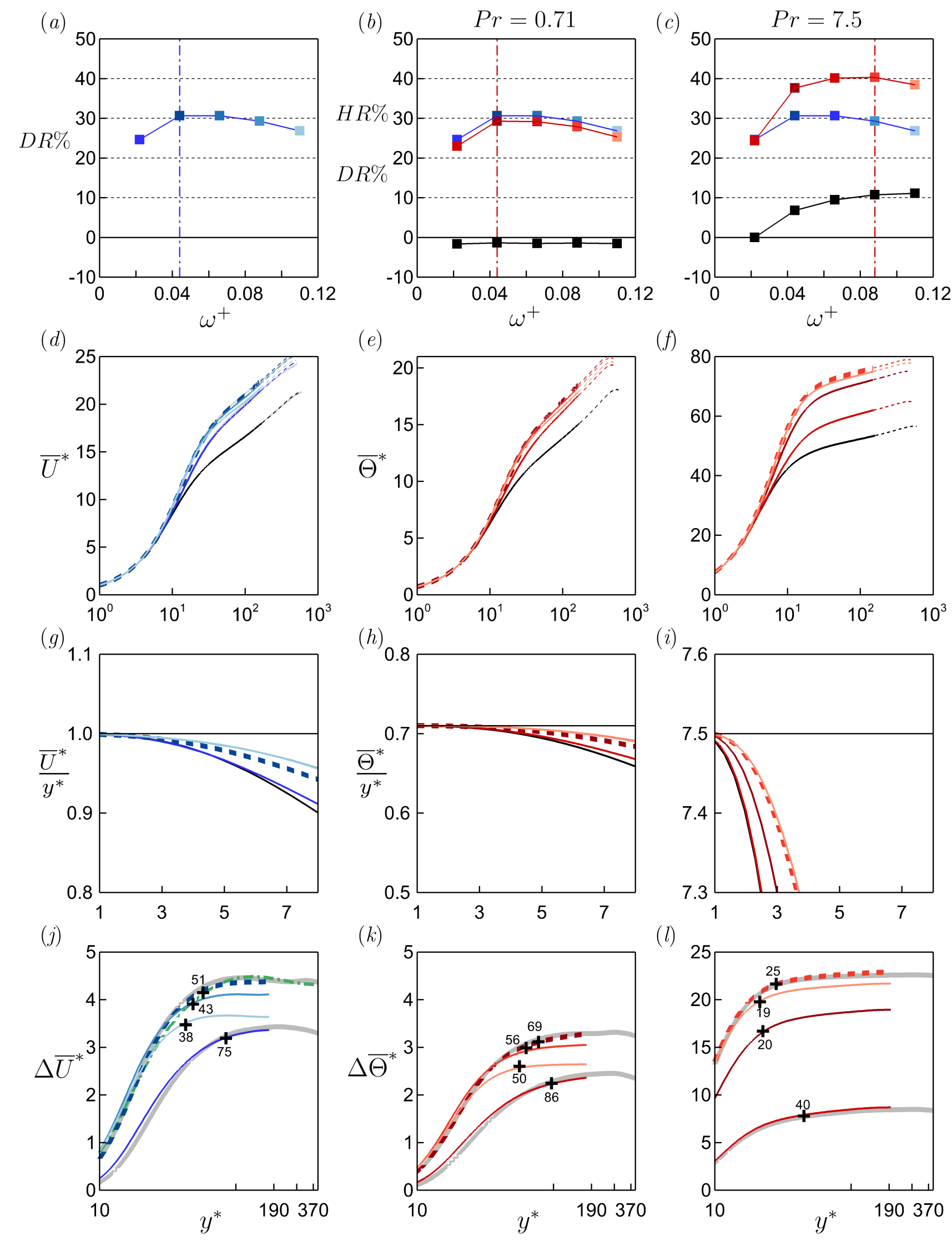}
  \caption{Travelling wave actuation with $A^+ = 12$, $\kappa^+_x = 0.0014$. Left column: $DR\%$ and $\overline{U}^*$ profiles, where $\Delta \overline{U}^* = \overline{U}^* - \overline{U}^*_0$ (results independent of $Pr$). Middle and right columns correspond to $Pr=0.71$ and $7.5$, respectively: $DR\%$ (blue line), $HR\%$ (red line), and $\overline{\Theta}^*$ profiles, where $\Delta \overline{\Theta}^* = \Theta^* - \Theta^*_0$. In the panels (\textit{d--i}), the black line is the reference (non-actuated) case, the blue lines and the red lines show the effects of increasing $\omega^+$ on $\overline{U}^*$ and $\overline{\Theta^*}$, respectively, and the thick dashed line corresponds to $DR_\mathrm{max}$ or $HR_\mathrm{max}$. In (\textit{d,e,f}), the thin dotted lines for $y^* \gtrsim 170$ are the reconstructed profiles following \S\ \ref{sec:calculate_cf_ch}. In (\textit{j,k,l}), we plot the reduced-domain profiles (red/blue) up to $y^*_\mathrm{res} \simeq 170$, and the thick grey profiles are from the full-domain cases ($L_x \times L_z = 7.6 h \times \pi h$) from Appendix~\ref{sec:grid_domain}; the dashed-dotted green profile in (\textit{j}) is at matched $A^+ = 12, \kappa^+_x = 0.0014, \omega^+ = 0.044$ but at $Re_{\tau_0} = 950$ from \cite{rouhi2022turbulent}. The black cross symbols mark the distance $y^*$ where $\Delta \overline{U}^* = 0.9 \Delta \overline{U}^*_{170}$ ($\Delta \overline{\Theta}^* = 0.9 \Delta \overline{\Theta}^*_{170}$).}
  \label{fig:profiles_mean}
\end{figure}

\subsection{Mean profiles and turbulence statistics}\label{sec:mean_profiles}

\begin{figure}
  \centering
 \includegraphics[width=1.0\textwidth,trim={{0.0\textwidth} {0.0\textwidth} {0.0\textwidth} {0.0\textwidth}},clip]{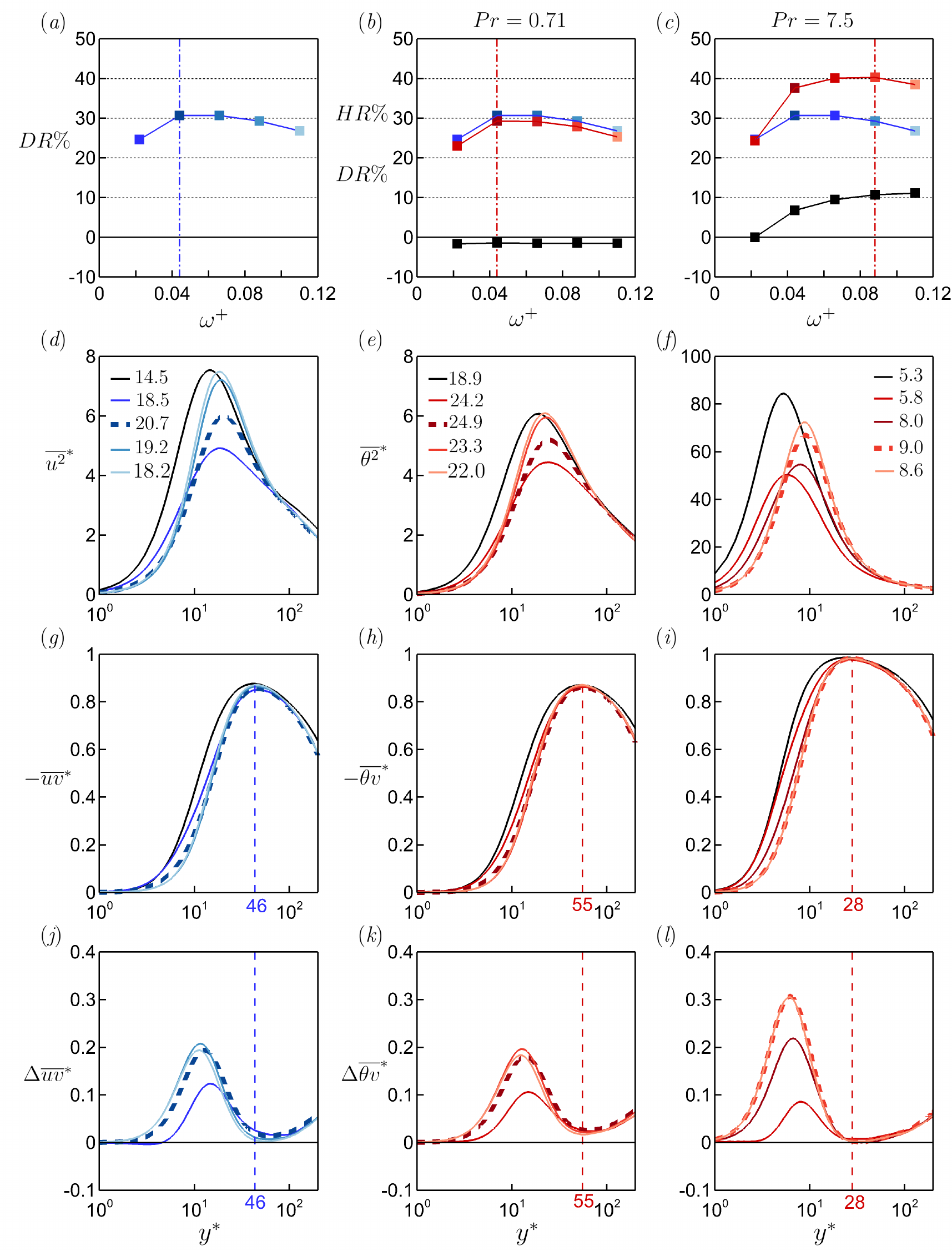}
  \caption{Travelling wave actuation with $A^+ = 12$, $\kappa^+_x = 0.0014$.  Notation and symbols as in  figure~\ref{fig:profiles_mean}.  Here, $\Delta \overline{ u v}^* = \overline{ u v }^* - \overline{ u v }^*_0$, and $\Delta \overline{ \theta v }^* = \overline{ \theta v }^* - \overline{ \theta v }^*_0$. In (\textit{d,e,f}), the numbers give the $y^*$ location of the inner-peak in the $\overline{ u^2 }^*$ and $\overline{\theta^2 }^*$ profiles. In (\textit{g--l}), the vertical dashed line locates the peak of non-actuated $\overline{ u v }^*_0$ (\textit{g,j}) and $\overline{ \theta v }^*_0$ (\textit{h,i,k,l}).}
  \label{fig:profiles_stress}
\end{figure}

We now consider the distributions of the mean velocity $\overline{U}^*$ ($=\overline{U}/u_\tau$), the mean temperature $\overline{\Theta}^*$ ($= \overline{\Theta}/\theta_\tau$)  (figure~\ref{fig:profiles_mean}), and the velocity and temperature statistics (figure~\ref{fig:profiles_stress}) as we vary $\omega^+$ and $Pr$. We focus on the travelling wave case, but since the observed trends are consistent with those from the plane oscillation case (figures~\ref{fig:profiles_mean_plane} and \ref{fig:profiles_stress_plane} in Appendix~\ref{sec:plane_stats}), our conclusions are applicable to both types of actuation.

It is well known from the literature that drag reduction by either plane oscillation or travelling wave coincides with the thickening of the viscous sublayer~\citep{choi1998turbulent,choi2001mechanism,choi2002near,di2002particle,touber2012,hurst2014,gatti2016,chandran2022turbulent,rouhi2022turbulent}. This is evident also from our $\overline{U}^*/y^*$ profiles (figure~\ref{fig:profiles_mean}\textit{g}),  where the actuated profiles (blue profiles) depart from unity  farther from the wall compared to the non-actuated profile (black profile). The conductive sublayer also thickens  with wall oscillation, but it then thins substantially with increasing Prandtl number, so that at $Pr = 7.5$  the conductive sublayer is substantially thinner than the viscous sublayer (figure~\ref{fig:profiles_mean}\textit{i}). 

Viscous sublayer and conductive sublayer thickening due to the wall oscillation shifts the $\overline{U}^*$ and $\overline{\Theta}^*$ profiles. This is shown in figure~\ref{fig:profiles_mean}(\textit{j,k,l}) where we plot the differences $\Delta \overline{U}^* = \overline{U}^* - \overline{U}^*_0$ and $\Delta \overline{\Theta}^* = \overline{\Theta}^* - \overline{\Theta}^*_0$ between the actuated and the non-actuated cases.  The cases with the maximum $DR$ (or $HR$) have the highest $\Delta \overline{U}^*$ (or $\Delta \overline{\Theta}^*$), as highlighted by the thick dashed line in the plots. The profiles in colour are from the production runs with the reduced domain (table~\ref{tab:runs}), and are plotted up to $y^*_\mathrm{res} \simeq 170$. As a reference, the profiles in grey are from the full-domain cases from Appendix~\ref{sec:grid_domain}. There is a good agreement between the reduced-domain and the full-domain profiles up to $y^*_\mathrm{res}$. Furthermore, the variations of $\Delta \overline{U}^*$ and $\Delta \overline{\Theta}^*$ beyond $y^*_\mathrm{res}$ are within $3\%$. Therefore, it is reasonable to consider $\Delta \overline{U}^*$ and $\Delta \overline{\Theta}^*$ at $y^*_\mathrm{res} = 170$ ($\Delta \overline{U}^*_{170}$ and $\Delta \overline{\Theta}^*_{170}$) as their asymptotic values. According to \cite{rouhi2022turbulent}, the distance $y^*$ where $\Delta \overline{U}^*$ reaches a plateau indicates the extent to which the Stokes layer disturbs the $\overline{U}^*$ profile.  In figure~\ref{fig:profiles_mean}(\textit{j,k,l}), therefore, we mark each profile at the $y^*$ location where $\Delta \overline{U}^* = 0.9 \Delta \overline{U}^*_{170}$ ($\Delta \overline{\Theta}^* = 0.9 \Delta \overline{\Theta}^*_{170}$). For the present profiles with fixed $Re_{\tau_0}, A^+$ and $\kappa^+_x$, the distance to reach the plateau in $\Delta \overline{\Theta}^*$  depends on $\omega^+$ and $Pr$ (compare figures~\ref{fig:profiles_mean}\textit{k,l}), where the plateau is reached at a lower $y^*$ with increasing Prandtl number. We expect that the $\Delta \overline{\Theta}^*$ profiles and the distances to their plateaus would also depend on $A^+$ and $\kappa^+_x$, but not on $Re_{\tau_0}$. Our conjecture is based on the behaviour of its analogue $\Delta \overline{U}^*$ (figure~\ref{fig:profiles_mean}\textit{j}), that for a fixed set of actuation parameters is almost identical between $Re_{\tau_0} = 590$ (thick blue dashed line) and $Re_{\tau_0} = 950$ (green dashed-dotted line). The Reynolds number independence of the plateau in $\Delta \overline{U}^*$ for the travelling wave is further confirmed by \cite{gatti2016} and \cite{gatti2024turbulent}.  


In figure~\ref{fig:profiles_stress}, we plot the turbulence statistics for the same cases shown in figure~\ref{fig:profiles_mean}. Drag reduction and viscous sublayer thickening (figures~\ref{fig:profiles_mean}\textit{a,d,g}) coincide with the near-wall attenuation of $\overline{ u^2}^*$ and the shift in its inner peak (figure~\ref{fig:profiles_stress}\textit{d}), as known from previous work~\citep{quadrio2000numerical,ricco2004effects,quadrio2011,touber2012,ricco2021review,rouhi2022turbulent}. Similarly, we find that at each $Pr$, heat-transfer reduction and conductive sublayer thickening (figures~\ref{fig:profiles_mean}\textit{b,c,e,f,h,i}) coincide with the near-wall attenuation of $\overline{ \theta^2 }^*$ and the shift in its inner peak (figures~\ref{fig:profiles_stress}\textit{e,f}). Interestingly, $DR_\mathrm{max}$ occurs when the inner peak in $\overline{ u^2}^*$ is located farthest from the wall, and a similar connection exists between $HR_\mathrm{max}$ and the location of the inner peak in $\overline{ \theta^2}^*$ (compare the listed numbers in figures~\ref{fig:profiles_stress}\textit{d,e,f}).  

\begin{figure}
  \centering
 \includegraphics[width=1.0\textwidth,trim={{0.05\textwidth} {0.02\textwidth} {0.05\textwidth} {0.0\textwidth}},clip]{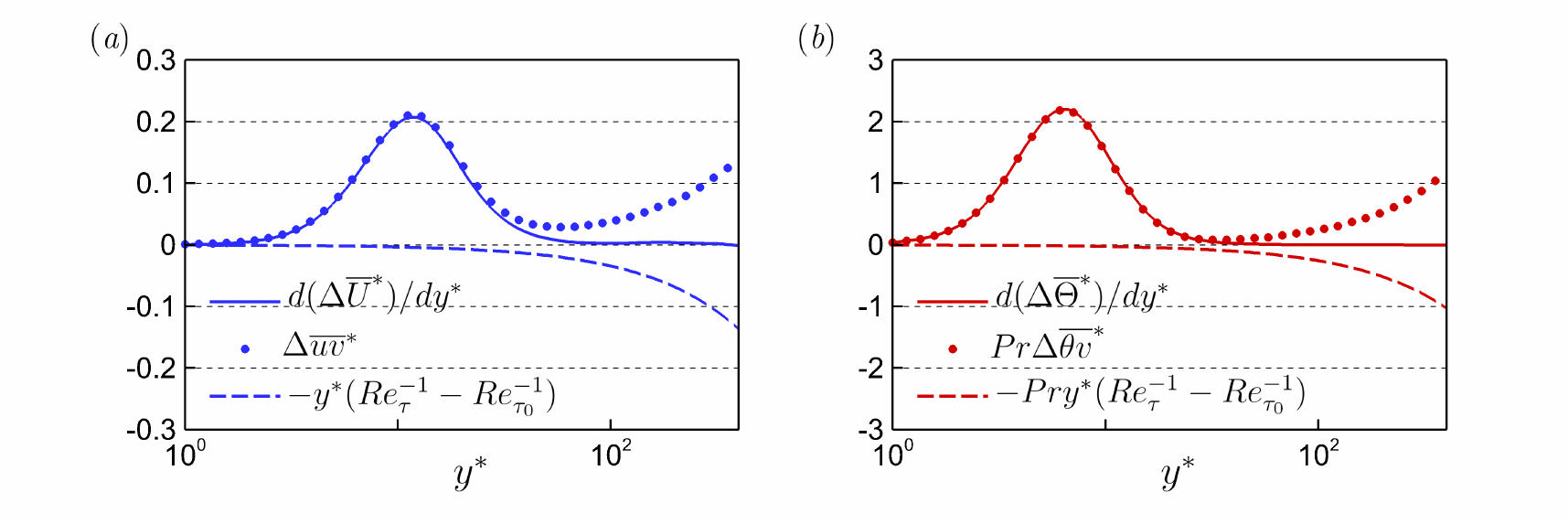}
  \caption{Travelling wave actuation with $A^+ = 12$, $\kappa^+_x = 0.0014$, $\omega^+ = 0.088$ at $Pr = 7.5$.  (\textit{a}) Terms in the averaged streamwise momentum equation \eqref{eq:dDUp}. (\textit{b}) Terms in the averaged temperature equation \eqref{eq:dDCp}. 
  }
  \label{fig:budgets_Dup_Dcp}
\end{figure}

In addition, the attenuation of the turbulent shear-stress $\overline{ u v }^*$ (figure~\ref{fig:profiles_stress}\textit{g}) and the wall-normal turbulent temperature-flux $\overline{ \theta v }^*$ (figures~\ref{fig:profiles_stress}\textit{h,i}) are directly linked to $DR$ and $HR$, respectively. 
In figure~\ref{fig:profiles_stress}(\textit{j,k,l}), we plot the differences between the actuated and non-actuated cases $\Delta \overline{ u v }^* = \overline{ u v }^* - \overline{ u v }^*_0$ and $\Delta \overline{ \theta v }^* = \overline{ \theta v }^* - \overline{ \theta v }^*_0$.  Most of the attenuation occurs near the wall, with the maxima occurring near $y^* \simeq 10$. For $y^* > 10$, the attenuation declines, and  $\Delta \overline{ u v }^*$ and $\Delta \overline{ \theta v }^*$ reach local minima at points that coincide approximately with the locations of the peaks in $\overline{ u v }^*_0$ and $\overline{ \theta v }^*_0$. 


By integrating the plane- and time-averaged streamwise momentum (\ref{eq:mom}) and temperature (\ref{eq:scalar}) equations from zero to $y^*$, we obtain
\begin{align}
 \frac{d \overline{U}^*}{dy^*} = \overline{ u v }^* + \left( 1 - y^* Re_\tau^{-1} \right), \qquad
  \frac{d \overline{\Theta}^*}{dy^*} = Pr \, \left[ \overline{ \theta v }^* + \left( 1 - y^* Re_\tau^{-1} \right) + \mbox{Res}_{\Theta} \right], \tag{3.2\textit{a,b}} 
\end{align}
where $\mbox{Res}_\Theta \equiv Re^{-1}_\tau \int_0^{y^*} (1 - \overline{U}^*/U^*_b)dy'$. Hence,
\begin{align}
 \frac{d (\Delta \overline{U}^*)}{dy^*} &= \Delta \overline{ u v }^* - y^*\left( {Re_\tau^{-1}} - {Re_{\tau_0}^{-1}} \right), \tag{3.3\textit{a}} \label{eq:dDUp} \\
  \frac{d(\Delta \overline{\Theta}^*)}{dy^*} &= Pr \, \left[ \Delta \overline{ \theta v }^* - y^*\left( {Re_\tau^{-1}} - {Re_{\tau_0}^{-1}} \right) + \underbrace{\Delta \mbox{Res}_{\Theta}}_{\simeq 0} \right]. \tag{3.3\textit{b}} \label{eq:dDCp}
\end{align}
The difference $\Delta \mbox{Res}_{\Theta}$ between the actuated and the non-actuated cases is small and can be neglected.  The remaining terms in (\ref{eq:dDUp}) and (\ref{eq:dDCp}) are plotted in figure~\ref{fig:budgets_Dup_Dcp} for the travelling wave case at $Pr = 7.5$ and $\omega^+=0.088$. 
We see that the right-hand-side of \eqref{eq:dDUp} and \eqref{eq:dDCp} are dominated by $\Delta \overline{uv}^*$ and $\Delta \overline{\theta v}^*$, up to their respective minima, but farther from the wall these terms are cancelled by the term containing the difference in Reynolds numbers $\left( {Re_\tau^{-1}} - {Re_{\tau_0}^{-1}} \right)$.  That is, the net contributions to $\Delta \overline{U}^*$ and $\Delta \overline{\Theta}^*$, hence $DR$ and $HR$, come from $\Delta \overline{uv}^*$ and $\Delta \overline{\theta v}^*$, and then only up to the points where $\Delta \overline{uv}^*$ and $\Delta \overline{\theta v}^*$ reach their minimum values. 

\subsection{Source of inequality between $HR$ and $DR$}
\label{sec:hr_dr_break}

Integrating (\ref{eq:dDUp}) and (\ref{eq:dDCp}) once more with respect to $y^*$ gives 
\begin{align}
 \Delta \overline{U}^*_{y^*} =\int_0^{y^*}  \Delta^*_{uv} dy', \qquad  
 \Delta \overline{\Theta}^*_{y^*} =  Pr \int_0^{y^*}  \Delta^*_{\theta v} dy', \tag{3.4\textit{a,b}} \label{eq:uv_cv_int}
\end{align}
where
\begin{align}
 \Delta^*_{uv} & \equiv \Delta \overline{ u v }^* - y^*(Re_\tau^{-1} - Re_{\tau_0}^{-1}), \tag{3.4\textit{c}} \label{eq:uv_cv_net1} \\[1mm] \Delta^*_{\theta v} & \equiv \Delta \overline{ \theta v }^* - y^*(Re_\tau^{-1} - Re_{\tau_0}^{-1}). \tag{3.4\textit{d}} \label{eq:uv_cv_net2}
\end{align}
By using \eqref{eq:uv_cv_int} we can relate $\Delta^*_{uv}$ and $\Delta^*_{\theta v}$ to $\Delta \overline{U}^*$ and $\Delta \overline{\Theta}^*$, hence to $DR$ and $HR$, and so establish the connection between the drag and heat-transfer reduction and the turbulence attenuation. For the drag reduction, \cite{gatti2016} derived the relation between $DR$ ($=1-R_f$) and the asymptotic value of $\Delta \overline{U}^*$ in the log region \eqref{eq:dr_model1}. In Appendix \ref{sec:gq_model_hr}, we derive a similar relation between $HR$ ($=1-R_h$) and the asymptotic value of $\Delta \overline{\Theta}^*$ in the log region \eqref{eq:hr_model1}.  That is, we have
\begin{align}
\Delta \overline{U}^*_{170} &  = 
\sqrt{\frac{2}{C_{{f_0}}}}\left[ \frac{1}{\sqrt{R_f}} - 1 \right] - \frac{1}{2\kappa_u}\ln{R_f}   = \underbrace{\int^{170}_0 \Delta^*_{uv} dy'}_{I_{uv}} \tag{3.5\textit{a}} \label{eq:dr_model1} \\
\Delta \overline{\Theta}^*_{170} & =   \frac{\sqrt{C_{{f_0}}/2}}{C_{{h_0}}}\left[ \frac{\sqrt{R_f}}{R_h} - 1 \right] - \frac{1}{2\kappa_\theta}\ln{R_f}  = Pr \underbrace{\int^{170}_0 \Delta^*_{\theta v} dy'}_{I_{\theta v}} \tag{3.5\textit{b}} \label{eq:hr_model1}
\end{align}
These derivations assume that the profiles of $\overline{U}^*$ and $\overline{\Theta}^*$ have well-defined log regions with slopes that are not affected by the wall oscillation, which is the case for our results (figure~\ref{fig:profiles_mean}).  
The integration was performed up to $y^*_\mathrm{res} = 170$, where the  profiles of $\Delta \overline{U}^*$ and $\Delta \overline{\Theta}^*$ reach their asymptotic levels (figures~\ref{fig:profiles_mean}\textit{j,k,l}) and their derivatives are zero (figure~\ref{fig:budgets_Dup_Dcp}). On the right-hand-sides of \eqref{eq:dr_model1} and \eqref{eq:hr_model1} we have the integrals of attenuation in the turbulent shear-stress ($I_{uv}$) and the turbulent temperature-flux ($I_{\theta v}$). For a fixed set of viscous-scaled actuation parameters, $I_{uv} = \Delta \overline{U}^*_{170}$ is constant~\citep{gatti2016,rouhi2022turbulent} (also shown in figure~\ref{fig:profiles_mean}\textit{j}), but $I_{\theta v} = \Delta \overline{\Theta}^*_{170}/Pr$ depends on $Pr$ (figures~\ref{fig:profiles_stress}\textit{k,l}). 

\begin{figure}
  \centering
 \includegraphics[width=0.9\textwidth,trim={{0.0\textwidth} {0.0\textwidth} {0.0\textwidth} {0.0\textwidth}},clip]{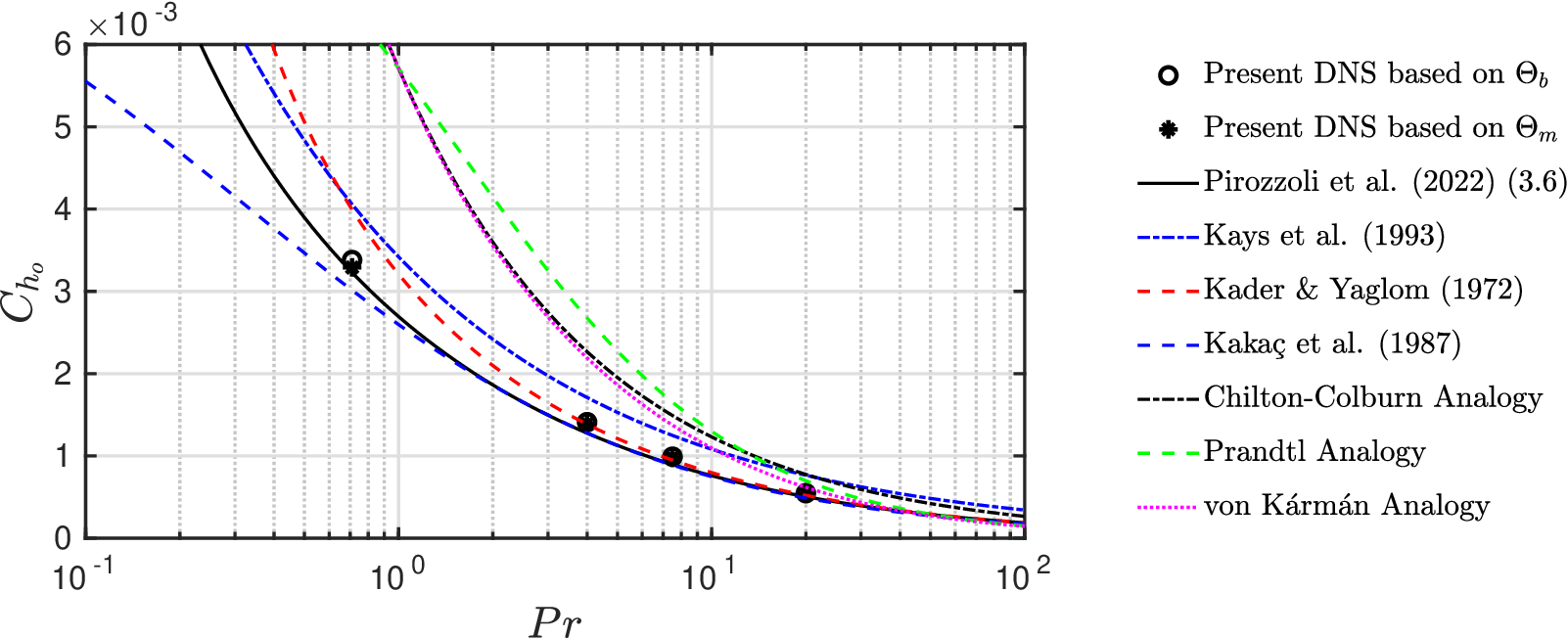}
  \caption{Comparison of the non-actuated $C_{h_0}$ between our DNS cases (left side of table~\ref{tab:runs}) and several empirical relations. For the DNS data, $C_{h_0}$ is calculated either based on the bulk temperature $\Theta_b$ (empty symbols) or the mixed-mean temperature $\Theta_m$ (filled symbols). The empirical relations are by \cite{pirozzoli2022dns} (\ref{eq:ch_priozzoli}), \cite{kays1993}, \cite{kader1972heat}, \cite{kakacc1987handbook}, Chilton-Colburn Analogy~\citep{chilton1934mass,colburn1964method}, Prandtl Analogy (see 11 in \citealt{schlunder1998analogy}), and von K\'arm\'an Analogy (see 4 in \citealt{li2010modeling}).}
  \label{fig:Ch0}
\end{figure}

To predict $DR$ and $HR$ from \eqref{eq:dr_model1} and \eqref{eq:hr_model1}, we only need $\Delta \overline{U}^*_{170}$ and $\Delta \overline{\Theta}^*_{170}$ (that is, $I_{uv}$ and $I_{\theta v}$) as the inputs; the other parameters are associated with the non-actuated channel flow for which semi-empirical relations exist in the literature.  We choose $\kappa_u = 0.4$ and $\kappa_\theta = 0.46$~\citep{pirozzoli2016passive}; the suitability of these choices is confirmed in \S\ref{sec:calculate_cf_ch} and Appendix~\ref{sec:grid_domain}. \cite{dean1978}'s correlation is used for the non-actuated skin-friction coefficient $C_{f_0} = 0.073(2Re_b)^{-1/4}$, which could be re-expressed as $C_{{f_0}} = 0.037Re^{-2/7}_{\tau_0}$; this correlation agrees well with the DNS data~\citep{macdonald2019}, and yields less than $2\%$ difference with our DNS result $C_{f_0} = 0.0057$ at $Re_{\tau_0} = 590$. For the non-actuated $C_{h_0}$, in figure~\ref{fig:Ch0} we compare our DNS data at $Re_{\tau_0} = 590$ and different $Pr$ (cases on the left side of table~\ref{tab:runs}) with several empirical relations. These relations define $C_{h_0}$ either based on the bulk temperature $\Theta_b$, or the mixed-mean temperature $\Theta_m$. For our DNS data, there is a negligible difference between the two definitions (empty versus the filled circles). Among the relations, the latest correlation by \cite{pirozzoli2022dns} agrees well with our DNS data (solid black line). This relation is an improvement to \cite{kader1972heat}'s formula, and its accuracy is supported by the recent DNS of \cite{pirozzoli2023prandtl} for $\mathcal{O}(10^{-2}) \lesssim Pr \lesssim \mathcal{O}(10^{1})$. It is formulated as
\begin{align}
 \frac{1}{C_{{h_0}}} = \frac{\kappa_u}{\kappa_\theta}\frac{2}{C_{{f_0}}} + \left( \beta_{CL} - \beta_2 - \frac{\kappa_u}{\kappa_\theta}B \right)\sqrt{\frac{2}{C_{{f_0}}}} + \beta_3. \tag{3.6} \label{eq:ch_priozzoli}
\end{align}
Here $\beta_{CL}(Pr) = B_\theta(Pr) + 3.504 - 1.5/\kappa_\theta$, $\beta_2 = 4.92$, $\beta_3 = 39.6$, $B = 1.23$, and $B_\theta(Pr)$ is the log-law additive constant for $\overline{\Theta}^*$~\citep{kader1972heat}, as introduced in Appendix~\ref{sec:gq_model_hr}. 

\begin{figure}
  \centering
 \includegraphics[width=1.0\textwidth,trim={{0.0\textwidth} {0.0\textwidth} {0.1\textwidth} {0.0\textwidth}},clip]{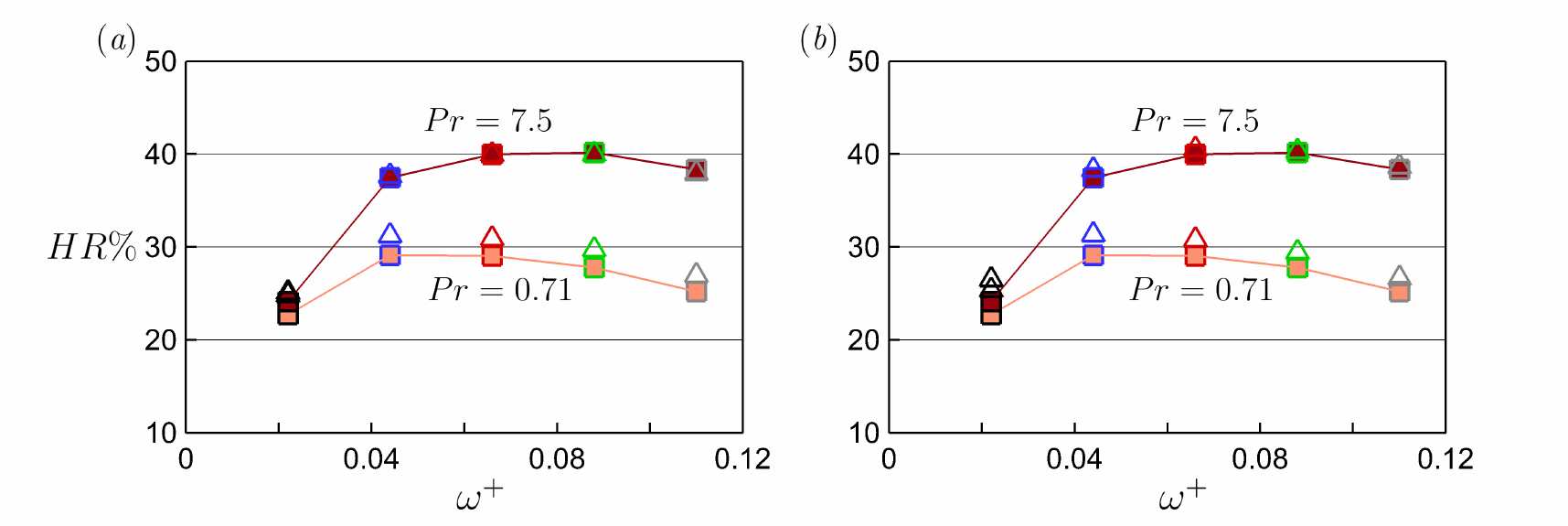}
  \caption{Travelling wave actuation with $A^+ = 12$, $\kappa^+_x = 0.0014$.  Comparison between the direct $HR$ from DNS (filled squares) and the predicted $HR$ (empty triangles) by solving \eqref{eq:dr_model1} and \eqref{eq:hr_model1}.  In (\textit{a}) for predicting $HR$ we use $I_{uv}$ and $I_{\theta v}$ directly from DNS. In (\textit{b}) for predicting $HR$ we use a power-law estimate for $I_{\theta v} = I_{uv}/Pr^\gamma$, with the values of $\gamma$ reported in figure~\ref{fig:Pr_scaling}(\textit{b}).}
  \label{fig:model_HR_DR}
\end{figure}

\begin{figure}
  \centering
 \includegraphics[width=1.0\textwidth,trim={{0.08\textwidth} {0.05\textwidth} {0.11\textwidth} {0.0\textwidth}},clip]{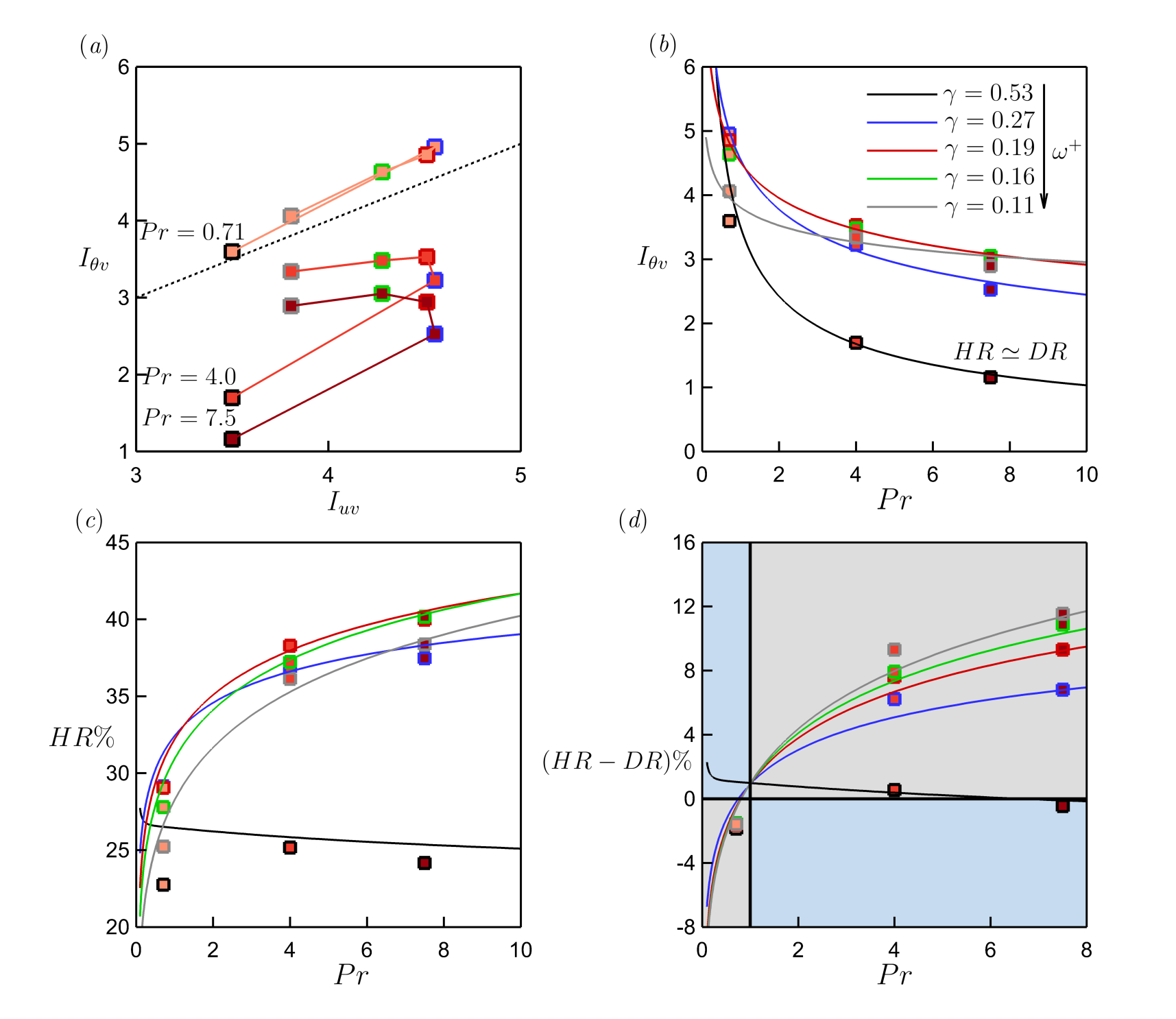}
 \put(-1.7,1.6){\scriptsize{$\gamma > 0.5$}}
 \put(-1.7,2.3){\scriptsize{$\gamma < 0.5$}}
  \caption{Travelling wave at $Pr = 0.71, 4.0$ and $7.5$ (same data as in figure~\ref{fig:averaged_HR_DR_Pr}\textit{a,c,e}).  The outline colour of each data point indicates its $\omega^+$; $\omega^+ = 0.022$ (black), $0.044$ (blue), $0.066$ (red), $0.088$ (green) and $0.110$ (grey). (\textit{a}) $I_{\theta v}$ versus $I_{uv}$; the dotted line is for $I_{\theta v}=I_{uv}$. (\textit{b}) $I_{\theta v}$ versus $Pr$; the solid curves are the power-law fit $I_{\theta v} =  I_{u v}/Pr^{\gamma}$, with $\gamma$ as shown. (\textit{c,d}) $HR$ and $HR-DR$ versus $Pr$; the solid curves are the prediction of \eqref{eq:hr_model1} using $I_{\theta v} =  I_{u v}/Pr^{\gamma}$. In (\textit{d}) blue zone marks $\gamma > 0.5$, grey zone marks $\gamma < 0.5$.  
  }
  \label{fig:Pr_scaling}
\end{figure}

In figure~\ref{fig:model_HR_DR}(\textit{a}), we compare the values of $HR$ from DNS with those from 
\eqref{eq:dr_model1} and \eqref{eq:hr_model1} using  $I_{uv}$ and $I_{\theta v}$ as obtained from DNS. The results support the accuracy of the model for relating $HR$ to the attenuation of the turbulent flux. In  figure~\ref{fig:model_HR_DR}(\textit{b}), we also show the agreement between the model and the DNS when we use a power law estimate for  $I_{\theta v}$ ($=I_{uv}/Pr^{\gamma}$) where only $I_{uv}$ (which is independent of $Pr$) is obtained from DNS. We justify the power law as follows.


From \eqref{eq:hr_model1}, we note that the Prandtl number dependence of $HR$ (hence $HR- DR$) is due to both $C_{{h_0}}$ and  $ I_{\theta v}$. According to the Reynolds analogy, we would expect 
$I_{\theta v} = I_{u v}$ when $Pr = 1.0$.   Figure~\ref{fig:Pr_scaling}(\textit{a}) indicates that, regardless of the actuation frequency $\omega^+$, $I_{\theta v} > I_{u v}$ for $Pr = 0.71$, and  $I_{\theta v} < I_{u v}$ for $Pr > 1$, with the difference increasing with $Pr$.  In other words, increasing $Pr$ beyond unity leads to a lower attenuation of $\overline{\theta v}$ compared to $\overline{u v}$, even though in this regime $HR > DR$ (figure~\ref{fig:averaged_HR_DR_Pr}\textit{c--f}). Figure~\ref{fig:Pr_scaling}(\textit{b}) shows how $I_{\theta v}$ depends on $Pr$, and that $I_{\theta v} = I_{uv}/Pr^\gamma$ is a good approximation to the data. For the present data with a fixed $A^+$ and $\kappa^+_x$, $\gamma$ depends on $\omega^+$ but it is independent of $Pr$.
When we use this power law in (\ref{eq:hr_model1}) and solve for $HR$ by using $I_{uv}$ from DNS, figures~\ref{fig:Pr_scaling}(\textit{c,d}) show that the model closely matches the data.  Furthermore, for $\gamma < 0.5$ we can achieve $HR > DR$ for $Pr > 1$, and  
the smaller $\gamma$ is the higher $HR$ is compared to $DR$. From figures~\ref{fig:Pr_scaling}(\textit{b,d}), as $\omega^+$ increases from $0.022$ (black curve) to $0.110$ (grey curve), $\gamma$ decreases from about $0.5$ to $0.1$, and $HR-DR$ increases from almost zero to $10\%$. Therefore, we expect $HR<DR$ for $Pr > 1.0$ if $\gamma > 0.5$. This occurs for the plane oscillation at $\omega^+ = 0.022$, where $\gamma = 2.43$ and $HR- DR \simeq -5\%$ (figure~\ref{fig:Pr_scaling_plane}). 

\begin{figure}
  \centering
 \includegraphics[width=1.0\textwidth,trim={{0.0\textwidth} {0.12\textwidth} {0.15\textwidth} {0.0\textwidth}},clip]{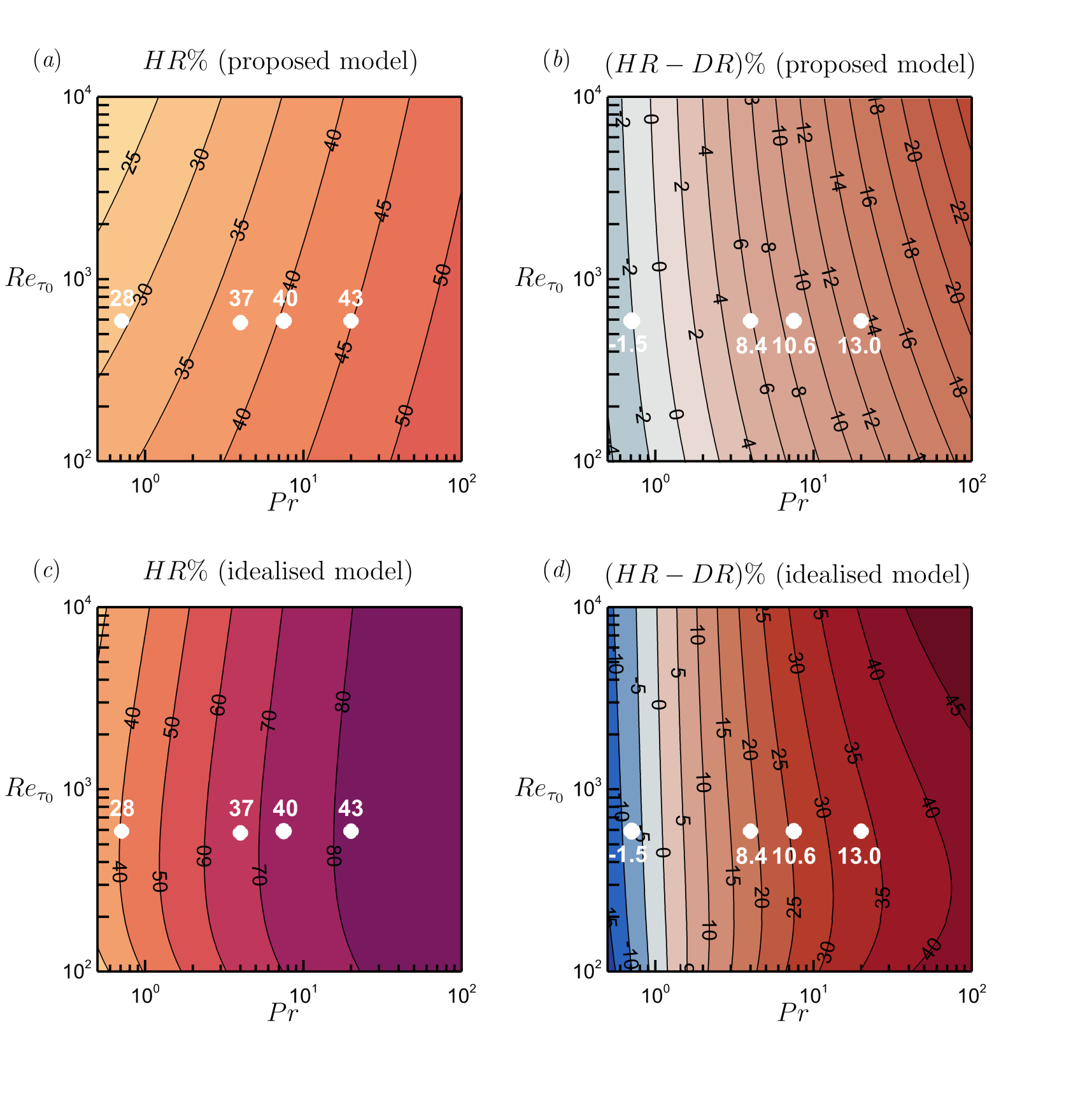}
  \caption{Predicted maps of (\textit{a,c}) $HR$ and (\textit{b,d}) $HR-DR$ for the travelling wave with $A^+ = 12, \kappa^+_x = 0.0014$ and $\omega^+ = 0.088$. (\textit{a,b}) are obtained from our proposed model $I_{\theta v} = I_{uv}/Pr^{\gamma}$ with $I_{uv} = 4.28, \gamma = 0.16$ obtained from DNS, and by solving (\ref{eq:dr_model1}, \ref{eq:hr_model1}). (\textit{c,d}) are obtained from the simplified model (\ref{eq:dUp_ideal}, \ref{eq:dCp_ideal}) and by solving (\ref{eq:dr_model1}, \ref{eq:hr_model1}). The white bullets represent DNS results at $Re_{\tau_0} = 590$ for $Pr = 0.71, 4.0, 7.5$ and $20$.}
  \label{fig:HR_prediction}
\end{figure}

\subsection{Heat transfer reduction at higher Prandtl number and Reynolds number}\label{sec:hr_model}

\cite{gatti2016} used \eqref{eq:dr_model1} to extrapolate their low Reynolds number $DR$ data to higher Reynolds numbers, and the accuracy of this extrapolation was corroborated by \cite{rouhi2022turbulent} up to $Re_{\tau_0} = 4000$ using LES, and by \cite{gatti2024turbulent} up to $Re_{\tau_0} = 6000$ using DNS. For a fixed set of actuation parameters $(A^+,\kappa^+_x,\omega^+)$, $I_{uv}$ is Reynolds and Prandtl number  independent, and so the only Reynolds number dependency of $DR$ is through $C_{f_0}$. Hence, \cite{gatti2016} could solve (\ref{eq:dr_model1}) and predict $DR$ at any Reynolds number.  We can now do the same for $HR$ by using the power law relationship $I_{\theta v} = I_{uv}/Pr^{\gamma}$. By knowing $I_{uv}$ and $\gamma$ for a fixed set of actuation parameters, we can then solve (\ref{eq:dr_model1}) and (\ref{eq:hr_model1}) to map $HR$ and $HR- DR$ as functions of $Pr$ and $Re_{\tau_0}$.   

In figure~\ref{fig:HR_prediction}(\textit{a,b}), we show these predictions for the travelling wave case with $\omega^+ = 0.088$, where the DNS gives $I_{uv} = 4.28$ and $\gamma = 0.16$ (data points with green outline in figure~\ref{fig:Pr_scaling}). The predictions agree well with the DNS data points, including the point at $Pr = 20$ ($HR = 43\%, HR - DR = 13\%$)  (see table~\ref{tab:act_pr}). 
At a given Reynolds number, $HR$ increases almost logarithmically with Prandtl number. For instance, at $Re_{\tau_0} = 590$  $HR$ increases from $30\%$ to $40\%$ as the Prandtl number increases from 0.71 to 7.5, and $HR = 50\%$ would be obtained at a Prandtl number of about 75. In addition, $HR$ decreases very slowly with increasing $Re_{\tau_0}$. For instance, at $Pr = 7.5$ $HR$ decreases from $40\%$ to about $36\%$ as $Re_{\tau_0}$ increases from $590$ to $5900$. Such a slow decrease with $Re_{\tau_0}$ is similarly observed in the predictive model for $DR$ (\ref{eq:dr_model1})  \citep{marusic2021,rouhi2022turbulent,gatti2024turbulent}. 

In figure~\ref{fig:HR_prediction}(\textit{c,d}), we predict $HR$ and $HR-DR$ for the same actuated case as in figure~\ref{fig:HR_prediction}(\textit{a,b}), but with a simplified analytical approach, similar to that proposed by \cite{iwamoto2005friction}. The approach assumes that turbulence is completely damped in the near-wall region, which here we consider that to be up to the Stokes layer protrusion height $\ell^*_{0.01}$ (figure~\ref{fig:profiles_phase}\textit{a}). Therefore, for $0 \le y^* \le \ell^*_{0.01}$ the profiles are assumed to be laminar $\overline{U}^* = y^*[1-y^*/(2Re_\tau)], \overline{\Theta}^* = Pr y^*[1-y^*/(2Re_\tau)]$, and for $y^* > \ell^*_{0.01}$ the profiles follow the turbulent formulations (\ref{eq:up_log}, \ref{eq:cp_log}). By matching the laminar and turbulent profiles at $y^* = \ell^*_{0.01}$, we obtain
\begin{align}
 \Delta \overline{U}^*_{170} &= \ell^*_{0.01} \left( 1 - \frac{\ell^*_{0.01}}{ 2Re_\tau} \right) - \left( \frac{1}{\kappa_u}\ln \ell^*_{0.01} + B_u \right) \tag{3.7\textit{a}} \label{eq:dUp_ideal} \\
 \Delta \overline{\Theta}^*_{170} &= Pr \ell^*_{0.01} \left( 1 - \frac{\ell^*_{0.01}}{ 2Re_\tau} \right) - \left( \frac{1}{\kappa_\theta}\ln \ell^*_{0.01} + B_\theta \right). \tag{3.7\textit{b}} \label{eq:dCp_ideal}
\end{align}
We substitute for $\Delta \overline{U}^*_{170}, \Delta \overline{\Theta}^*_{170}$ in (\ref{eq:dr_model1}, \ref{eq:hr_model1}) to predict $HR$ and $HR-DR$. With this approach, the predicted $HR$ and $HR-DR$ increase with $Pr$ (figure~\ref{fig:HR_prediction}\textit{c,d}), consistent with our proposed model and DNS (figure~\ref{fig:HR_prediction}\textit{a,b}). However, quantitatively, this approach over-predicts $HR$ and $HR-DR$ by two times compared to the DNS data and our proposed model, which is of no surprise. The Stokes layer does not completely attenuate the near-wall turbulence, even down to $y^* \sim \mathcal{O}(1)$ (figure~\ref{fig:profiles_stress}). The level of attenuation results from the complex interaction between the Stokes layer and the near-wall velocity and thermal fields, as we discuss next. 



\subsection{Stokes layer interaction with the velocity and thermal fields: integral parameters}
\label{sec:stokes}
For $Re_{\tau_0} \lesssim \mathcal{O}(10^3)$, the primary source of drag reduction is the protrusion of the Stokes layer due to the wall oscillation that modifies the near-wall velocity field~\citep{choi1998turbulent,choi2001mechanism,choi2002near,choi2002drag,quadrio2000numerical,ricco2004modification,quadrio2011,ricco2021review}, and we would expect that a similar interaction between the Stokes layer and the near-wall temperature field leads to $HR$. As a result, the near-wall $\overline{u^2}^*$ profiles are modified (figure~\ref{fig:profiles_stress}\textit{d}), consistent with the literature~\citep{jung1992suppression,baron1995turbulent,choi1998turbulent,choi2001mechanism,quadrio2000numerical,ricco2004effects,quadrio2011,touber2012,ricco2012changes,rouhi2022turbulent}, and similar trends are seen in the $\overline{\theta^2}^*$ profiles (figures~\ref{fig:profiles_stress}\textit{e,f}). To determine how the Stokes layer affects $I_{uv}$ and $I_{\theta v}$ as the Prandtl number changes, we first 
identify the Stokes layer characteristics through the harmonic component of the spanwise velocity $\overline{\tilde{w}^2}$, where $\tilde{w}$ is obtained by applying triple decomposition
\begin{align}
 W(x,y,z,t) = \overline{W}(y) + \tilde{w}(x,y,t) + w(x,y,z,t), \tag{3.8\textit{a}} \label{eq:tri_decom} \\
 \tilde{w}(x,y,t) = \underbrace{\frac{1}{N}\sum_{n=0}^{N-1} W(x,y,t+nT_{osc})}_{\left< W \right>(x,y,t)} - \overline{W}(y). \tag{3.8\textit{b}} \label{eq:harmonic}
\end{align}
Here, $\overline{W}$ is the plane- and time-averaged component of the spanwise velocity, $\tilde{w}$ is the harmonic component, $w$ is the turbulent component, and $\left< W \right>$ is the spanwise and phase averaged value of $W$.  For the streamwise and wall-normal  velocities and the temperature field, the harmonic components are negligible compared the turbulent components. 
Following \cite{rouhi2022turbulent}, we quantify the Stokes layer protrusion height $\ell^*_{0.01}$ as the wall distance where $\overline{\tilde{w}^2}^* = 0.01$, and we locate the Stokes layer thickness $\delta^*_S$ where $\overline{ \tilde{w}^2}^* = \frac{1}{2}{A^*}^2\mathrm{e}^{-2}$. We also calculate the production due to the Stokes layer according to $P^*_{33} \equiv -2 \overline{\left< w v \right>\partial \tilde{w}/\partial y }^*-2 \overline{\left< w u \right>\partial \tilde{w}/\partial x}^*$, where $P^*_{33}$ is the only external source term due to the Stokes layer that injects energy into the turbulent stress budgets \citep{touber2012,umair2022reynolds}. 

\begin{figure}
  \centering
 \includegraphics[width=1.0\textwidth,trim={{0.12\textwidth} {0.02\textwidth} {0.05\textwidth} {0.0\textwidth}},clip]{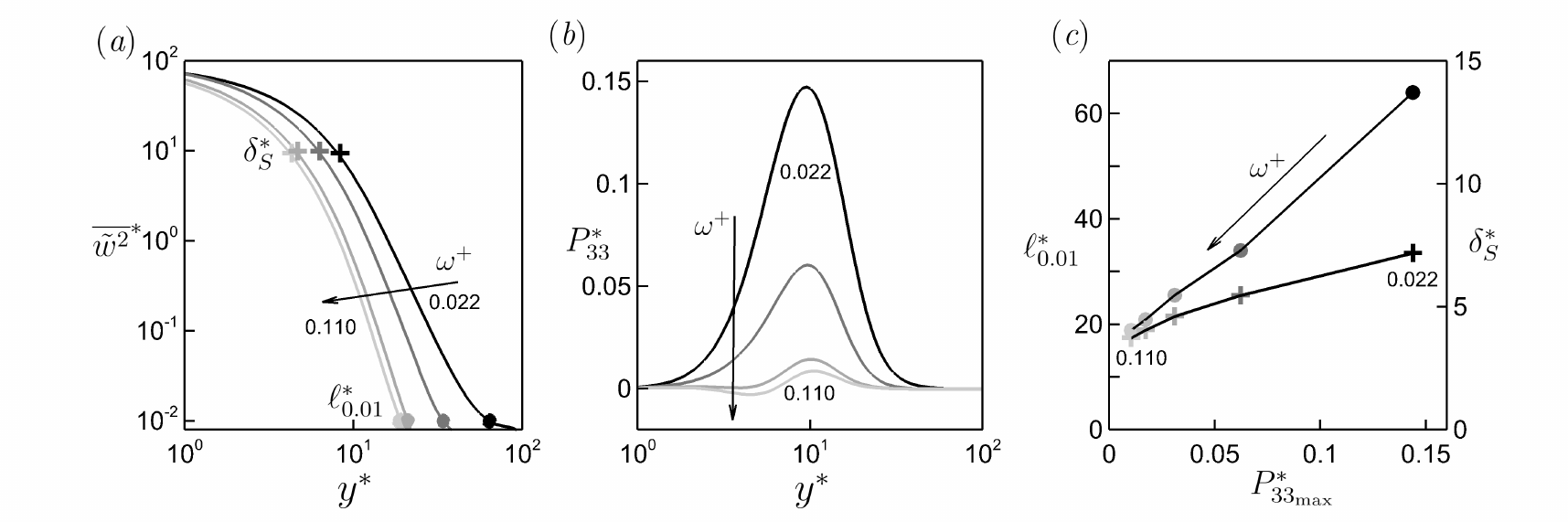}
  \caption{Characteristics of the Stokes layer for the travelling wave with $A^+ = 12, \kappa^+_x = 0.0014$ and $0.022 \le \omega^+ \le 0.110$ (same cases as in figure~\ref{fig:averaged_HR_DR_Pr}\textit{a,c,e}).  Results are independent of Prandtl number.  The colour of the profiles and data points change from black at $\omega^+ = 0.022$ to light grey at $\omega^+ = 0.110$. (\textit{a}) Profiles of $\overline{ \tilde{w}^2 }^*$. Stokes layer protrusion height $\ell^*_{0.01}$~\citep{rouhi2022turbulent} marked at $\overline{ \tilde{w}^2}^* = 0.01$ (filled bullets); laminar Stokes layer thickness $\delta^*_S$ marked at $\overline{ \tilde{w}^2}^* = \frac{1}{2}{A^*}^2\mathrm{e}^{-2}$ (cross symbols). (\textit{b}) Production due to the Stokes layer $P^*_{33}$. (\textit{c}) $\ell^*_{0.01}$ (filled bullets) and $\delta^*_S$ (cross symbols) versus the maximum value of $\tilde{P}^*_{33}$.}
  \label{fig:profiles_phase}
\end{figure}

The behaviour of these Stokes layer parameters is given in figure~\ref{fig:profiles_phase}, for the same travelling wave cases shown in figure~\ref{fig:averaged_HR_DR_Pr}(\textit{a,c,e}). The profiles do not depend on $Pr$, that is, at a fixed $\omega^+$ the Stokes layer structure remains unchanged as the Prandtl number changes. When $\omega^+$ decreases, $\ell^*_{0.01}$ and the maximum production increase proportionately (figure~\ref{fig:profiles_phase}\textit{c}). In other words, the Stokes layer becomes more protrusive while injecting more energy into the turbulent field. We note that $\ell^*_{0.01}$ is almost linearly proportional to  $P^*_{{33}_\mathrm{max}}$, but $\delta^*_S$ is less responsive to the rise in the production. Therefore, $\ell^*_{0.01}$ better quantifies the Stokes layer protrusion and strength, as sensed by the turbulent field, in agreement with the observations by \cite{rouhi2022turbulent}.   

\begin{figure}
  \centering
 \includegraphics[width=1.0\textwidth,trim={{0.06\textwidth} {0.0\textwidth} {0.12\textwidth} {0.0\textwidth}},clip]{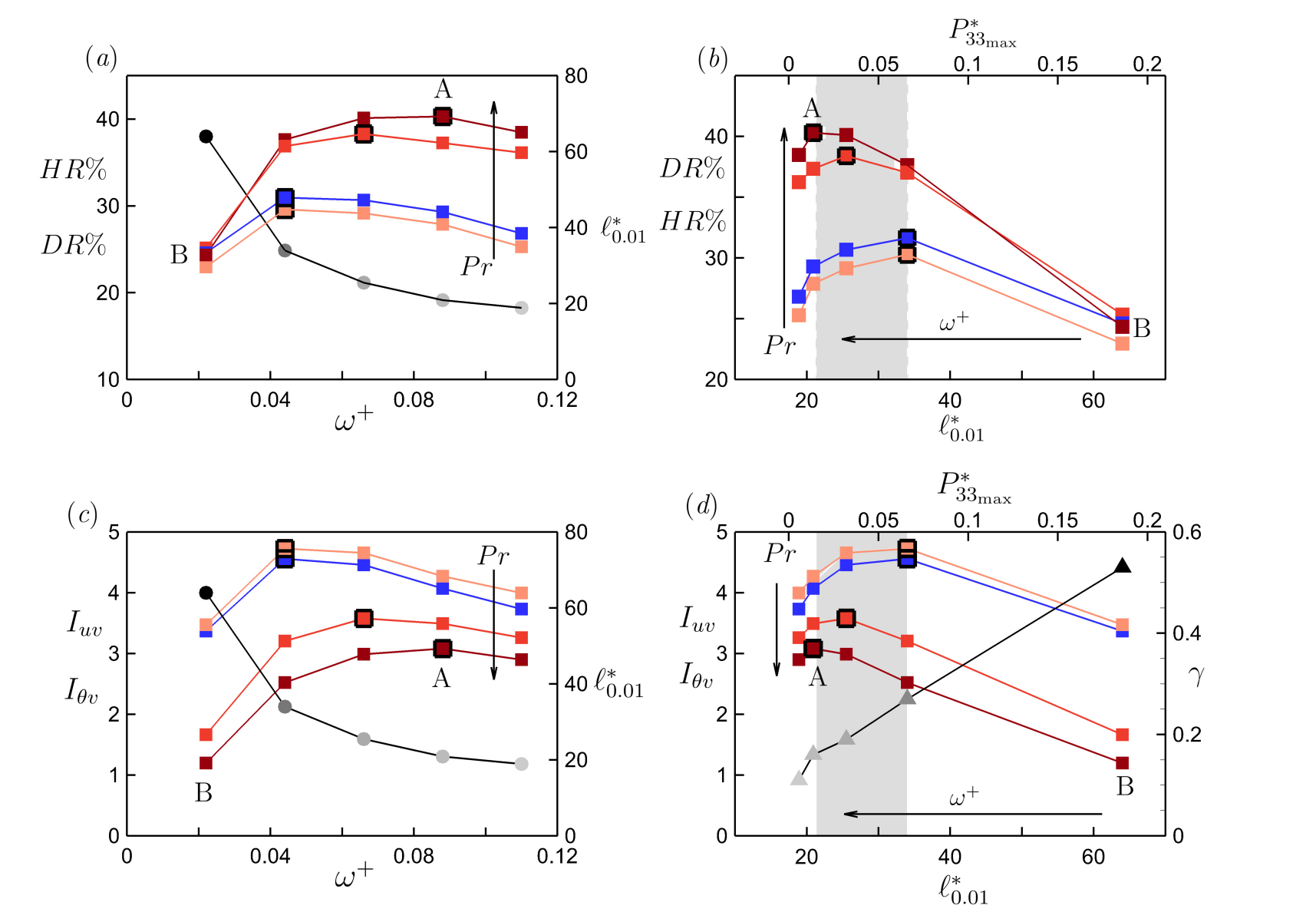}
  \caption{Variation of the integral parameters  with the Stokes layer characteristics $(\ell^*_{0.01}, P^*_{{33}_\mathrm{max}})$ for the same travelling wave cases as in figure~\ref{fig:profiles_phase}. (\textit{a})  $DR$ (blue squares), $HR$ (light orange to brick red squares),  $\ell^*_{0.01}$ (black to grey bullets) versus $\omega^+$. (\textit{b}) Same data as in (\textit{a}) versus $\ell^*_{0.01}$ (bottom axis) and $P^*_{{33}_\mathrm{max}}$ (top axis).  (\textit{c,d}) correspond to (\textit{a,b}) for $I_{uv}$ (blue squares) and $I_{\theta v}$ (light orange to brick red squares); in (\textit{d}) we overlay $\gamma$ (grey to black triangles). In all plots, the colour of $HR$ and $I_{\theta v}$ changes from light orange at $Pr=0.71$ to brick red at $Pr = 7.5$. The points associated with $DR_\mathrm{max}$ and $HR_\mathrm{max}$ are highlighted with a larger symbol size and black outline. In (\textit{b,d}), the grey region shades the range of $DR_\mathrm{max}$ and $HR_\mathrm{max}$. In all plots we highlight two cases that we further analyse in figures~\ref{fig:specy_specx} and \ref{fig:flowviz_wu_wc}: Case A at $\omega^+ = 0.088$ with $\gamma = 0.16$, and   Case B at $\omega^+ = 0.022$ with $\gamma = 0.53$. 
  }
  \label{fig:DR_HR_duw_dcw}
\end{figure}

In figure~\ref{fig:DR_HR_duw_dcw}, we assess the relation between the integral quantities $DR$, $HR$, $I_{\theta v}$, $I_{uv}$, and the Stokes layer characteristics.  \cite{rouhi2022turbulent} showed that $DR$ increases with $\ell^*_{0.01}$ up to an optimal value for maximum $DR$, corresponding to an optimal level of  $P^*_{{33}_\mathrm{max}}$. For our present cases, this optimal point is $\ell^*_{0.01} = 34 $ (figure~\ref{fig:DR_HR_duw_dcw}\textit{b}), where protrusion beyond this point causes $DR$ to decrease. We observe a similar trend in $HR$ (figure~\ref{fig:DR_HR_duw_dcw}\textit{b}), but the corresponding optimal value of $\ell^*_{0.01}$  for maximum $HR$ decreases from 34 to 21 as the Prandtl number increases from 0.71 to 7.5.  Figures~\ref{fig:DR_HR_duw_dcw}(\textit{c,d}) show that the variations of $I_{uv}$ and $I_{\theta v}$ with $\omega^+$ and $\ell^*_{0.01}$ are consistent with the trends in $DR$ and $HR$. In terms of variations with $Pr$, however, $HR$ increases even as $I_{\theta v}$ decreases. From $Pr=0.71$ to $7.5$, the maximum value of  $I_{\theta v}$ shifts to smaller values of $\ell^*_{0.01}$,  corresponding to a shift in $\omega^+$ from 0.044 ($\ell^*_{0.01} = 34$) to 0.088 ($\ell^*_{0.01} = 21$).  Finally, it appears that the exponent $\gamma$ in the power law increases almost linearly with $\ell^*_{0.01}$ and $P^*_{33_\mathrm{max}}$ (figure~\ref{fig:DR_HR_duw_dcw}\textit{d}).

To summarise, we find that for $Pr > 1$, achieving maximum $HR$ requires a less protrusive Stokes layer than achieving maximum $DR$. As the Stokes layer becomes more protrusive (energetic), it loses its efficacy in attenuating $\overline{\theta v}^*$ relative to $\overline{u v}^*$, that is, $\gamma$ increases as $\ell^*_{0.01}$ increases. 


\subsection{Stokes layer interaction with the velocity and thermal fields: scale-wise analysis}\label{sec:stokes2}

To better understand our observations on the integral parameters shown in figure~\ref{fig:DR_HR_duw_dcw}, we now examine the  spectrograms of $\overline{u v}^*$ and $\overline{\theta v}^*$  and visualise the distributions of the instantaneous $u v$ and $\theta v$ near the wall. We focus on two cases at $Pr = 7.5$, Case A at $\omega^+ = 0.088$, where $HR$ is maximum and $DR$ is near maximum ($\gamma = 0.16$), and Case B at $\omega^+ = 0.022$, where $HR$ and $DR$ drop below their maximum values, owing to a highly protrusive Stokes layer ($\gamma = 0.53$).

\begin{figure}
  \centering
 \includegraphics[width=1.0\textwidth,trim={{0.06\textwidth} {0.0\textwidth} {0.05\textwidth} {-0.05\textwidth}},clip]{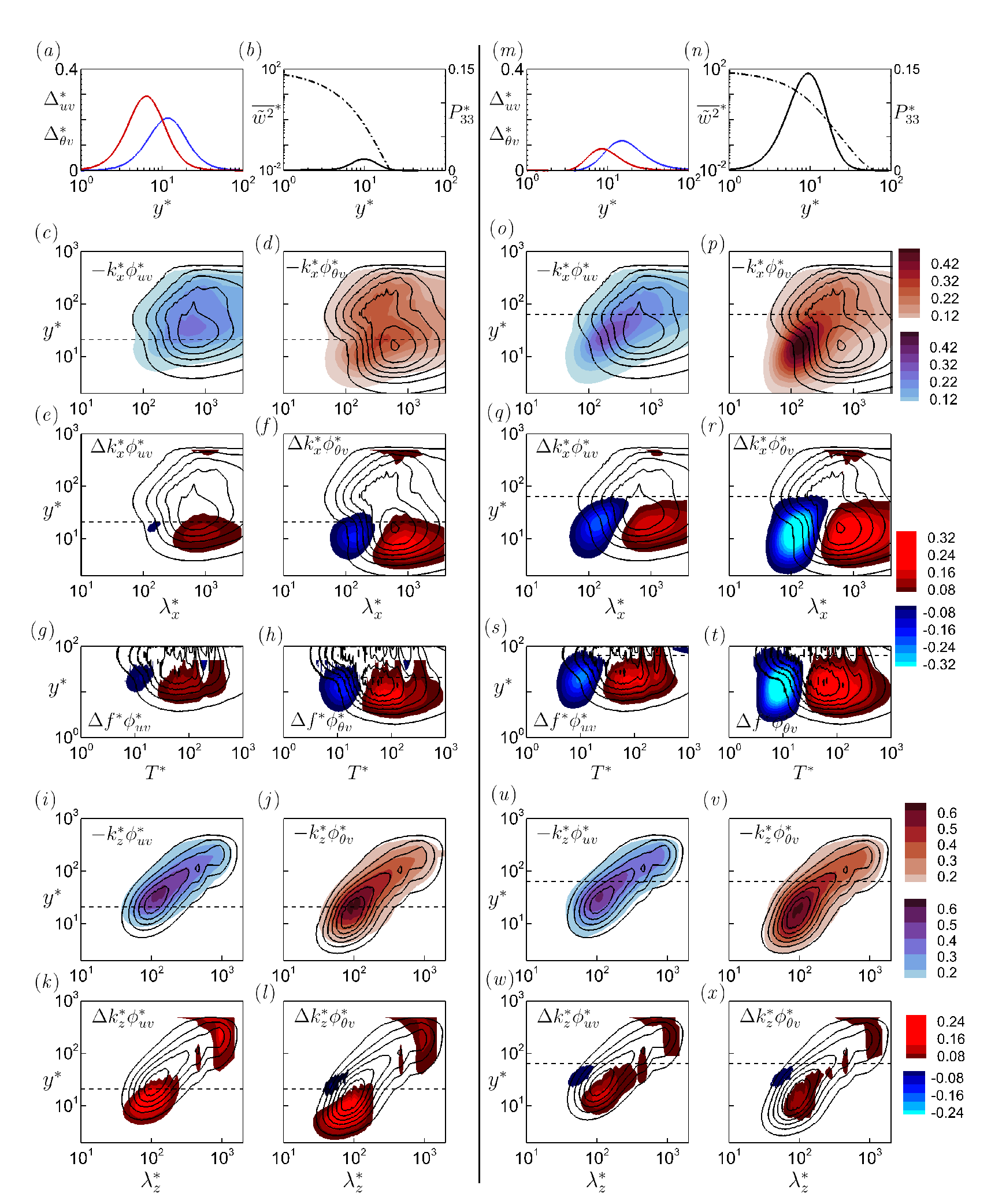}
 \put(-9,14.1){Case A}
 \put(-3.9,14.1){Case B}
  \caption{Comparisons between Case A (left two columns) and Case B (right two columns. (\textit{a,m}) Profiles of $\Delta^*_{uv}$ (blue) and $\Delta^*_{\theta v}$ (red). (\textit{b,n}) Profiles of $\overline{ \tilde{w}^2}^*$ (dashed-dotted line) and $P^*_{33}$ (solid line). Streamwise spectrograms: (\textit{c,o}) $-k^*_x\phi^*_{uv}$;  (\textit{d,p}) $-k^*_x\phi^*_{\theta v}$ (contour lines and contour fields represent non-actuated and actuated cases, respectively). Difference between the actuated and the non-actuated cases for the wavenumber spectra (\textit{e,f,q,r}), and frequency spectra (\textit{g,h,s,t}). Spanwise spectrograms: (\textit{i,u}) $-k^*_z\phi^*_{uv}$; (\textit{j,v}) $-k^*_z\phi^*_{\theta v}$. Difference between the actuated and the non-actuated cases (\textit{k,l,w,x}). In all the spectrgrams, the horizontal dashed line marks $\ell^*_{0.01}$. }
  \label{fig:specy_specx}
\end{figure}

\begin{landscape}
\begin{figure}
  \centering
 \includegraphics[width=1.5\textwidth,trim={{0.0\textwidth} {0.02\textwidth} {0.0\textwidth} {0.0\textwidth}},clip]{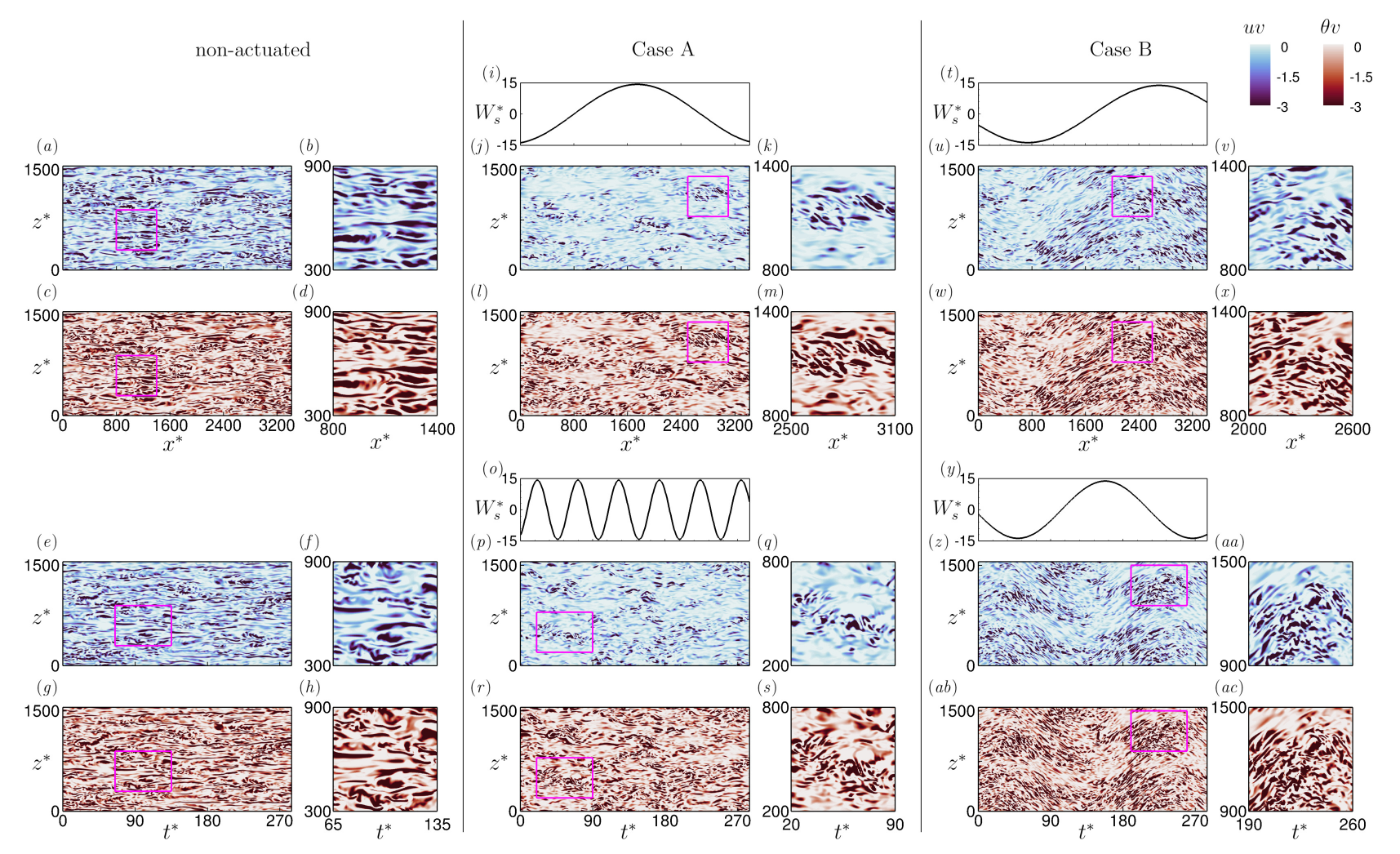}
  \caption{Visualisations of the instantaneous fields of $u v$ (blue intensity fields) and $\theta v$ (brick intensity fields) at $y^+ = 14$ for the same cases as in figure~\ref{fig:specy_specx}. Next to each field, we magnify the small square outlined in pink. Left: non-actuated case; middle: actuated Case A; right: actuated Case B. The top two rows visualise $u v$ and $\theta v$ at one time over the $x^*$-$z^*$ plane. (\textit{i,t}) plot the corresponding spanwise wall velocity $W^*_s$ over $x^*$ at the same time.   The bottom two rows visualise $u v$ and $\theta v$ at one $x^*$-location over $z^*$ and time $t^*$. (\textit{o,y}) plot $W^*_s$ over $t^*$ at the same $x^*$-location.}
  \label{fig:flowviz_wu_wc}
\end{figure}
\end{landscape}

Figure~\ref{fig:specy_specx} displays the various pre-multiplied spectrograms for these two cases, where $\phi$ is the spectral density, $k_x$ is the streamwise wavenumber,  $k_z$ is the spanwise wavenumber, and $f$ is the frequency.  The differences between the actuated and the non-actuated cases are denoted by, for example, $\Delta k^*_x \phi^*_{uv} = k^*_x \phi^*_{uv} - k^*_x \phi^*_{{uv}_0}$.  The extent up to which $\Delta k^*_x \phi^*_{uv}$ and $\Delta k^*_x \phi^*_{\theta v}$ are non-zero (red and blue fields in figures~\ref{fig:specy_specx}\textit{e,f,q,r}) coincides with the height of the Stokes layer protrusion  $\ell^*_{0.01}$ (horizontal dashed line); this supports the robustness of $\ell^*_{0.01}$ to measure the extent up to which the Stokes layer modifies the turbulent field. 
Overall, for both Cases A and B, and $k^*_x \phi^*_{uv}$ and $ k^*_x \phi^*_{\theta v}$, the Stokes layer protrusion (production) leads to two outcomes: (1) large-scale energy attenuation as a favourable outcome (positive $\Delta k^*_x \phi^*_{uv}, \Delta k^*_x \phi^*_{\theta v}$ for $\lambda^*_x \gtrsim 350$, shown as the red fields in figures~\ref{fig:specy_specx}\textit{e,f,q,r}), and (2) small-scale energy amplification as an unfavourable outcome (negative $\Delta k^*_x \phi^*_{uv}, \Delta k^*_x \phi^*_{\theta v}$ for $\lambda^*_x \lesssim 350$, shown as the blue fields). The large-scale energy attenuation corresponds to the attenuation of the near-wall streaks with $\lambda^*_x \simeq 10^3$ (see the energetic peaks in the non-actuated spectrograms). The small-scale energy amplification corresponds to the emergence of smaller scales with $\lambda^*_x \simeq 10^2$ (figures~\ref{fig:flowviz_wu_wc}\textit{j--ac}). The drastic changes in the streamwise spectrograms (figure~\ref{fig:specy_specx}\textit{e,f,q,r}) are echoed in the response of the frequency spectrograms (figures~\ref{fig:specy_specx}\textit{g,h,s,t}). Changing the actuation frequency $\omega^+$, hence changing the Stokes layer time-scale, modifies the near-wall turbulence time-scale $T^*$, which in turn modifies the streamwise length-scale $\lambda^*_x$ (of course, $T^*$ is related to $\lambda^*_x$ through a convection speed $\sim \mathcal{O}(10)$ according to Taylor's  hypothesis).

Further insight can be gained by comparing the near-wall instantaneous fields of $u v$ and $\theta v$, as shown in  figure~\ref{fig:flowviz_wu_wc}.  The response of $u v$ to the Stokes layer (blue intensity fields) largely follows the trends seen in $\Delta k^*_x \phi^*_{uv}$ and $ \Delta f^* \phi^*_{uv}$. 
For Case A, we see that the large scales associated with the near-wall streaks are attenuated. At the same time, sparse patches of smaller scales with $\lambda^*_x \simeq 10^2$ ($T^* \simeq 10$) emerge. For Case B, the emerging smaller scales possess a similar structure and size to those in Case A, but with a noticeably larger population; they appear as large and closely spaced patches that cover a significant area of the near-wall region. Consistently, negative regions of $\Delta k^*_x \phi^*_{uv}$ and $ \Delta f^* \phi^*_{uv}$ are small for Case A (figure~\ref{fig:specy_specx}\textit{e,g}), and  large for Case B (figure~\ref{fig:specy_specx}\textit{q,s}). Thus, $I_{uv}$ and $DR$ are greater for Case A than for Case B (figure~\ref{fig:DR_HR_duw_dcw}). Attenuation of the large scales associated with the near-wall streaks is considered to be a primary source of drag reduction, as extensively reported in the literature~\citep{baron1995turbulent,di2002particle,karniadakis2003mechanisms,ricco2004modification,quadrio2009,touber2012,ricco2021review,marusic2021,rouhi2022turbulent}. However, amplification of the smaller scales as a source of drag increase, is reported to a lesser extent. \cite{touber2012} observed such amplification in the frequency spectrograms $f^*\phi^*_{uu}$, $f^*\phi^*_{vv}$, and $f^*\phi^*_{ww}$. Consistent with our figure~\ref{fig:specy_specx}, they noted that as  $\omega^+$ changes from 0.06 to 0.03, and the Stokes layer production (protrusion) increases, energy accumulates in the scales with $T^* \simeq 10$, and $DR$ decreases. Similarly, energy amplification at $T^* \simeq 10$ is observed in $f^*\phi^*_{\tau \tau}$ by \cite{chandran2022turbulent}, and in the near-wall $f^* \phi^*_{uu}$ by \cite{deshpande2023relationship}. 

As to the $\theta v$ field (brick intensity fields in figure~\ref{fig:flowviz_wu_wc}), we see that with $Pr > 1.0$ the Stokes layer modifies the ${\theta v}$ field more than the ${u v}$ field. 
In Case A, for example, the scales with $\lambda^*_x \simeq 10^2$ and $T^* \simeq 10$ are more frequent and more densely populated in the $\theta v$ fields (figures~\ref{fig:flowviz_wu_wc}\textit{l,m,r,s}) than in the $u v$ fields (figures~\ref{fig:flowviz_wu_wc}\textit{j,k,p,q}). Consistently, $\Delta k^*_x\phi^*_{\theta v}$ (figure~\ref{fig:specy_specx}\textit{f}) is more positive than $\Delta k^*_x\phi^*_{u v}$ (figure~\ref{fig:specy_specx}\textit{e}) for $\lambda^*_x \gtrsim 350$, and more negative for $\lambda^*_x \lesssim 350$.  We can make the same observations with respect to Case B (figures~\ref{fig:specy_specx}\textit{q,r}). As a result, at each $\omega^+$, $I_{\theta v}$ falls below $I_{uv}$ when $Pr > 1$ (figure~\ref{fig:DR_HR_duw_dcw}\textit{c,d}).

\begin{figure}
  \centering
 \includegraphics[width=\textwidth,trim={{0.0\textwidth} {0.07\textwidth} {0.0\textwidth} {0.0\textwidth}},clip]{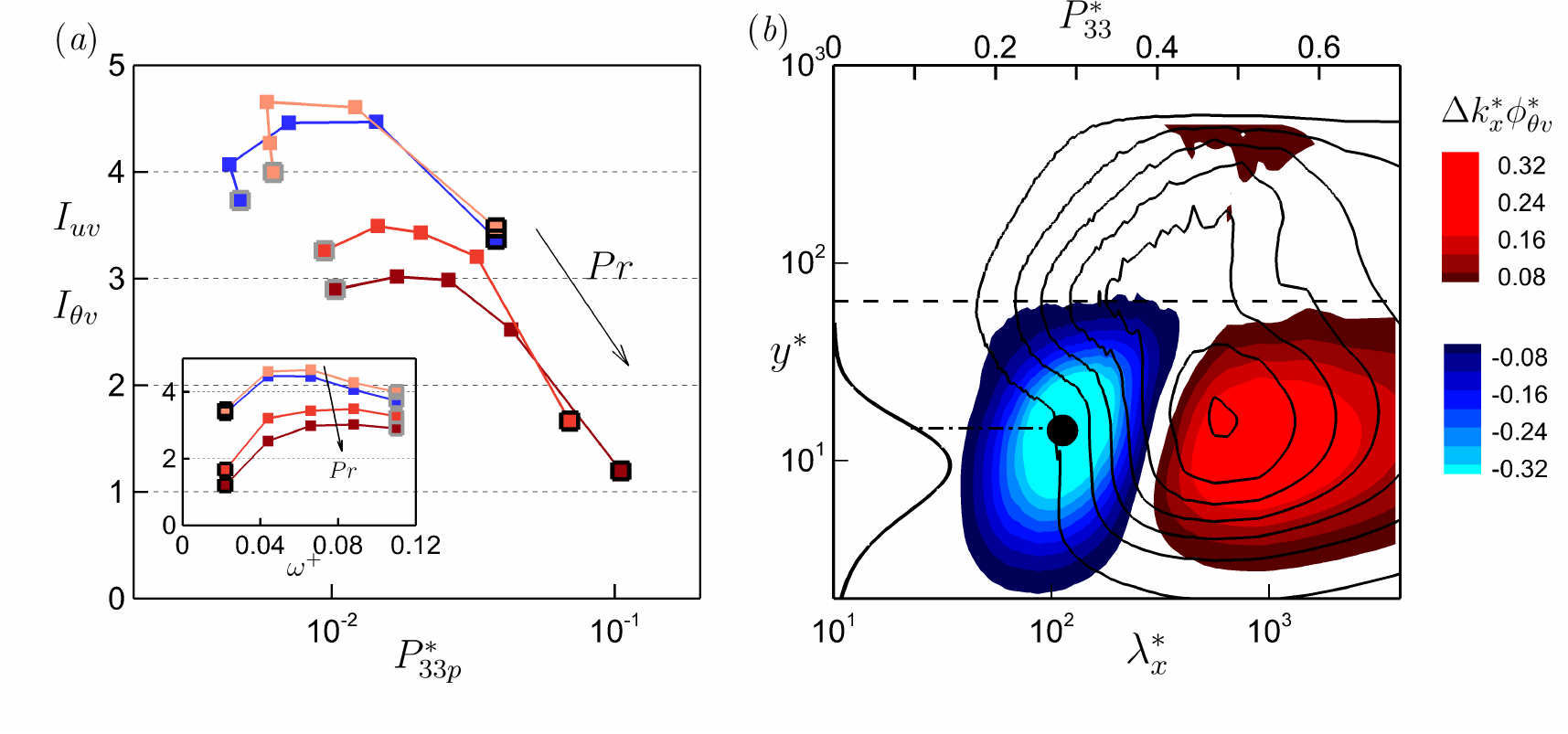}
  \caption{Assessment of the relation between $I_{uv}$ and $I_{\theta v}$, and the local Stokes layer production $P^*_{33p}$ at the negative peaks of $\Delta k^*_x \phi^*_{uv}$ and $\Delta k^*_x \phi^*_{\theta v}$. (\textit{a}) Same data of $I_{uv}$ and $I_{\theta v}$ as in figure~\ref{fig:DR_HR_duw_dcw}(\textit{c}), but versus $P^*_{33p}$; to ease the inspection, the inset shows figure~\ref{fig:DR_HR_duw_dcw}(\textit{c}). The data points at $\omega^+ = 0.022$ and $0.110$ have black and grey outlines, respectively. (\textit{b}) Illustrates obtaining of $P^*_{33p}$ for Case B (figure~\ref{fig:specy_specx}\textit{r}), by intersecting the negative peak of $\Delta k^*_x \phi^*_{\theta v}$ with the $P^*_{33}$ profile (on the left axis).}
  \label{fig:P33_z_peak}
\end{figure}

At a fixed $\omega^+$, $u v$ and $\theta v$ are exposed to an identical Stokes layer. However, the larger change in $\Delta k^*_x \phi^*_{\theta v}$ compared to $\Delta k^*_x \phi^*_{u v}$ for $Pr > 1$ implies that the energetic near-wall scales of $\theta v $ are locally exposed to a stronger Stokes layer production. We establish this connection by evaluating the local Stokes layer production $P^*_{33p}$ at the negative peaks of $\Delta k^*_x \phi^*_{\theta v}$ and $\Delta k^*_x \phi^*_{u v}$; these peaks are associated with the energetic small scales with $\lambda^*_x \simeq 10^2$, that contribute to the drop in $I_{uv}$ and $I_{\theta v}$. Figure~\ref{fig:P33_z_peak}(\textit{b}) demonstrates this process for Case B (figure~\ref{fig:specy_specx}\textit{r}), where the negative peak of $\Delta k^*_x \phi^*_{\theta v}$ is intersected with the $P^*_{33}$ profile. With increasing $Pr$, the conductive sublayer thins, and the negative peak in $\Delta k^*_x \phi^*_{\theta v}$ falls closer to the wall compared to its counterpart from $\Delta k^*_x \phi^*_{u v}$. This means that at each $\omega^+$, the energetic small scales of $\theta v$ are exposed to a larger Stokes layer production, hence have a higher energy. This leads to the drop in $I_{\theta v}$ compared to $I_{uv}$, as demonstrated in figure~\ref{fig:P33_z_peak}(\textit{a}). At $Pr = 0.71$, the local production $P^*_{33p}$ is close between $uv$ and $\theta v$, and $I_{uv}$ and $I_{\theta v}$ are close to each other. However, at $Pr = 4.0$ and $7.5$, $P^*_{33p}$ at each $\omega^+$ is several times higher for $\theta v$ than $uv$, and $I_{\theta v}$ drops below $I_{u v}$. In addition, 
  \noindent   at $\omega^+ = 0.022$, $P^*_{33p}$ increases by almost $3$ times from $Pr = 0.71$ to $7.5$, whereas at $\omega^+ = 0.110$, $P^*_{33p}$ increases by $1.5$ times. As a result, $I_{\theta v}$ has a milder decay rate at $\omega^+ = 0.110$ ($\gamma = 0.11$) than at $\omega^+ = 0.022$ ($\gamma = 0.53$).


We have seen how the behaviour of the Stokes layer leads to $I_{\theta v}<I_{u v}$ by examining the streamwise and frequency spectrograms.  This connection is more difficult to find in the spanwise spectrograms $k^*_z \phi^*_{u v}$ and $ k^*_z \phi^*_{\theta v}$ (figures~\ref{fig:specy_specx}\textit{i--l}, \textit{u--x}). The Stokes layer predominantly modifies the time scale $T^*$, hence the streamwise length-scale $\lambda^*_x$, of the near-wall turbulence. However, $k^*_z \phi^*_{u v}$ and $ k^*_z \phi^*_{\theta v}$ are integrated over $T^*$ and $\lambda^*_x$, and so the smaller-scale energy amplification is cancelled by the large-scale energy attenuation. Thus, even though the Stokes layer protrusion (production) increases from Case A to Case B, the distributions of $k^*_z \phi^*_{u v}$ and $ k^*_z \phi^*_{\theta v}$ suggest that the near-wall turbulence is distorted to a lesser extent. 

\begin{figure}
  \centering
 \includegraphics[width=\textwidth,trim={{0.0\textwidth} {0.02\textwidth} {0.0\textwidth} {0.0\textwidth}},clip]{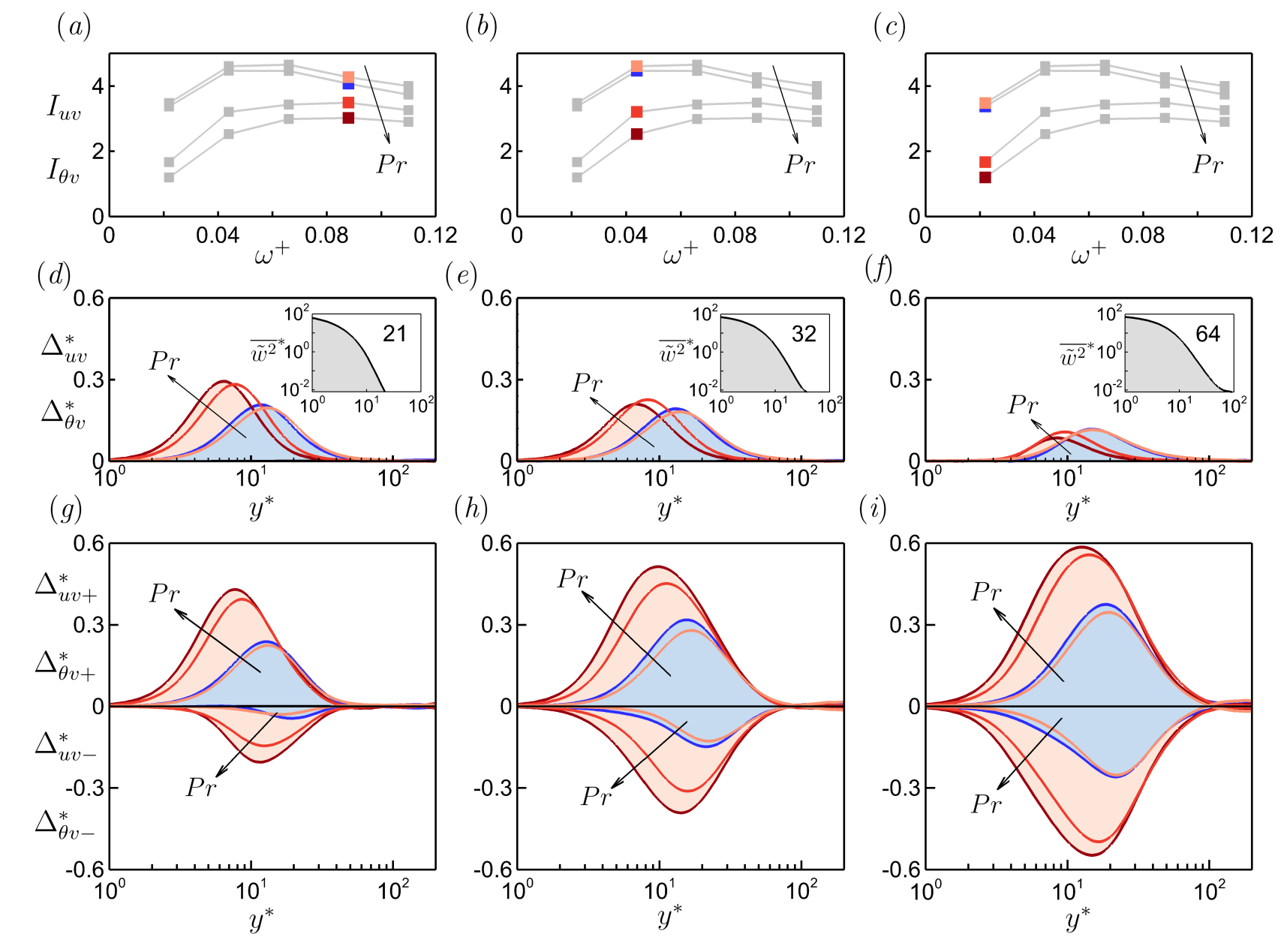}
  \caption{ Large- and small-scale contributions to  $\Delta^*_{uv}, \Delta^*_{\theta v}$ and their integrals $I_{uv}, I_{\theta v}$ for the travelling wave case.   Left column $\omega^+ = 0.088$; middle $\omega^+ = 0.044$; right  $\omega^+ = 0.022$.  Blue $uv$ profiles; for $\theta v$ profiles light orange $Pr = 0.71$, orange 4.0, brick red 7.5. (\textit{a,b,c}) $I_{uv}, I_{\theta v}$ versus $\omega^+$. (\textit{d,e,f}) $\Delta^*_{uv}, \Delta^*_{\theta v}$ versus $y^*$; the insets plot $\overline{\tilde{w}^2}^*$ representing the Stokes layer where the bold number reports $\ell^*_{0.01}$. (\textit{d,e,f}) Total  $\Delta^*_{uv} \Delta^*_{\theta v}$; (\textit{g,h,i})  decomposed $\Delta^*_{uv\pm}, \Delta^*_{\theta v\pm}$. In (\textit{d--i}), we shade under the profiles of $\Delta^*_{uv}$ and $\Delta^*_{\theta v}$ at $Pr = 7.5$, as well as their decomposition.}
  \label{fig:int_specx_specf}
\end{figure}

Our observations regarding the large-scale attenuation and small-scale amplification can be extended by applying scale-wise decomposition to $\Delta^*_{uv}$ and $ \Delta^*_{\theta v}$. For instance, we can write $\Delta^*_{uv} = \Delta^*_{uv+} + \Delta^*_{uv-}$, where 
\begin{align}
\Delta^*_{uv+} &= \int_{\lambda^*_x > 350} \Delta \phi^*_{uv}(k^*_x,y^*) dk^*_x - y^*\left(Re^{-1}_\tau - Re^{-1}_{\tau_0}\right), \tag{3.9a} \label{eq:decomposition1} \\
\Delta^*_{uv-} &= \int_{\lambda^*_x \le 350} \Delta \phi^*_{uv}(k^*_x,y^*) dk^*_x. \tag{3.9b} \label{eq:decomposition2}
\end{align}
We set the threshold of $\lambda^*_x = 350$ for partitioning because the spectrograms for all cases show  $\Delta k^*_x \phi^*_{uv} >0$ for $\lambda^*_x \gtrsim 350$, and  $\Delta k^*_x \phi^*_{uv}<0$ for $\lambda^*_x \lesssim 350$ (see figures~\ref{fig:specy_specx}\textit{e,q}).  As noted earlier (figure~\ref{fig:budgets_Dup_Dcp}\textit{a} and \ref{eq:dDUp}), we need to subtract the term $y^* \left( Re^{-1}_\tau - Re^{-1}_{\tau_0} \right)$ from $\Delta \overline{u v}^*$  to obtain its net contribution $\Delta^*_{uv}$.  
Considering figure~\ref{fig:specy_specx}(\textit{e,q}), this contribution appears in the large scales of $\Delta k^*_x \phi^*_{uv}$ ($\lambda^*_x > 350$), and therefore we subtract $y^* \left( Re^{-1}_\tau - Re^{-1}_{\tau_0} \right)$ from the large-scale integration (\ref{eq:decomposition1}). 
The same reasoning is applied to the partitioning of  $\Delta^*_{\theta v}$, and we calculate $\Delta^*_{\theta v+}$ and $\Delta^*_{\theta v-}$ in a manner that is identical to that expressed by (\ref{eq:decomposition1}) and (\ref{eq:decomposition2}).

Figure~\ref{fig:int_specx_specf} shows how changing $\omega^+$ or $Pr$ modifies the balance between the large-scale attenuation ($\Delta^*_{uv+}, \Delta^*_{\theta v+}$) and the small-scale amplification ($\Delta^*_{uv-}, \Delta^*_{\theta v-}$). The net attenuations $\Delta^*_{uv}$ and $\Delta^*_{\theta v}$ (figure~\ref{fig:int_specx_specf}\textit{d,e,f}) depend on the disparity between their decomposed parts (figure~\ref{fig:int_specx_specf}\textit{g,h,i}). Decreasing $\omega^+$ (increasing the Stokes layer protrusion) simultaneously increase $ \Delta^*_{u v+}, \Delta^*_{\theta v+}$ and $\Delta^*_{u v-}, \Delta^*_{\theta v-}$. However, below the optimal $\omega^+$, $ \Delta^*_{\theta v-}, \Delta^*_{u v-}$ increase more than $ \Delta^*_{\theta v+}, \Delta^*_{u v+}$, leading to the drop in $I_{uv}, I_{\theta v}$. We demonstrate this by shading the areas under $\Delta^*_{\theta v}$ and its decompositions at $Pr=7.5$ light orange. From the optimal $\omega^+ = 0.088$ (figure~\ref{fig:int_specx_specf}\textit{g}) to $0.022$ (figure~\ref{fig:int_specx_specf}\textit{i}), $\Delta^*_{\theta v-}$ increases more than $\Delta^*_{\theta v+}$, and $I_{\theta v}$ decreases. At each $\omega^+$, increasing $Pr$ increases $\Delta^*_{\theta v-}$ more than $\Delta^*_{\theta v+}$, hence the drop in $I_{\theta v}$. This is due to the thinning of the conductive sublayer and the increase in the local Stokes layer production, as discussed with respect to figure~\ref{fig:P33_z_peak}.\\

\subsection{Comparison between the travelling wave and the plane oscillation}\label{sec:plane_oscillation}

\begin{figure}
  \centering
 \includegraphics[width=1.0\textwidth,trim={{0.12\textwidth} {0.06\textwidth} {0.12\textwidth} {0.0\textwidth}},clip]{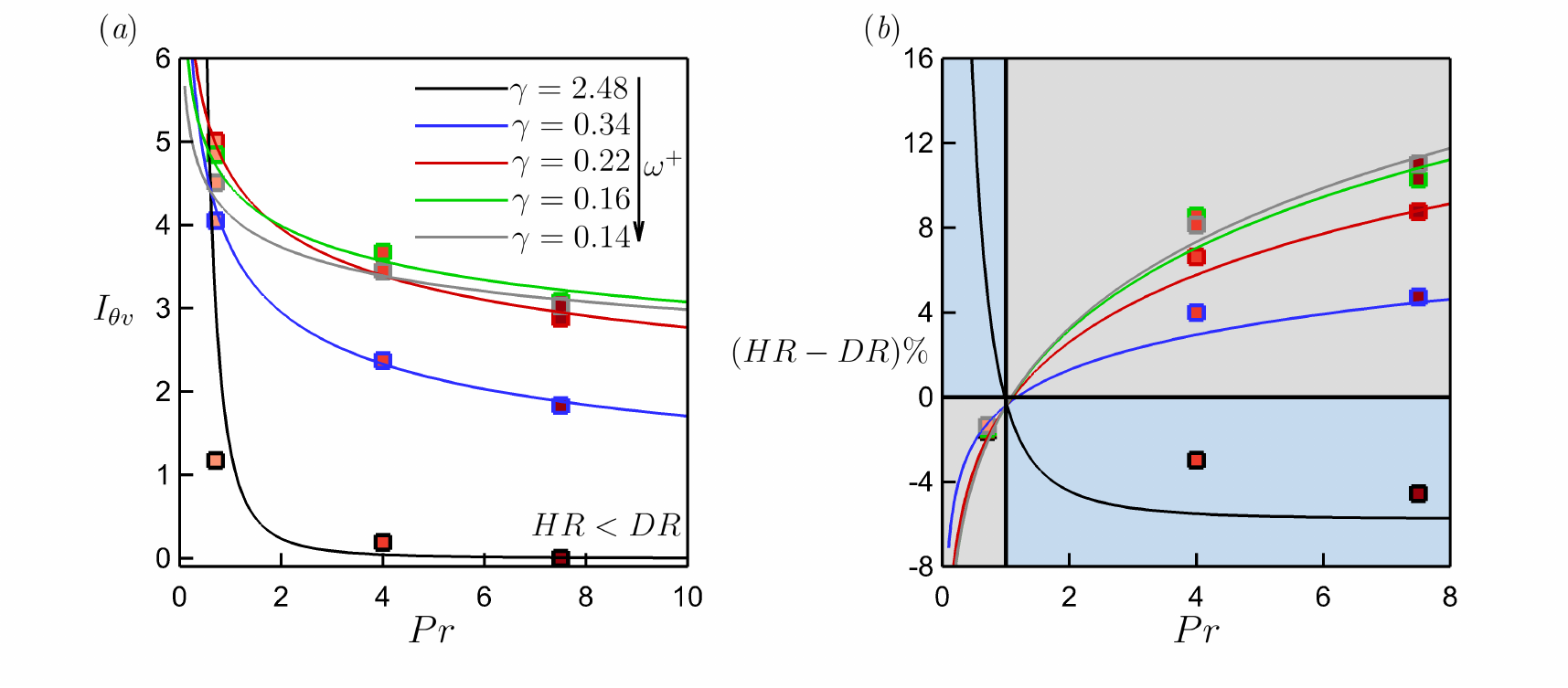}
 \put(-1.6,1.6){\scriptsize{$\gamma > 0.5$}}
 \put(-1.6,2.2){\scriptsize{$\gamma < 0.5$}}
  \caption{Similar plots as in figure~\ref{fig:Pr_scaling}(\textit{b,d}), but for the plane oscillation with $A^+ = 12$ (figure~\ref{fig:flowviz_setup}\textit{c}). The outline colour of each data point indicates its $\omega^+$; $\omega^+ = 0.022$ (black), $0.044$ (blue), $0.066$ (red), $0.088$ (green) and $0.110$ (grey).}
  \label{fig:Pr_scaling_plane}
\end{figure}

\begin{figure}
  \centering
 \includegraphics[width=1.0\textwidth,trim={{0.06\textwidth} {0.0\textwidth} {0.12\textwidth} {0.0\textwidth}},clip]{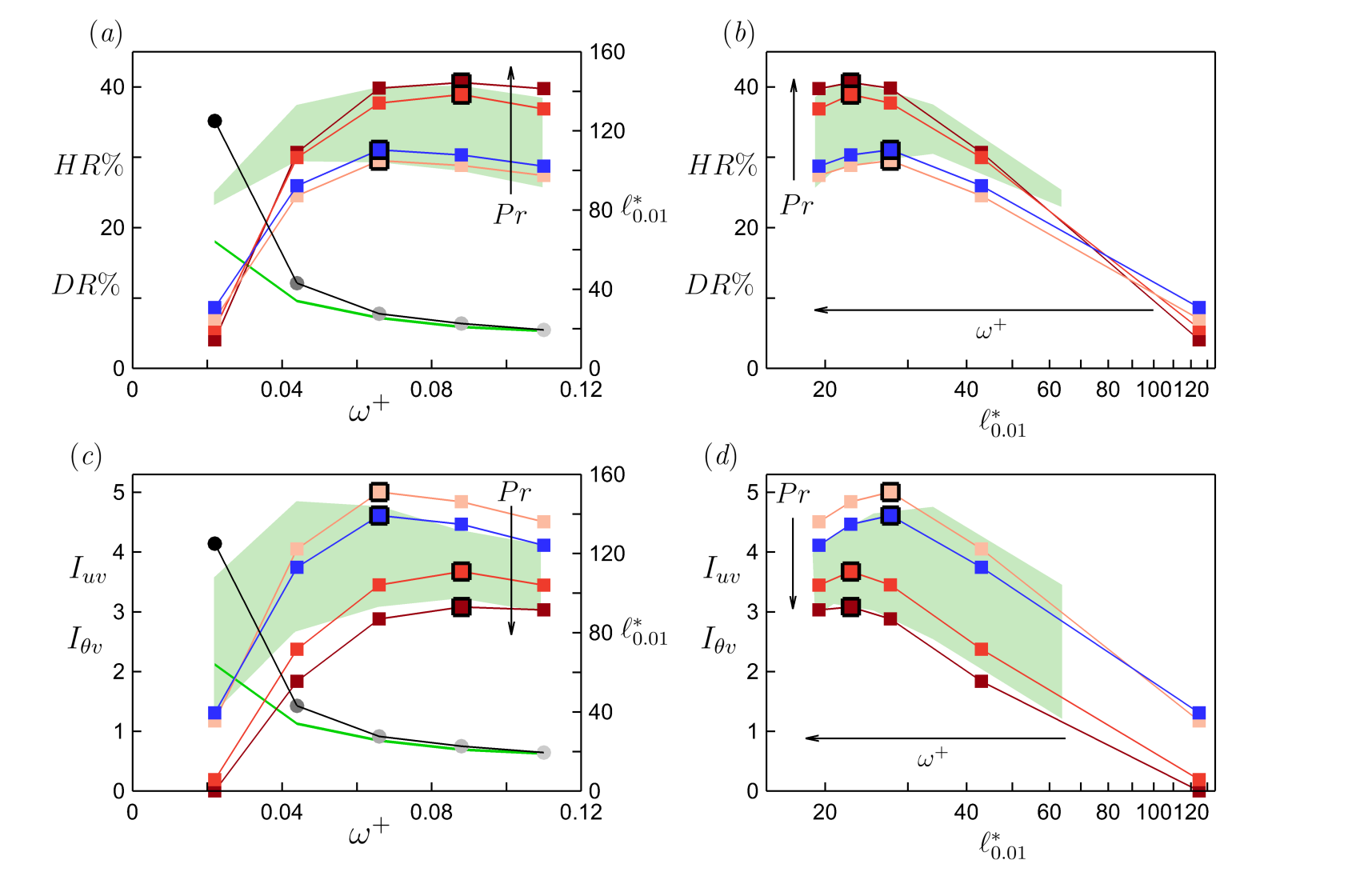}
  \caption{Same plots as in figure~\ref{fig:DR_HR_duw_dcw}, except the data points (filled squares and circles) are from the plane oscillation with $A^+ = 12$. The range of variations in the travelling wave data points are shaded in green. In (\textit{a,c}), the green lines plot $\ell^*_{0.01}$ versus $\omega^+$ for the travelling wave.
  }
  \label{fig:DR_HR_duw_dcw_plane}
\end{figure}

All our findings based on the travelling wave motion (\S\ref{sec:mean_profiles} to \S\ref{sec:stokes2}) remain broadly valid for the case of plane oscillation.  By comparing figures~\ref{fig:Pr_scaling_plane} and \ref{fig:DR_HR_duw_dcw_plane} with figures~\ref{fig:Pr_scaling} and \ref{fig:DR_HR_duw_dcw}, we see that, for example, the power-law relation $I_{\theta v} = I_{uv}/Pr^{\gamma}$ fits well with the plane oscillation data (figure~\ref{fig:Pr_scaling_plane}\textit{a}), and the division $HR> DR$ ($\gamma < 0.5$) and $HR < DR$ ($\gamma > 0.5$) is also valid (figure~\ref{fig:Pr_scaling_plane}\textit{b}). The value of $\gamma$ at each $\omega^+$ is close to its counterpart for the travelling wave case, with the exception of $\omega^+ = 0.022$, where $\gamma = 2.48$ for the plane oscillation (black lines in figure~\ref{fig:Pr_scaling_plane}), almost five times larger the the value of 0.53 found for the travelling wave motion. Such a strong decay rate drops $I_{\theta v}$ to almost zero by $Pr = 7.5$ for the plane oscillation, leading to $HR-DR \simeq -5\%$.  This strong decay is associated with the appearance of a highly protrusive Stokes layer (figures~\ref{fig:DR_HR_duw_dcw_plane}\textit{a,c}), in agreement with the discussion given in \S\ref{sec:stokes2}.  

Figure~\ref{fig:DR_HR_duw_dcw_plane} shows that for $0.066 \le \omega^+ \le 0.110$, $DR$, $HR$, $I_{u v}$,  $I_{\theta v}$, and $\ell^*_{0.01}$ for the plane oscillation are close to the values found for the travelling wave.  However, as $\omega^+$ decreases to $ 0.044$ and then to $0.022$, significant differences appear. For the plane oscillation, the Stokes layer becomes highly protrusive, reaching up to $\ell^*_{0.01} \simeq 120$ at $\omega^+ = 0.022$. As discussed in \S\ref{sec:stokes2}, such a protrusive Stokes layer loses its efficacy in producing a net attenuation of the near-wall turbulence. As a result, $I_{uv}$ and $I_{\theta v}$ significantly drop, and $\gamma$ rises significantly.  \cite{rouhi2022turbulent} showed that when $\ell^*_{0.01} \sim \mathcal{O}(10^2)$, the Stokes layer departs from its laminar-like structure, the mean velocity and temperature profiles become highly disturbed (figure~\ref{fig:profiles_mean_plane}\textit{j,k,l}), and the predictive relations for $DR$ \eqref{eq:dr_model1} and $HR$ \eqref{eq:hr_model1} begin to fail. Similarly, we observe that at $\omega^+ = 0.022$ the predictive relation for $HR$ \eqref{eq:hr_model1} is much less accurate compared to its performance at higher frequencies (figure~\ref{fig:Pr_scaling_plane}\textit{b}).

\begin{figure}
  \centering
 \includegraphics[width=1.0\textwidth,trim={{0.10\textwidth} {0.02\textwidth} {0.08\textwidth} {0.0\textwidth}},clip]{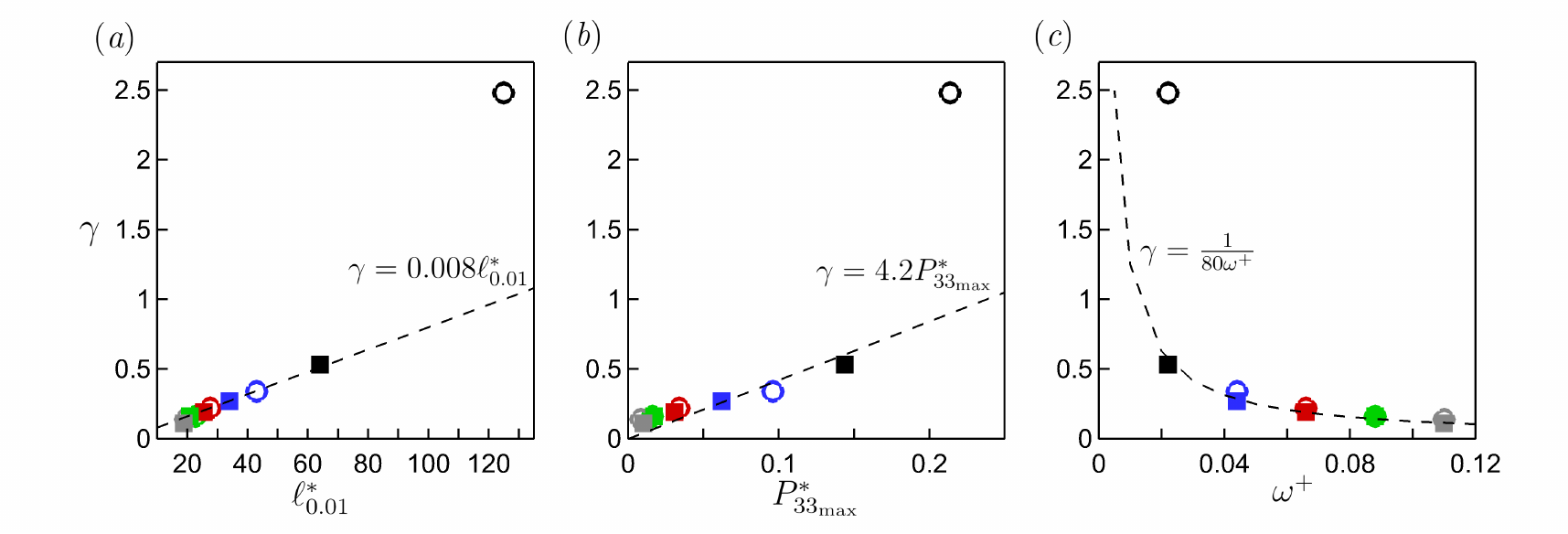}
  \caption{Assessment of the relations (\ref{eq:gamma}) between $\gamma$, $P^*_{33_\mathrm{max}}$, $\ell^*_{0.01}$ and $\omega^+$ for the travelling wave (filled squares) and the plane oscillation (empty circles); $\omega^+ = 0.022$ (black), $0.044$ (blue), $0.066$ (red), $0.088$ (green) and $0.110$ (grey). 
  }
  \label{fig:gamma_P33_l}
\end{figure}

As a final note, it appears that $\gamma$, a $Pr$-independent quantity, increases almost linearly with $\ell^*_{0.01}$ and $P^*_{{33}_\mathrm{max}}$, and decreases inversely with increasing $\omega^+$ following
\begin{equation}
\gamma \simeq 4.2 P^*_{{33}_\mathrm{max}} \simeq 0.008 \ell^*_{0.01} \simeq {1}/({80\omega^+}). \tag{3.10} \label{eq:gamma}
\end{equation}
This is supported by the results shown in figure~\ref{fig:gamma_P33_l}, where we compile all our data for the travelling wave and the plane oscillation cases (with the exception of the plane oscillation at $\omega^+ = 0.022$ where the Stokes layer is particularly protrusive). Our present data are at a fixed $A^+ = 12$ for a plane oscillation ($\kappa^+_x = 0$) and a travelling wave ($\kappa^+_x = 0.0014$). The relation for $\gamma$ (\ref{eq:gamma}) is expected to also depend on $A^+$ and $\kappa^+_x$. However, for the upstream travelling wave with $\omega^+ \gtrsim 0.04$, we anticipate $\gamma$ to weakly depend on $\kappa^+_x$. Our conjecture is partly based on the validity of (\ref{eq:gamma}) for $\kappa^+_x = 0$ and $\kappa^+_x = 0.0014$ for $\omega^+ \gtrsim 0.04$. Furthermore, $\gamma$ relates $\Delta \overline{\Theta}^*_{170}$ to $\Delta \overline{U}^*_{170}$, which $\Delta \overline{U}^*_{170}$ is seen to weakly depend on $\kappa^+_x$ for the upstream travelling wave with $\omega^+ \gtrsim 0.04$ (figure~13 in \citealt{gatti2016}), and we anticipate $\Delta \overline{\Theta}^*_{170}$ to behave similarly.

\section{Conclusions}\label{sec:conclusions}


Direct numerical simulations of turbulent half channel flow with forced convection were performed at a friction Reynolds number of 590 for Prandtl numbers of 0.71 (air), 4.0, 7.5 (water), and 20 (molten salt). Spanwise wall forcing was applied either as a plane wall oscillation or a streamwise in-plane travelling wave with wavenumber $\kappa^+_x = 0.0014$ ($\lambda^+ \simeq 4500$). For both oscillation mechanisms, we fix the amplitude $A^+ = 12$ and change the frequency from $\omega^+ = 0.022$ to $0.110$ (upstream travelling waves only). 

\begin{figure}
  \centering
 \includegraphics[width=1.0\textwidth,trim={{0.0\textwidth} {0.0\textwidth} {0.0\textwidth} {0.0\textwidth}},clip]{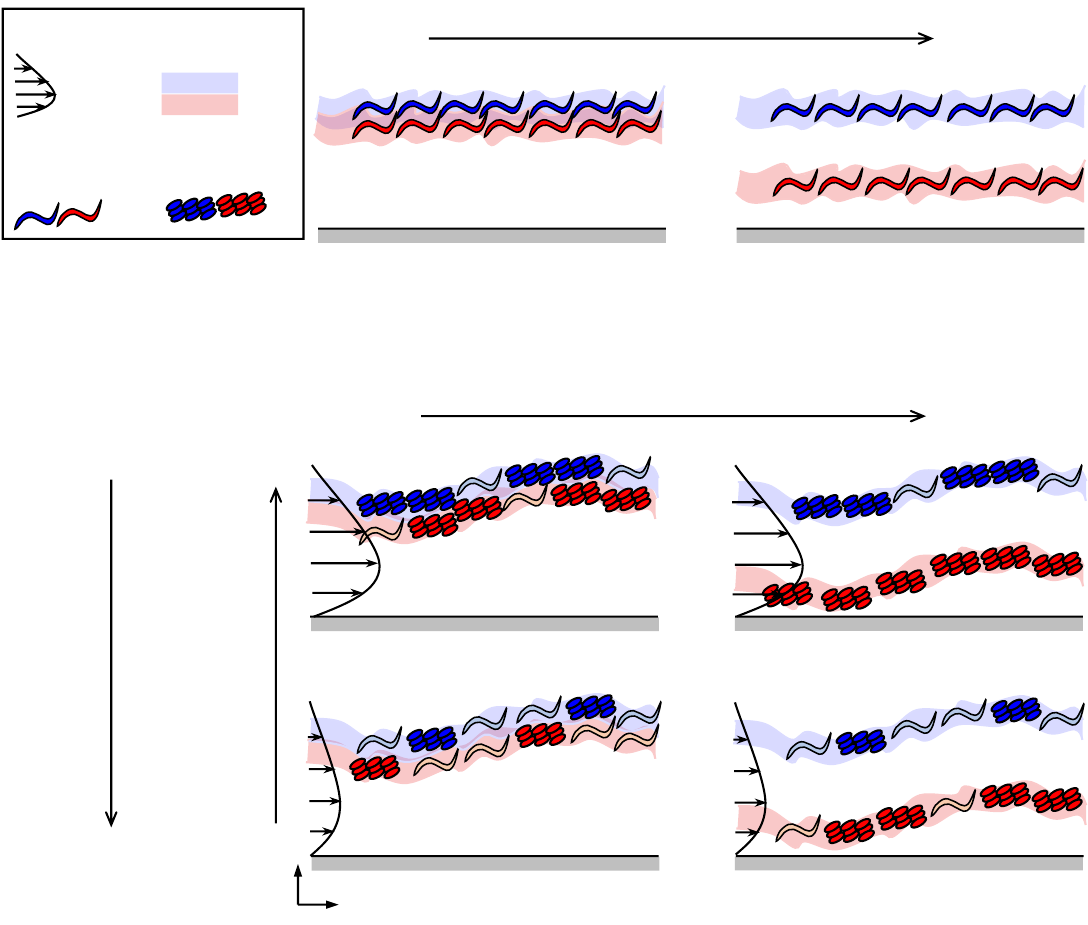}
 \put(-4.7,9.4){\scriptsize{non-actuated}}
 \put(-6.6,9.3){\scriptsize{$Pr \simeq 1$}}
 \put(-2.6,9.3){\scriptsize{$Pr > 1$}}
 \put(-7.9,8.8){(\textit{a})}
 \put(-3.7,8.8){(\textit{b})}
 \put(-10.2,9.2){\scriptsize{Streams of energetic}}
 \put(-9.5,8.95){\scriptsize{$uv, \theta v$}}
 \put(-11.2,9.2){\scriptsize{$P^*_{33}$}}
 \put(-9.7,7.9){\scriptsize{Small scales}}
 \put(-9.5,7.65){\scriptsize{$\lambda^+_x \simeq 100$}}
  \put(-10.8,7.9){\scriptsize{Streaks}}
 \put(-11.0,7.6){\scriptsize{$\lambda^+_x \simeq 1000$}}
 \put(-4.7,6.3){\scriptsize{actuated}}
 \put(-5.3,5.95){\scriptsize{$HR - DR \propto Pr^{(0.5 - \gamma)} - 1$}}
 \put(-5.0,5.55){\scriptsize{$I_{\theta v} = I_{uv}/Pr^{\gamma}$}}
  \put(-6.6,5.4){\scriptsize{$Pr \simeq 1$}}
 \put(-2.6,5.4){\scriptsize{$Pr > 1$}}
  \put(-7.9,5.0){(\textit{c})}
 \put(-3.7,5.0){(\textit{d})}
 \put(-6.0,3.3){\scriptsize{$HR-DR \simeq 0$}}
 \put(-1.5,3.3){\scriptsize{$HR-DR \lesssim 0$}}
   \put(-7.9,2.5){(\textit{e})}
 \put(-3.7,2.5){(\textit{f})}
 \put(-6.0,0.85){\scriptsize{$HR-DR \simeq 0$}}
 \put(-1.5,0.85){\scriptsize{$HR-DR > 0$}}
 \put(-7.8,0.05){\scriptsize{$x$}}
 \put(-8.35,0.5){\scriptsize{$y$}}
 \put(-11.0,5.1){\scriptsize{$\gamma \simeq 1/(80 \omega^+) \simeq 4.2 P^*_{33_\mathrm{max}}$}}
 \put(-11.3,4.1){\scriptsize{$\omega^+ \lesssim 0.025$}}
 \put(-9.2,4.1){\scriptsize{$\gamma \gtrsim 0.5$}}
 \put(-9.9,3.75){\scriptsize{$P^*_{33_\mathrm{max}} \gtrsim 0.12$}}
  \put(-11.3,1.6){\scriptsize{$\omega^+ > 0.025$}}
 \put(-9.2,1.6){\scriptsize{$\gamma < 0.5$}}
 \put(-9.9,1.25){\scriptsize{$P^*_{33_\mathrm{max}} < 0.12$}}
  \caption{Schematic illustration of our findings from \S\ref{sec:hr_dr_break} to \S\ref{sec:plane_oscillation}. The meaning of each sketch is written in the top-left framed area. (\textit{a--f}) Near-wall illustration of the energetic scales of $uv$ and $\theta v$ for (\textit{a,b}) non-actuated cases, and (\textit{c--f}) actuated cases. (\textit{a,c,e}) correspond to $Pr \simeq 1$, and (\textit{b,d,f}) correspond to $Pr > 1$. For the actuated cases, (\textit{c,d}) represent $\omega^+ \lesssim 0.025$, and (\textit{e,f}) represent $\omega^+ > 0.025$. The Stokes layer production profile is drawn for each actuated case.}
  \label{fig:schematic_conclusion}
\end{figure}

The key finding of the present work is that for $\omega^+ > 0.022$ and $Pr > 1$, we achieve $HR > DR$; especially, for $\omega^+ \gtrsim 0.066$ and $Pr \gtrsim 4.0$, there is a significant disparity between $HR \simeq 40\%$ and $DR \simeq 30\%$. On the other hand, for $\omega^+ \le 0.022$ and $Pr > 1$, $HR \lesssim DR$. These results apply to the travelling wave motion and the plane oscillation. To help understand these results, we derived explicit relations between $HR$ and $ DR$ and the integrals of the attenuation in the turbulent shear-stress ($I_{uv}$) and turbulent temperature-flux ($I_{\theta v}$). Figure~\ref{fig:schematic_conclusion} summarises how $I_{uv}$ and $I_{\theta v}$ change with $Pr$ and $\omega^+$, and illustrates the underlying physics. For the cases considered here, we find that $I_{\theta v} \simeq  I_{uv}/Pr^{\gamma}$, where $\gamma \simeq 1/(80\omega^+) \simeq 4.2 P^*_{33_\mathrm{max}}$ is proportional to the Stokes layer production $P^*_{33}$. Therefore, we expect $I_{\theta v}$ to fall below $I_{uv}$ for $Pr > 1$ and for increasing $\gamma$ (decreasing $\omega^+$ and increasing $P^*_{33}$). Stokes layer production attenuates the energy of the streak-like scales with $\lambda^+_x \simeq 1000$ (hence the rise in $I_{uv}, I_{\theta v}$), but generates small scales with $\lambda^+_x \simeq 100$ and amplifies their energy (hence the drop in $I_{uv}, I_{\theta v}$). At a fixed $\omega^+$ ($\gamma$), increasing $Pr$ towards $Pr > 1$, thins the conductive sublayer, and the energetic small scales of $\theta v$ travel closer to the wall, and locally are exposed to a larger Stokes layer production compared to those of $uv$. Larger Stokes layer production means larger amplification of energy in the small scales of $\theta v$ compared to $uv$, leading to $I_{\theta v} < I_{uv}$ (figure~\ref{fig:schematic_conclusion}\textit{d,f}). In other words, for $Pr > 1$, the energy attenuation of $\theta v$ is less than $uv$. Nevertheless, by exploiting $I_{\theta v} \simeq  I_{uv}/Pr^{\gamma}$, we arrive at $HR - DR \propto Pr^{(0.5-\gamma)} - 1$, as $HR$ depends on $\Delta \overline{\Theta}^* = Pr I_{\theta v} = Pr^{(1-\gamma)} I_{uv}$, and the non-actuated Stanton number that decays with $Pr$. Therefore, we can achieve $HR-DR > 0$ if $\gamma \lesssim 0.5$ ($\omega^+ \gtrsim 0.025$). These findings and the relations between $HR-DR$, $\omega^+$, $\gamma$ and $P^*_{33_\mathrm{max}}$ agree well with our DNS data for both the travelling wave and the plane oscillation. 

The time scale of the near-wall energetic temperature scales $\mathcal{T}^+_\theta$ was found to be almost insensitive to $Pr$, and it remains close to its counterpart for the velocity scales $\mathcal{T}^+_u \simeq 100$. As a result, for the plane oscillation, the optimal value of $\omega^+$ for the maximum $HR$ is close to the one for the maximum $DR$, that is, $\omega^+ \simeq 2\pi/100 \simeq 0.066$. However, for the travelling wave, the optimal frequency for $HR$ changes from $0.044$ at $Pr = 0.71$ to $0.088$ at $Pr = 7.5$ because it depends on the advection speed of the energetic temperature scales. The thinning of the conductive sublayer decreases the advection speed, and so the optimal frequency of actuation increases.

From the Prandtl number scaling of the thermal statistics, we derived a predictive model for $HR$. By knowing $\gamma$ and $I_{uv}$ for a fixed set of actuation parameters $(A^+,\kappa^+_x,\omega^+)$ at a given Reynolds and Prandtl number, the model predicts $HR$ at other Prandtl and Reynolds numbers. Using this model, we generated a map $HR(Pr, Re_{\tau_0})$ for the optimal case with $\omega^+ = 0.088$, and found good agreement with the computations for Prandtl numbers up to 7.5.  
To further evaluate the model, we conducted an extra simulation at $\omega^+ = 0.088$ and $Pr = 20, Re_{\tau_0} = 590$. Again, good agreement with the model was found (DNS $HR = 43\%, HR - DR = 13\%$ compared with the model $HR = 45\%, HR - DR = 14\%$).


As an early systematic work on turbulent heat-transfer control by spanwise wall forcing, we focused on $Pr$ and $\omega^+ > 0$, due to their significant effect on $HR$, and $HR - DR$. Our findings are encouraging to pursue this work towards investigating the effects of $A^+, \kappa^+_x, Re_{\tau_0}$ and $\omega^+ < 0$ (downstream travelling wave). We anticipate that our discoveries could be applied to a wider parameter space. Our conjecture is partly supported by the validity of our findings for both the travelling wave and plane oscillation in the present study. Also, similar findings for $DR$ are shown to be applicable to a wide parameter space of $(A^+, \omega^+, \kappa^+_x, Re_{\tau_0})$. The prediction model for $DR$ by \cite{gatti2016}, the analogue of our model for $HR$, is shown to agree well with the DNS~\citep{gatti2024turbulent} and LES~\citep{rouhi2022turbulent} data for $1000 \lesssim Re_{\tau_0} \lesssim 6000$ over the actuation parameter space where $DR$ significantly changes ($5 \le A^+ \le 12, 0 \le \kappa^+_x \le 0.02, -0.2 \le \omega^+ \le 0.2$). Also, the findings related to the Stokes layer protrusion and $DR$ are found to be valid over the range of $\kappa^+_x$ considered so far. For instance, up to the Stokes layer thickness $\delta^*_S \simeq 6$ (protrusion height $\ell^*_{0.01} \simeq 25$), there is a linear relation between $DR, \ell^*_{0.01}$ and $ \delta^*_S$, regardless of the value of $\kappa^+_x$~\citep{quadrio2011,rouhi2022turbulent}. This is because the modification of the velocity scales by the Stokes layer is similar between the plane oscillation~\citep{touber2012} and the travelling wave at different values of $\kappa^+_x$~\citep{rouhi2022turbulent}.

\vspace{0.5cm}

\noindent \textbf{Acknowledgements}\\
The research was funded through the Deep Science Fund of Intellectual Ventures (IV). We acknowledge Dr.\ Daniel Chung for providing his DNS solver. We thank EPSRC for the computational time made available on ARCHER2 via the UK Turbulence Consortium (EP/R029326/1), and IV for providing additional computational time on ARCHER2.\\

\noindent \textbf{Declaration of interests.} The authors report no conflict of interest.

\appendix

\begin{table}
\centering
\begin{tabular}{ccccccccc}
 setup  & $Re_{\tau_0}$ & $Pr$ & $L_x, L_z$ &  $ \Delta^+_x , \Delta^+_z $ & $\Delta^+_y$ & $C_{f_0}$ &  $Nu_0$  & Legend  \\ \\  
 full fine  & $590$ & $7.5$ & $2 \pi h , \pi h$ &   $ 3.6 , 1.8 $ & $0.12- 3.8$ & $0.0058$ &  $162$ & {\color{gray}\solid} \\
 reduced coarse  & $590$ & $7.5$ & $2.7 h , 0.85 h$ &  $ 8.3 , 3.9 $ & $0.28 - 8.5$ & $0.0057$ &  $162$ & \dashed \\
 MKM99  & $590$ & - & $2\pi h , \pi h$ &  $ 9.7 , 4.8 $ & $0.09 - 7.2$ & $0.0057$ &  - & \filsqr \\
 AH21  & $500$ & $7.0$ & $2\pi h , \pi h$ &  $ 8.2 , 4.1 $ & $0.27 - 3.9$ & - &  $132$ & \circle  
  \end{tabular}
\caption{Validation simulations for the reduced coarse setup. Non-actuated half channel flow at  $Pr = 7.5$. Statistics for these cases are shown in figure~\ref{fig:grid_domain_non_actuated}. The reference cases are MKM99~\citep{moser1999direct} and AH21~\citep{alcantara2021directb}.} \label{tab:non_act}
\end{table}

\section{Validation of the grid and the domain size}\label{sec:grid_domain}
The production runs were performed in a reduced-domain half channel flow ($L_x \gtrsim 2.7h, L_z \simeq 0.85 h$) with  grid resolution $\Delta^+_x \times \Delta^+_z \simeq 8 \times 4$; we call this setup ``reduced coarse''.  To check that this setup gives accurate $DR, HR$ and statistics of interest over our parameter space, calculations with a finer grid resolution of $\Delta^+_x \times \Delta^+_z \simeq 4 \times 2$ were conducted in a full domain ($L_x \gtrsim 2\pi h, L_z \simeq \pi h$); we call this setup ``full fine''.  Two additional  setups are also used, called ``full coarse'' ($L_x \gtrsim 7.6 h, L_z \simeq \pi h, \Delta^+_x \times \Delta^+_z \simeq 8 \times 4$) and ``full intermediate'' ($L_x \gtrsim 7.6 h, L_z \simeq \pi h, \Delta^+_x \times \Delta^+_z \simeq 6 \times 3$). For all simulations, $Re_{\tau_0}=590$ and $A^+=12$. For all the reduced domain cases, we reconstruct the $\overline{U}^*$ and $\overline{\Theta}^*$ profiles beyond $y^*_\mathrm{res} = 170$, and calculate $DR$ and $HR$ following our approach in \S\ref{sec:calculate_cf_ch}.

\begin{figure}
  \centering
 \includegraphics[width=.95\textwidth,trim={{0.0\textwidth} {0.0\textwidth} {0.0\textwidth} {0.0\textwidth}},clip]{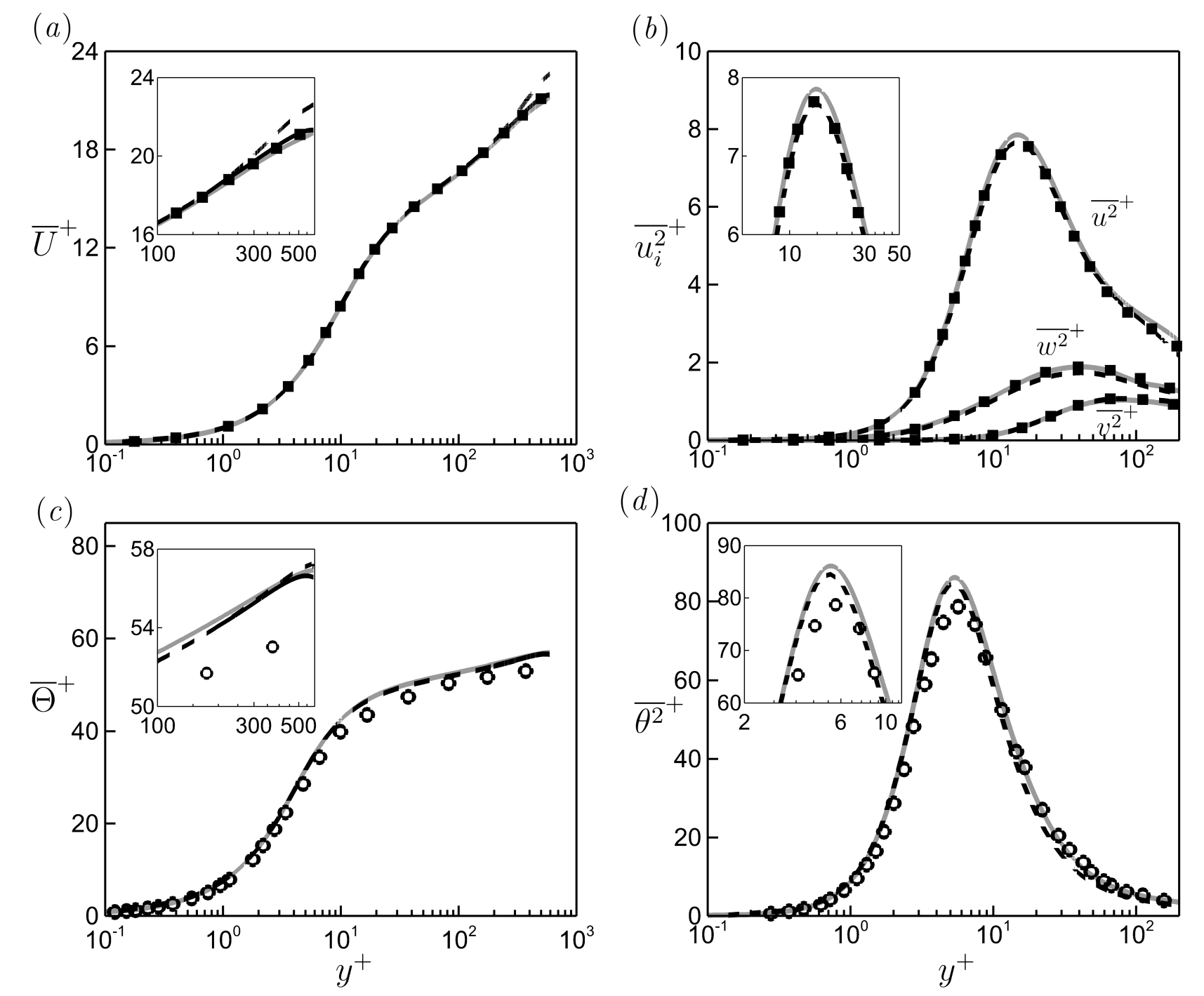}
  \caption{Comparing the results for the three setups given in table~\ref{tab:non_act}. (\textit{a}) Mean velocity profiles $\overline{U}^+$. (\textit{b}) Turbulent stresses $\overline{u^2}^+, \overline{v^2}^+, \overline{w^2}^+$. (\textit{c}) Mean temperature profiles $\overline{\Theta}^+$. (\textit{d}) Mean square of turbulent temperature $\overline{ \theta^2 }^+$. Solid grey line, ``full fine'' setup. Dashed line: ``reduced coarse'' setup. Solid black line in (\textit{a,c}), reconstructed profile for $y^+ \ge 200$ as in \S\ \ref{sec:calculate_cf_ch}.  Square symbols: DNS of \cite{moser1999direct}. Circle symbols: DNS of \cite{alcantara2021directb}.}
  \label{fig:grid_domain_non_actuated}
\end{figure}

For the non-actuated half channel flow at $Pr = 7.5$, table~\ref{tab:non_act} shows that for the full fine and reduced coarse setups, the Nusselt numbers $Nu_0$ are identical, and the skin-friction coefficients $C_{f_0}$ differ by less than $2\%$. The profiles of $\overline{U}^+, \overline{\Theta}^+$, turbulent stresses $\overline{u^2}^+, \overline{v^2}^+, \overline{w^2}^+$, and mean square of turbulent temperature $\overline{\theta^2}^+$ are also in excellent agreement (figure~\ref{fig:grid_domain_non_actuated}). Compared to the reference data, our obtained $C_{f_0}$ from the reduced coarse case is identical to the one from \cite{moser1999direct} (table~\ref{tab:non_act}); similarly, our velocity statistics up to $y^+_\mathrm{res} \simeq 200$ are in excellent agreement (figure~\ref{fig:grid_domain_non_actuated}). In terms of the temperature statistics $\overline{\Theta}^+, \overline{\theta^2}^+$, the small differences between our results and those from \cite{alcantara2021directb} are most likely due to the differences in $Re_{\tau_0}$ and $Pr$ (590 compared to 500, and 7.5 compared to 7).

\begin{table}
\centering
\begin{tabular}{cccccccc|c}
 case  & $\omega^+$ & $L_x, L_z$ &  $ \Delta^+_x , \Delta^+_z $ &  $ \Delta^+_y $ & $DR\%$ &  $HR\%$  &  &  \\ \\ 
 full coarse & $0.022$ & $7.6h, \pi h$ &  $ 7.4 , 3.9 $ &  $ 0.20 - 6.0 $ & $24.5$ &  $23.3$  &   & Set 1\\
reduc. coarse & $0.022$ & $7.6h, 0.85 h$ &  $ 8.8 , 3.9 $ &  $ 0.28 - 8.5 $ & $25.2$ &  $23.6$  &  & $Pr = 0.71$  \\
   &  &   &  &  & &   &   & $\kappa^+_x = 0.0014$  \\
full coarse & $0.044$ & $7.6h, \pi h$ &  $ 7.4 , 3.9 $ &  $ 0.20 - 6.0 $ & $30.4$ &  $29.1$  &  & ${\epsilon_{DR}}_\mathrm{max}, {\epsilon_{HR}}_\mathrm{max}$ \\
reduc. coarse  & $0.044$ & $7.6h, 0.85 h$ &  $ 8.8 , 3.9 $ &  $ 0.28 - 8.5 $ & $30.6$ &  $29.3$  &  & $0.7\%, 0.6\%$ \\
   &  &   &  &   &  &  &   & \\
full coarse  & $0.088$ & $7.6h, \pi h$ &  $ 7.4 , 3.9 $ &  $ 0.20 - 6.0 $ & $29.9$ &  $28.5$  &  &  \\
reduc. coarse & $0.088$ & $7.6h, 0.85 h$ &  $ 8.8 , 3.9 $ &  $ 0.28 - 8.5 $ & $29.3$ &  $27.9$  &  & \\ \hline
   &  &   &  &   &  &  & Line   & \\
full fine  & $0.022$ & $7.6h, \pi h$ &  $ 3.9 , 1.8 $ &  $ 0.12 - 3.8 $ & $26.2$ &  $25.4$  & {\color{gray}\solid}  & Set 2  \\
reduc. coarse  & $0.022$ & $7.6h, 0.85 h$ &  $ 8.8 , 3.9 $ &  $ 0.28 - 8.5 $ & $25.6$ &  $24.0$  & \dashed  & $Pr = 7.5$ \\ 
   &  &   &  &  &  &  &   & $\kappa^+_x= 0.0014$ \\
full fine  & $0.044$ & $7.6h, \pi h$ &  $ 3.9 , 1.8 $ &  $ 0.12 - 3.8 $ & $31.4$ &  $38.6$  & {\color{gray}\solid}  & \\
full inter. & $0.044$ & $7.6h, \pi h$ &  $ 5.8 , 2.7 $ &  $ 0.18 - 5.7 $ & $30.9$ &  $37.7$  & \dasheddotted & ${\epsilon_{DR}}_\mathrm{max}, {\epsilon_{HR}}_\mathrm{max}$ \\
full coarse  & $0.044$ & $7.6h, \pi h$ &  $ 8.8 , 4.0 $ &  $ 0.28 - 8.5 $ & $31.0$ &  $37.0$  & \dotted  & $1.5\%, 1.6\%$ \\
reduc. coarse  & $0.044$ & $7.6h, 0.85 h$ &  $ 8.8 , 3.9 $ &  $ 0.28 - 8.5 $ & $31.8$ &  $37.7$  & \dashed & \\
   &  &   &  &  &   &  &   & \\
full fine   & $0.088$ & $7.6h, \pi h$ &  $ 3.9 , 1.8 $ &  $ 0.12 - 3.8 $ & $29.0$ &  $39.8$  & {\color{gray}\solid}  & \\
reduc. coarse  & $0.088$ & $7.6h, 0.85 h$ &  $ 8.8 , 3.9 $ &  $ 0.28 - 8.5 $ & $30.5$ &  $40.4$  & \dashed & \\ \hline
full fine  &  $0.044$ & $2\pi h, \pi h$ &  $ 3.6 , 1.8 $ &  $ 0.12 - 3.8 $ & $25.2$ &  $28.9$  &  & Set 3  \\
reduc. coarse  &  $0.044$ & $2.7h, 0.85 h$ &  $ 8.3 , 3.9 $ &  $ 0.28 - 8.5 $ & $26.5$ &  $30.8$  &  & $Pr = 7.5$ \\
   &  &   &  &  &  &    &   & $\kappa^+_x = 0$ \\
full fine  &  $0.088$ & $2\pi h, \pi h$ &  $ 3.6 , 1.8 $ &  $ 0.12 - 3.8 $ & $29.4$ &  $39.0$  &  & ${\epsilon_{DR}}_\mathrm{max}, {\epsilon_{HR}}_\mathrm{max}$  \\
reduc. coarse  &  $0.088$ & $2.7h, 0.85 h$ &  $ 8.3 , 3.9 $ &  $ 0.28 - 8.5 $ & $30.8$ &  $40.7$  &  & $1.4\%, 1.9\%$
  \end{tabular}
\caption{Validation simulations for grid and domain size.} 
\label{tab:act}
\end{table}

For the actuated half channel flow cases at $Pr=0.71$ and $7.5$, table~\ref{tab:act} lists three sets of validation cases. 
For Set 1, we compare the ``reduced coarse'' setup with the ``full coarse'' setup because the grid resolution of $\Delta^+_x \times \Delta^+_z \simeq 8 \times 4$ is fine enough at this $Pr$~\citep{pirozzoli2016passive,alcantara2018dns,alcantara2021direct,alcantara2021directb}.  Sets 1 and 2 are travelling wave cases, and Set 3 is for the plane oscillation case. For each set, we see that reducing the domain size and coarsening the grid from the full fine setup to the reduced coarse setup changes $DR$ and $HR$ by a maximum of $1.5\%$ and $1.9\%$, respectively.

\begin{figure}
  \centering
 \includegraphics[width=\textwidth,trim={{0.25\textwidth} {0.11\textwidth} {0.04\textwidth} {0.0\textwidth}},clip]{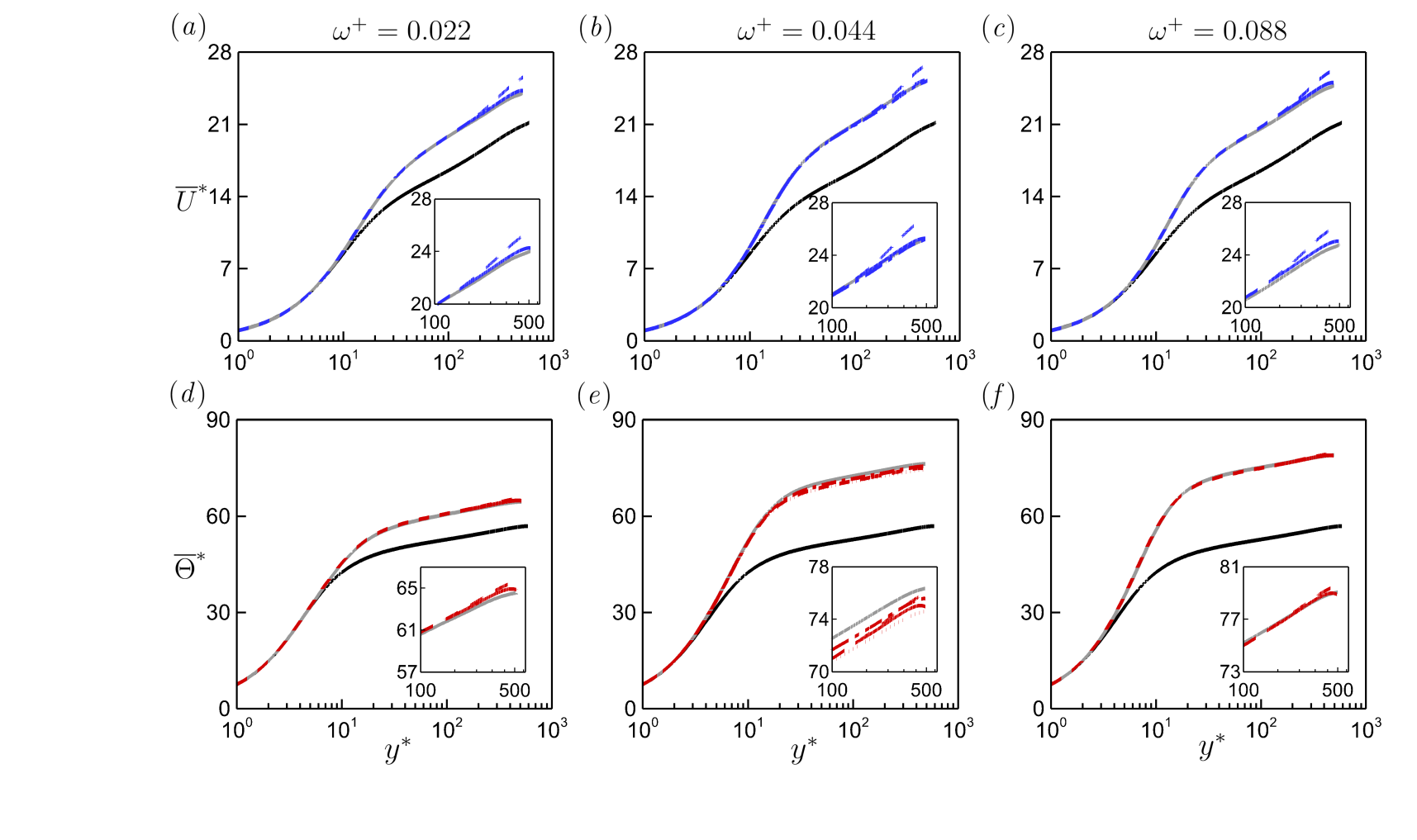}
  \caption{Comparison of the mean velocity (\textit{a,b,c}) and the mean temperature (\textit{d,e,f}) profiles for the travelling wave at $Pr = 7.5$ (Set 2, table~\ref{tab:act}). (\textit{a,d}) $\omega^+ = 0.022$, (\textit{b,e}) $\omega^+ = 0.044$ and (\textit{c,f}) $\omega^+ = 0.088$.  Solid grey line: ``full fine'' case.  Dashed-dotted line: ``full inter.'' case. Dotted line: full coarse case. Dashed line:``reduc.\ coarse'' case. Solid blue and red lines, reconstructed profiles for $y^* \ge 170$ as in \S\ \ref{sec:calculate_cf_ch}. Solid black line: non-actuated ``full fine'' case.}
  \label{fig:grid_domain_actuated}
\end{figure}
\begin{figure}
  \centering
 \includegraphics[width=\textwidth,trim={{0.02\textwidth} {0.02\textwidth} {0.0\textwidth} {0.0\textwidth}},clip]{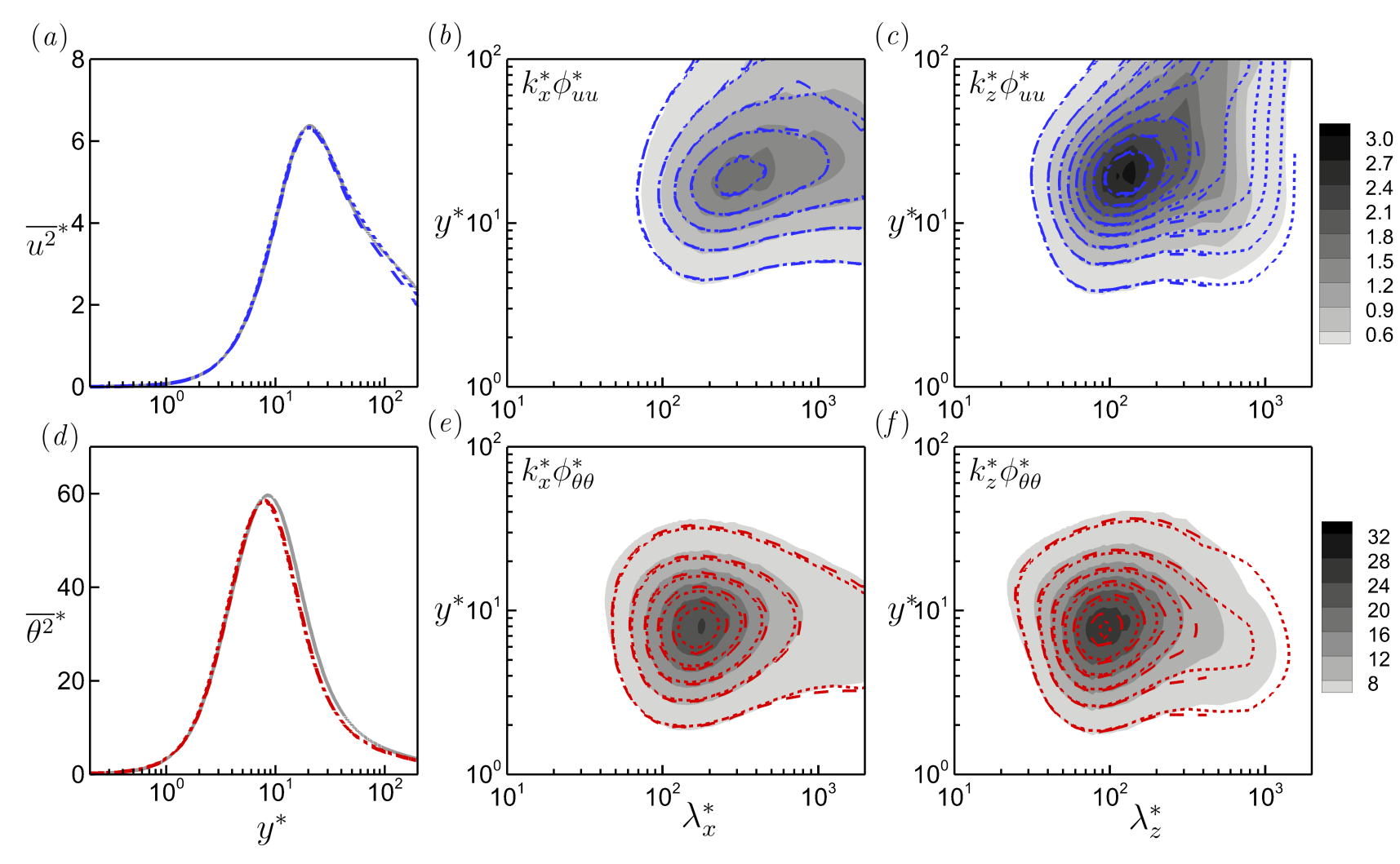}
  \caption{Comparison of (\textit{a}) streamwise turbulent stress $\overline{u^2}^*$ and (\textit{d}) mean square of turbulent temperature $\overline{\theta^2}^*$, as well as their pre-multiplied spectrograms $k^*_x \phi^*_{uu}, k^*_z \phi^*_{uu}, k^*_x \phi^*_{\theta \theta}, k^*_z \phi^*_{\theta \theta}$ (\textit{b,c,e,f}) for the travelling wave at $Pr = 7.5$, $\omega^+=0.044$ (Set 2, table~\ref{tab:act}). Solid grey line and grey-scale contour field: ``full fine'' case.  Dotted line and dotted contour lines: ``full coarse'' case.  Dashed line and dashed contour lines:``reduc.\ coarse'' case.}
  \label{fig:grid_stats_actuated}
\end{figure}

Set 2 contains the most challenging cases over the range $0.71 \le Pr \le 7.5$ in terms of their grid resolution requirement. In figure~\ref{fig:grid_domain_actuated}, we compare the mean velocity $\overline{U}^*$ and the mean temperature $\overline{\Theta}^*$ profiles for $Pr = 7.5$. In terms of the $\overline{U}^*$ profiles, the reduced coarse setup is almost identical to the full fine setup up to $y^*_\mathrm{res} \simeq 170$. Beyond $y^*_\mathrm{res}$, the reconstructed $\overline{U}^*$ profiles for the reduced coarse setup agree well with the full fine setup. In terms of the $\overline{\Theta}^*$ profiles, excellent agreement is obtained between the full fine setup and the reduced coarse setup at $\omega^+ = 0.022$ and $0.088$ (figures~\ref{fig:grid_domain_actuated}\textit{a,c}). At $\omega^+ = 0.044$, there is some sensitivity to the grid, where at $y^* = 200$, $\overline{\Theta}^* = 74.3$ for the full fine setup, and 73.0 for the reduced coarse setup. This difference (1.3 units) is small relative to the temperature difference $\Delta \overline{\Theta}^*_{200} \simeq 19.0$ between the actuated and the non-actuated profiles. In figure~\ref{fig:grid_stats_actuated}, we assess the accuracy of the reduced coarse setup in resolving $\overline{u^2}^*$ and $\overline{\theta^2}^*$, as well as their pre-multiplied spectrograms $k^*_x \phi^*_{uu}, k^*_z \phi^*_{uu}, k^*_x \phi^*_{\theta \theta}, k^*_z \phi^*_{\theta \theta}$ at $\omega^+ = 0.044$. The reduced coarse setup results agree well with the full fine setup results, especially in matching the energetic peaks in the spectrograms in terms of magnitude, location $y^*$, and length scales $\lambda^*_x, \lambda^*_z$.

Table~\ref{tab:act_pr} summarises our validation study at $Pr = 20.0$ (table~\ref{tab:runs}). For the non-actuated case (table~\ref{tab:act_pr}), $C_{f_0}$ and $C_{h_0}$ from the coarse grid results differ from the intermediate grid results  by a maximum of $0.5\%$, and for the travelling wave case, $DR$ and $HR$ differ by  $<1\%$.

Based on these results, we conclude that  $DR, HR$ and the statistics of interest can be accurately computed using the reduced coarse setup with the grid resolution of $\Delta^+_x \times \Delta^+_z \simeq 8 \times 4$, and the domain size of $L_x \times L_z \simeq 7.6 h \times 0.85 h$ for the travelling wave case ($\kappa^+_x = 0.0014$), and the domain size of $L_x \times L_z \simeq 2.7 h \times 0.85 h$ for the plane oscillation case ($\kappa^+_x = 0$). These grid and domain size prescriptions were used for our production calculations (table~\ref{tab:runs}).

\begin{table}
\centering
\begin{tabular}{cccccccccc}
 case  & $\kappa^+_x$ & $\omega^+$ & $L_x, L_z$ &  $ \Delta^+_x , \Delta^+_z $ &  $ \Delta^+_y $  & $C_{f_0}$ & $C_{h_0}$ & $DR\%$ &  $HR\%$ \\ \\ 
 reduc. coarse  & - & - & $2.7 h, 0.85 h$ &  $ 8.3 , 3.9 $ &  $ 0.28 - 8.4 $ & $0.0057$ &  $5.47 \times 10^{-4}$  & - &  - \\
 reduc. inter.  & - & - & $2.7 h, 0.85 h$ &  $ 6.2 , 2.6 $ &  $ 0.18 - 5.6 $ & $0.0057$ &  $5.44 \times 10^{-4}$  & - & - \\ \\
 reduc. coarse  & $0.0014$ & $0.088$ & $7.6 h, 0.85 h$ &  $ 8.8 , 3.9 $ &  $ 0.28 - 8.4 $ & - &  -  & $29.5$ & $43.5$ \\
 reduc. inter.  & $0.0014$ & $0.088$ & $7.6 h, 0.85 h$ &  $ 5.8 , 2.6 $ &  $ 0.18 - 5.6 $ & - &  -  & $28.6$ & $43.1$
  \end{tabular}
\caption{Validation cases at $Re_{\tau_0} = 590$ and $Pr = 20.0$.} 
\label{tab:act_pr}
\end{table}

\section{Relation between heat-transfer reduction and temperature difference}\label{sec:gq_model_hr}

The mean velocity $\overline{U}^*$ and temperature $\overline{\Theta}^*$ profiles in the log region and beyond can be expressed based on the following semi-empirical wall-wake relations~\citep{rouhi2022riblet}
\begin{align}
 \overline{U}^* &= \frac{1}{\kappa_u} \ln(y^*) + B_u + \frac{2\Pi_u}{\kappa_u}\mathcal{W}_u(y/h) + \Delta \overline{U}^*_\mathrm{log}, \tag{C1\textit{a}} \label{eq:up_log} \\
 \overline{\Theta}^* &= \frac{1}{\kappa_\theta} \ln(y^*) + B_\theta + \frac{2\Pi_\theta}{\kappa_\theta}\mathcal{W}_\theta(y/h) + \Delta \overline{\Theta}^*_\mathrm{log}, \tag{C1\textit{b}} \label{eq:cp_log}
\end{align}
where $\kappa_u, \kappa_\theta$ and $B_u$ are constants, but the offset $B_\theta$ depends on $Pr$~\citep{kader1972heat,kader1981temperature}. The wake profiles $(2\Pi_u/\kappa_u)\mathcal{W}_u(y/h)$ and $ (2\Pi_\theta/\kappa_\theta)\mathcal{W}_\theta(y/h)$ depend on the flow configuration, e.g.\ channel, pipe or boundary layer~\citep{pope2000,nagib2008}. The log-law shifts, $\Delta \overline{U}^*_\mathrm{log}$ and $\Delta \overline{\Theta}^*_\mathrm{log}$, are non-zero for the actuated cases. These log-law shifts are evaluated at $y^* = 170$, where $\Delta \overline{U}^*$ and $\Delta \overline{\Theta}^*$ approximately reach their asymptotic values (figures~\ref{fig:profile_reconstruction} and \ref{fig:profiles_mean}). For fixed operating parameters ($A^+, \kappa^+_x, \omega^+$), $\Delta \overline{U}^*_\mathrm{log}$ is constant for the upstream travelling wave ~\citep{hurst2014,gatti2016,rouhi2022turbulent}. However, $\Delta \overline{\Theta}^*_\mathrm{log}$ strongly depends on $Pr$~(figures~\ref{fig:profiles_mean}\textit{k,l}). We set the constants $\kappa_u = 0.4$, $\kappa_\theta = 0.46$ and $B_u = 5.2$ following \cite{pirozzoli2016passive} and \cite{rouhi2022riblet}, and obtain the offset $B_\theta (Pr) = 12.5 Pr^{2/3} + 2.12 \ln Pr - 5.3$ following \cite{kader1972heat}, which is reported to be accurate for $Pr \gtrsim 0.7$. The suitability of these choices is confirmed in \S\ref{sec:calculate_cf_ch} and Appendix~\ref{sec:grid_domain}. 

The bulk velocity $U^*_b$ and bulk temperature $\Theta^*_b$ are found by integrating (\ref{eq:up_log}) and (\ref{eq:cp_log}), so that
\begin{align}
 U^*_b &= \frac{1}{\kappa_u} \ln(Re_\tau) - \frac{1}{\kappa_u} + B_u + \frac{2\Pi_u}{\kappa_u}\int_0^h\mathcal{W}_u(y/h)dy + \Delta \overline{U}^*_\mathrm{log}, \tag{C2\textit{a}} \label{eq:ubp} \\
 \Theta^*_b &= \frac{1}{\kappa_\theta} \ln(Re_\tau) - \frac{1}{\kappa_\theta} + B_\theta + \frac{2\Pi_\theta}{\kappa_\theta}\int_0^h\mathcal{W}_\theta(y/h)dy +\Delta \overline{\Theta}^*_\mathrm{log}. \tag{C2\textit{b}} \label{eq:cbp} 
\end{align}
By subtracting the non-actuated $U^*_{b_0}$ and $\Theta^*_{b_0}$ from their actuated counterparts $U^*_b$ and $\Theta^*_b$, and by assuming that the wake profiles are not influenced by the wall oscillation, we obtain
\begin{align}
 U^*_b - U^*_{b_0} = \frac{1}{\kappa_u} \ln\left( \frac{Re_\tau}{Re_{\tau_0}} \right) + \Delta \overline{U}^*_\mathrm{log}, \qquad \Theta^*_b - \Theta^*_{b_0} = \frac{1}{\kappa_\theta} \ln\left( \frac{Re_\tau}{Re_{\tau_0}} \right) + \Delta \overline{\Theta}^*_\mathrm{log}. \tag{C3\textit{a,b}} \label{eq:dupb_dcbp}
\end{align}
When $Re_\tau$ is matched between the actuated and the non-actuated cases ($Re_\tau = Re_{\tau_0}$), the first term on the right-hand-side of (\ref{eq:dupb_dcbp}) disappears. However, when the bulk Reynolds number $Re_b$ is matched, as in our work, this term is retained because $Re_\tau \ne Re_{\tau_0}$.
By using (\ref{eq:dupb_dcbp}) and the relations $C_f = 2/{U^*_b}^2, \ C_h = 1/(U^*_b \Theta^*_b), \ C_f/C_{f_0} = R_f = 1-DR$, and $ C_h/C_{h_0} = R_h = 1-HR$, we can relate $\Delta \overline{U}^*_\mathrm{log}$ and $\Delta \overline{\Theta}^*_\mathrm{log}$ to $R_f$ and $R_h$ according to
\begin{align}
 \Delta \overline{U}^*_\mathrm{log} &= \sqrt{\frac{2}{C_{f_0}}}\left[ \frac{1}{\sqrt{R_f}} - 1 \right] - \frac{1}{2\kappa_u}\ln R_f , \tag{C4\textit{a}} \label{eq:dr_model} \\
 \Delta \overline{\Theta}^*_\mathrm{log} &= \frac{\sqrt{C_{f_0}/2}}{C_{h_0}}\left[ \frac{\sqrt{R_f}}{R_h} - 1 \right] - \frac{1}{2\kappa_\theta}\ln R_f . \tag{C4\textit{b}} \label{eq:hr_model}
\end{align}
Equation (\ref{eq:dr_model}) was first derived by \cite{gatti2016} (their equation 4.7). Equation (\ref{eq:hr_model}) is new.

\section{Statistics for the plane oscillation}
\label{sec:plane_stats}

\begin{figure}
  \centering
 \includegraphics[width=1.0\textwidth,trim={{0.0\textwidth} {0.04\textwidth} {0.0\textwidth} {0.0\textwidth}},clip]{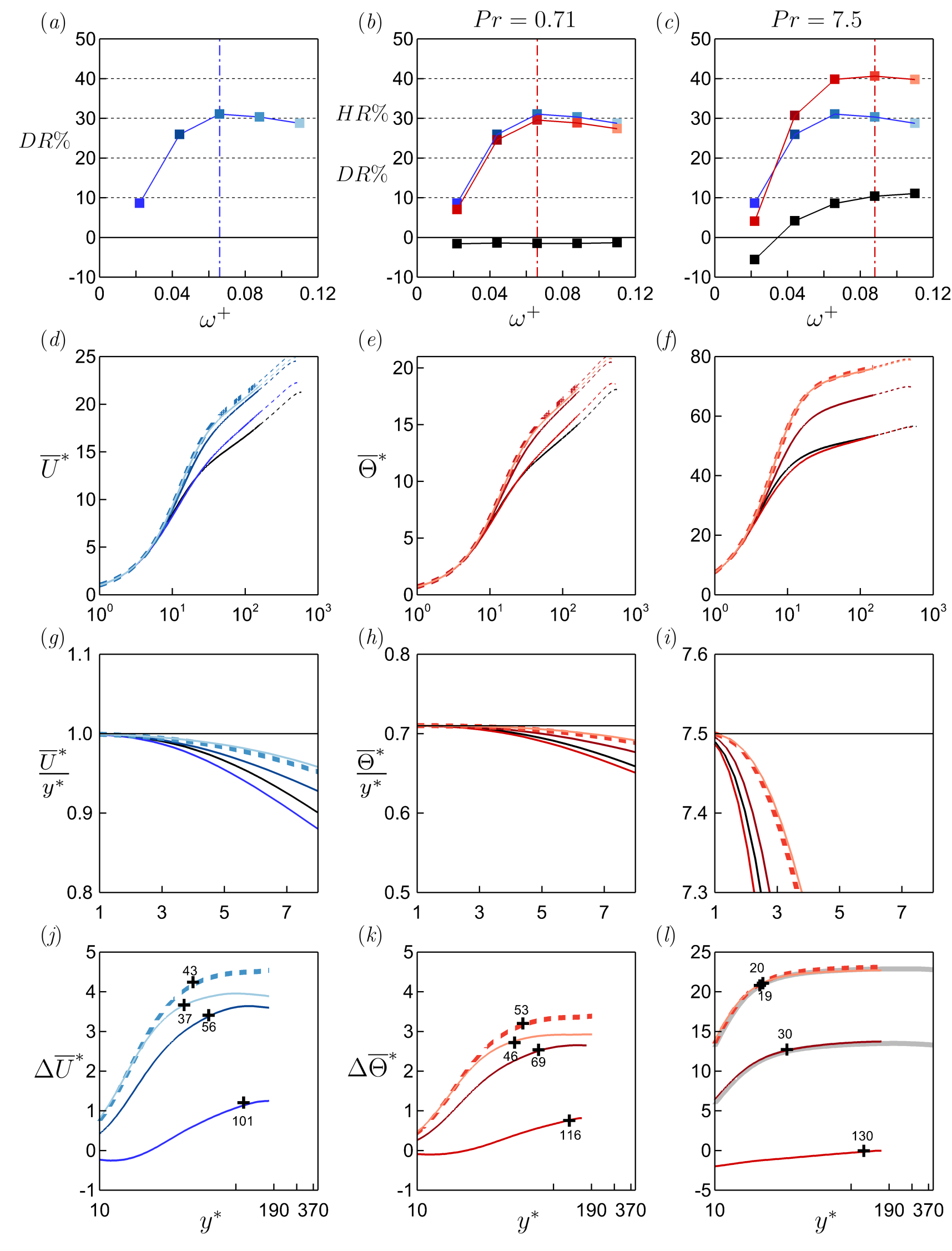}
  \caption{Same plots as in figure~\ref{fig:profiles_mean}, but for the plane oscillation case with $A^+ = 12$.}
  \label{fig:profiles_mean_plane}
\end{figure}

\begin{figure}
  \centering
 \includegraphics[width=1.0\textwidth,trim={{0.04\textwidth} {0.0\textwidth} {0.0\textwidth} {0.0\textwidth}},clip]{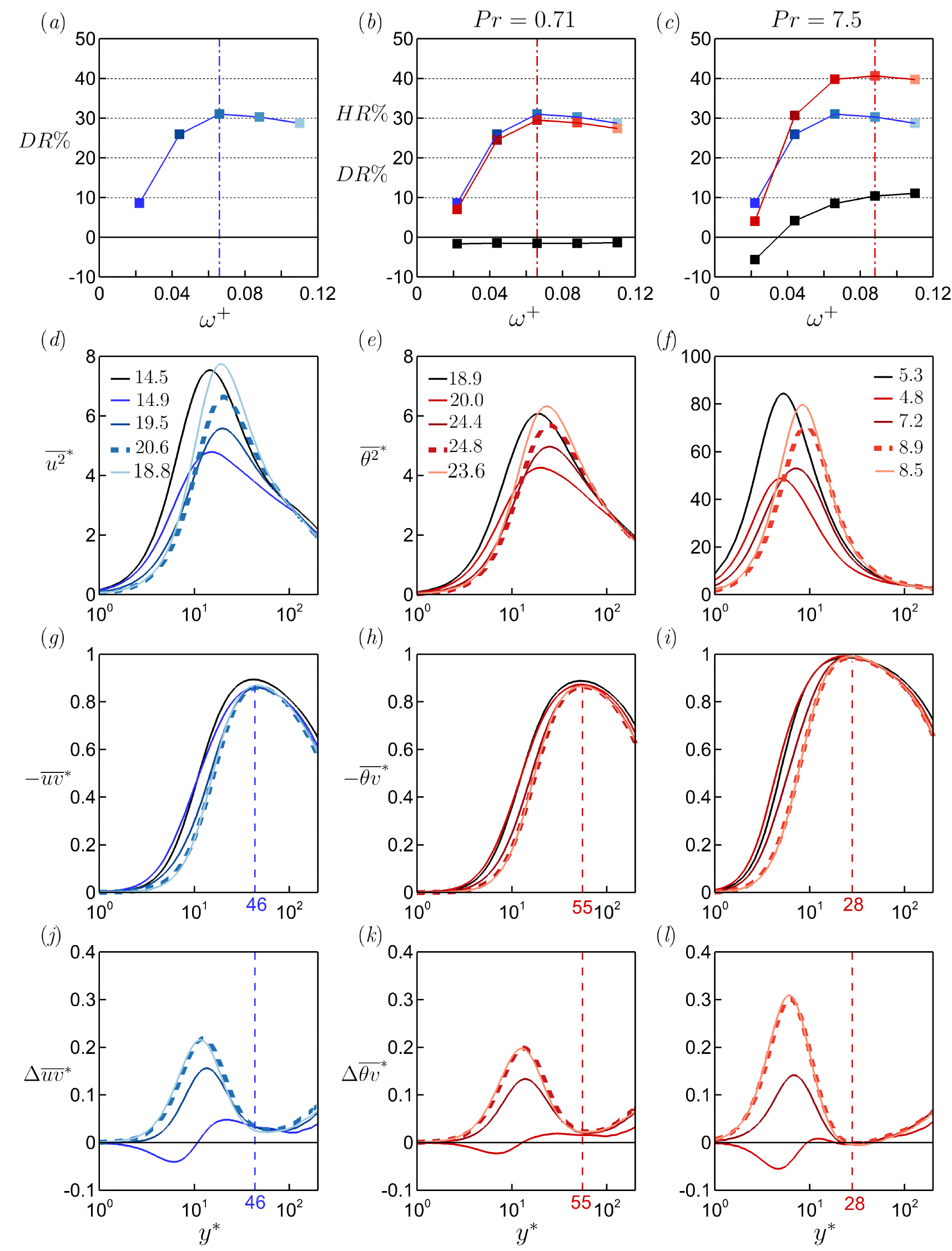}
  \caption{Same plots as in figure~\ref{fig:profiles_stress}, but for the plane oscillation case with $A^+ = 12$.}
  \label{fig:profiles_stress_plane}
\end{figure}

See figures~\ref{fig:profiles_mean_plane} and \ref{fig:profiles_stress_plane}.

\bibliography{references_dst}
\bibliographystyle{jfm}

\end{document}